\documentclass[11pt]{article}
\usepackage{url} 
\usepackage{graphicx}
\usepackage{xypic,epsf,amsmath,amssymb,latexsym,bbm,multicol}
\usepackage[english]{babel}
\usepackage{graphics}
\usepackage{latexsym}
\usepackage{amsmath}
\usepackage{amsfonts}
\usepackage{amssymb}
\usepackage{enumerate}
\usepackage{algorithm}
\usepackage{algorithmic}

\newcommand{\K}{{\mathcal K}}
\newcommand{\T}{{\mathcal T}}
\newcommand{\pK}{\Psi_{\K}}

\newcommand{\alg}{\mathcal} 

\newcommand{\A}{\alg{A}}
\newcommand{\B}{\alg{B}}
\newcommand{\Pred}{{\sf Pred}}

\long\def\ignore#1{}
\newcommand{\QED}{\hspace*{\fill}$\Box$}

\newtheorem{definition}{Definition}
\newtheorem{theorem}{Theorem}
\newtheorem{lemma}[theorem]{Lemma}

\newtheorem{remark}[theorem]{Remark}
\newtheorem{example}{Example}

\begin{document}
\setcounter{page}{1}

\title{\bf Parametric Systems: Verification and Synthesis} 

\author{Viorica Sofronie-Stokkermans \\ 
University Koblenz-Landau, Germany \\ 
sofronie@uni-koblenz.de } 
\date{}
\maketitle


\begin{abstract}
 In this paper we study possibilities of using 
hierarchical reasoning, symbol elimination and model generation  
for the verification of parametric  systems, 
where the parameters can be constants or functions. 
Our goal is to automatically provide guarantees 
that such systems satisfy certain safety or invariance conditions.
We analyze the possibility of automatically generating 
such guarantees in the form of constraints on parameters. 
We illustrate our methods on several examples.  
\end{abstract}


\section{Introduction}
Most of the applications in verification 
require reasoning about complex domains.  In this paper we 
identify several classes of verification problems 
for parametric reactive systems 
(modeled by transition constraints) and for 
several classes of parametric hybrid systems,  
and point out the reasoning tasks in the associated 
theories which need to be solved. 
The type of parametricity we consider   
refers to parametric data (including parametric change and environment)
specified using functions with certain properties and parametric topology,  
specified using data structures.

The first problem we address is to check whether a safety property
-- expressed by a suitable formula -- is an invariant, 
or holds for paths of bounded length, {\em for 
given instances of the parameters}, or 
{\em under given constraints on parameters}. 
For this type of problems, we aim at identifying situations in 
which decision procedures 
exist. We show that this is often 
the case, by investigating consequences of locality phenomena in verification.
If unsafety is detected, the method we use allows us to generate 
counterexamples to safety, i.e.\ concrete system descriptions 
which satisfy all the imposed constraints and are unsafe. 

We also analyze the dual problem -- related to system synthesis --  
of {\em deriving constraints between parameters} which guarantee that 
a certain safety property is an invariant of the system or holds 
for paths of bounded length. Such problems were studied before 
for the case when the parameters are constants 
\cite{Henzinger,frehse,farn-wang,Platzer09,sofronie-entcs}.  
We present a new approach which can be used also in the case when some 
of the parameters are allowed to be functional and show that sound and 
complete hierarchical reduction for SMT checking in local extensions 
allows to reduce the problem of checking that certain formulae are 
invariants to testing the satisfiability of certain formulae w.r.t.\
a standard theory. 
Quantifier elimination is used for generating constraints 
on the parameters of the system (be they data or functions) which 
guarantee safety. 
These constraints on the parameters 
may also be used to solve optimization problems (maximize/minimize some 
of the parameters) such that safety is guaranteed.
If we also express invariants in a parametric form, 
this method can also be used for identifying conditions which 
guarantee that formulae with a certain 
shape are invariants, and ultimately for generating invariants with 
a certain shape. 
There exist approaches to the verification of parametric reactive infinite 
state systems and timed automata 
(e.g.\ by Ghilardi et al. \cite{Ghilardi}, Hune et al. \cite{HuneRomijn}, Cimatti et al.\ \cite{Cimatti2}) 
and for parametric hybrid automata (e.g.\
by Henzinger et al. \cite{Henzinger}, 
Frehse, \cite{frehse}, Wang \cite{farn-wang}, and Cimatti et 
al. \cite{Cimatti}), but in most cases only situations in which the 
parameters are constants were considered. The idea of using hierarchical 
reasoning and quantifier elimination for obtaining constraints on 
the parameters (constants or functions) was first used in
\cite{sofronie-ijcar2010} and \cite{sofronie-cade13}. In this paper we
present the results in 
\cite{sofronie-ijcar2010} and \cite{sofronie-cade13} in a common
framework 
and extend them.

\smallskip
\noindent 
{\bf Structure of the paper.} The paper is structured as follows: In Section~\ref{local} we present 
existing results on local theory extensions which allow us to identify 
decidable theories interesting in verification. In Section~\ref{verif}
we present the type of systems we consider, namely parametric systems described
by transition constraint systems and parametric hybrid automata, and the verification problems
considered in this paper: invariant checking and bounded model
checking. 
In Section~\ref{sect-tcs} we identify situations in which 
decision procedures 
exist for invariant checking and bounded model checking of systems
modeled using transition constraints, as well 
as methods for obtaining constraints between the parameters which guarantee 
that certain properties are invariants. In Section~\ref{sect-ha} we
study similar problems for some classes of
parametric hybrid automata. Some ideas on defining and verifying
interconnected parametric hybrid automata are presented in Section~\ref{iha}.
In Section~\ref{conclusions} 
we draw conclusions. 

\subsection{Idea}
\label{illustration}

We illustrate the problems and the main ideas described in the paper on the following examples: 

\smallskip

\begin{example}
\label{ex1}
{\em Consider a discrete water level controller in which 
the inflow (${\sf in}$) 
in the interval of time for one step in the evolution of the system
is fixed.
 If the water level becomes greater than an alarm 
level $L_{\sf alarm}$  (below the overflow level $L_{\sf overflow}$) a valve is opened 
and a fixed quantity of water (${\sf out}$) is left out. 
Otherwise, the  valve remains closed. We want to check whether, assuming that we start from a state in which 
the water level $L$ satisfies $L \leq L_{\sf overflow}$, $L$ always
remains below $L_{\sf overflow}$. 

\

\

{\footnotesize {\noindent

\hspace{2.8cm}
\leavevmode
\epsfverbosetrue %
\def\epsfsize#1#2{0.35#1}              
~~~~~~~~~\epsffile{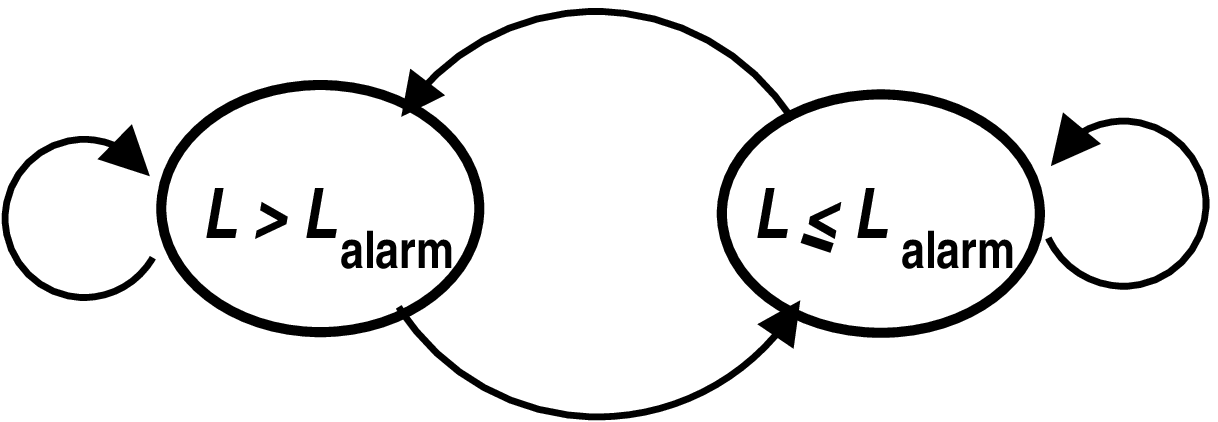} 
}

\medskip

\vspace{-2.2cm}
{\footnotesize \noindent $\begin{array}{lcl}
~~~~~~~~~~~~~& \!\!L' {:=} L {+} {\sf in} & \\[5ex]
~~~~~~~~~~~~~L' {:=} L {+} {\sf in} {-} {\sf out} ~~~& ~~~~~~~~~~~~~~~~~~~~~~~~~~~~~~~~~~~~~& L' {:=} L {+} {\sf in} \\[6ex]
~~~~~~~~~~~~~& \!\!\!\!\!\!\!\!L' {:=}  L {+} {\sf in} {-} {\sf out} & \\
\end{array}$ 
}}

\

\smallskip
\noindent 
Let ${\cal T}_S$ be ${\mathbb R}$, the theory of real
numbers\footnote{The theory of real numbers we consider here is the
  theory of the field of real numbers, i.e.\ the theory of real
  closed fields. Multiplication is denoted by $*$.}.
Assume that a set $\Gamma$ of  constraints 
on the parameters is given, e.g.\  
$\Gamma = \{ {\sf in} = {\sf out} {-} 10, {\sf in}  = L_{\sf overflow} $ 
${-} L_{\sf alarm} {-} 10, {\sf in} > 0, {\sf out} > 0, L_{\sf alarm}
< L_{\sf overflow} \}$. 
%
Then $L {\leq} L_{\sf overflow}$ is an inductive invariant iff $L \leq
L_{\sf overflow}$ holds in the initial state and the 
formulae (i), (ii) are unsatisfiable w.r.t.\ ${\cal T}_S {\cup} \Gamma$:
\begin{itemize}
\item[(i)] $\exists L, L' (L \leq L_{\sf overflow} \wedge  L > L_{\sf alarm} \wedge L' = L + {\sf in} - {\sf out} \wedge L' > L_{\sf overflow})$; 
\item[(ii)] $\exists L, L' (L \leq L_{\sf overflow} \wedge L \leq L_{\sf alarm} \wedge L' = L + {\sf in} \wedge L' > L_{\sf overflow})$.
\end{itemize}
It is easy to check that formulae (i) and (ii) above 
are unsatisfiable w.r.t.\  ${\cal T}_S \cup \Gamma$ 
by using a decision procedure for the theory of real numbers.

Assume now that fewer constraints 
on the parameters of the system are specified, i.e.\  that 
$\Gamma = \{ {\sf in} > 0, {\sf out} > 0, L_{\sf alarm} < L_{\sf  overflow} \}$. 
We still know that the safety condition 
is an invariant under updates iff the formulae in (i), (ii) are unsatisfiable w.r.t.\ 
${\cal T}_S$. We can eliminate the existentially quantified variables 
$L, L'$ using a method for quantifier elimination in ${\mathbb R}$
and thus show that 
(under the assumption that $L_{\sf alarm} < L_{\sf overflow}$ and ${\sf in} > 0$)  
the formula in (i) is equivalent to $({\sf in} > {\sf out})$ and 
the formula in (ii) is equivalent to 
$({\sf in} >  L_{\sf overflow} - L_{\sf alarm})$.
We can therefore conclude 
(under the assumption that $L_{\sf alarm} < L_{\sf overflow}$, ${\sf
  in} > 0$ and that
in the initial state $L \leq L_{\sf overflow}$)  
that $L \leq L_{\sf overflow}$ is an inductive invariant iff 
$({\sf in} \leq {\sf out}) ~ \wedge ~
({\sf in}  \leq L_{\sf overflow} {-} L_{\sf alarm})$.
}
\end{example}

\smallskip

\begin{example}
\label{ex2}
{\em 
Consider a variant of Example~\ref{ex1} in which the inflow varies in time. In all transitions, we will 
therefore replace ${\sf in}$ by ${\sf in}(t)$ (representing the inflow
between moment $t$ and moment $t+1$) and add the time change 
$t' = t + 1$. We have two choices for the theory ${\cal T}_S$: We can
choose 
${\cal T}_S = {\mathbb R} \cup {\mathbb Z}$, the many-sorted
combination of the theory of reals and integers (for modeling time) 
if time is considered to be discrete or ${\cal T}_S = {\mathbb R}$ if time is
considered to be continuous. Let ${\cal T}_S^{\sf \{ in, out \}}$ be
the combination of ${\cal T}_S$ with the uninterpreted function symbols
${\sf in}$ (unary) and ${\sf out}$ (a constant)). 

Assume that we describe the initial states 
using the formula ${\sf Init}(L) := L_a {\leq} L {\leq} L_b$,
where $L_a, L_b$ are parameters with $L_a {<} L_b$.
Then $L \leq L_{\sf overflow}$ is an inductive invariant 
iff the following formulae are unsatisfiable w.r.t.\ ${\cal T}_S$
resp.\ ${\cal T}_S^{\sf \{ in, out \}}$:  
\begin{enumerate}
\item[(1)] $\exists L (L_a \leq L \leq L_b \wedge  L > L_{\sf
    overflow})$ 
\item[(2)] Safety is invariant under transitions, i.e.\ the following
  formulae are unsatisfiable:
\begin{itemize}
\item[(i)] $\exists L, L', t, t' (L \leq L_{\sf overflow} \wedge  L >
  L_{\sf alarm} \wedge L' {=} L {+} {\sf in}(t) {-} {\sf out} \wedge
  t' {=} t {+} 1 \wedge L' {>} L_{\sf overflow})$; 
\item[(ii)] $\exists L, L', t, t' (L \leq L_{\sf overflow} \wedge L \leq L_{\sf alarm} \wedge L' {=} L
  {+} {\sf in}(t) \wedge t' {=} t {+} 1 \wedge L' {>} L_{\sf
    overflow})$.
\end{itemize}
\end{enumerate}
Under the assumption that $L_a < L_b$ we can prove (using 
quantifier elimination in the theory of reals \cite{tarski}) that 
(1) is unsatisfiable iff $L_{\sf overflow} < L_b$ is false, i.e.\ iff $L_b \leq
L_{\sf overflow}$ holds. 

\noindent It is not immediately clear how to eliminate the quantifiers 
in the formulae in (2)(i) and (2)(ii) because of the occurrences 
of the function ${\sf in}$. In this paper we identify situations 
in which the satisfiability problems can be reduced, in a sound and 
complete way, to satisfiability problems over the base theory, 
by using locality properties of these theories.
Locality allows us to perform a reduction to satisfiability checks w.r.t.\ 
${\mathbb R} \cup {\mathbb Z}$ (if we consider time to be discrete)
resp.\ ${\mathbb R}$ (for continuous time),  where we can eliminate all quantified 
variables except for the parameters and the 
variables which stand for arguments of the parametric functions;  
we then interpret the result back in the theory extension. 
This way we prove that: 
\begin{itemize}

\item (2)(i) holds iff
$\forall t ({\sf in}(t) {-} {\sf out} \leq 0)$  holds, and 
\item (2)(ii) holds iff 
$\forall t ({\sf in}(t)  \leq L_{\sf overflow}-L_{\sf alarm})$ holds. 
\end{itemize}
}
\end{example}

\smallskip

\begin{example}
\label{ex3}
{\em 
We can also model the water tank controller as a hybrid system, 
with two discrete states $s_1, s_2$ (state invariants 
$L \geq L_{\sf alarm}$ and $L < L_{\sf alarm}$) and changes 
described by jumps between these states and flows within each 
state.

\begin{center}
{\noindent
\epsfverbosetrue %
\def\epsfsize#1#2{0.54#1}              
~~~~~~~~~~~~~~\epsffile{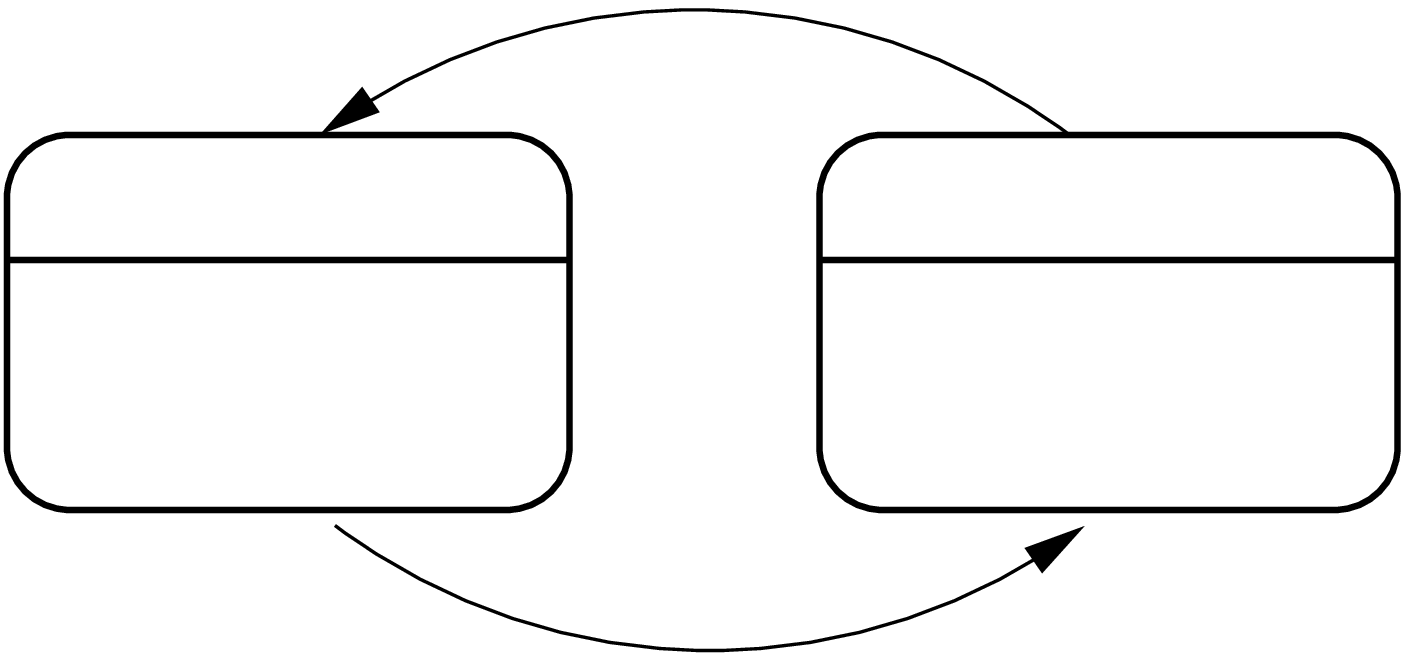} 
}

\vspace{-2.8cm} 
$~~~~~~~~~~~s_1 \quad L \geq L_{\sf alarm} ~~ \quad \quad \quad \quad
~~ s_2 \quad  L <
L_{\sf alarm}$

\vspace{2mm}
{\scriptsize 
$\quad  \quad \quad  \quad  \quad ~~~~\begin{array}{@{}lll}
~~~\dot{L} = {\sf in} {-} {\sf out} & & ~\dot{L} = {\sf in} \\
~~~\dot{{\sf infl}}(t) = {\sf in} & \quad  \quad \quad \quad  \quad \quad  \quad & ~\dot{{\sf infl}}(t) = {\sf in}\\
~~~\dot{{\sf outfl}}(t) = {\sf out} & & ~{\sf out}(t) = 0; \dot{{\sf  outfl}}(t) = 0
\end{array}$
}

\vspace{-3.2cm} $~~~~~~~~~~~~L \geq L_{\sf alarm}$

\vspace{3.8cm} $~~~~~~~~~~~~L < L_{\sf alarm}$
\end{center}


\noindent We model inflow and outflow by functions ${\sf infl}, {\sf outfl}$, where
${\sf infl}(t)$ (${\sf outfl}(t)$) is the inflow (resp.\ outflow) in time $t$.  
Assume that the inflow and outflow rates are constant 
and equal to ${\sf in}$, resp.\ ${\sf out}$ (i.e.\ the derivative of
${\sf infl}$ is equal to ${\sf in}$ at every point in time $t$ and the derivative of
${\sf outfl}$ is equal to ${\sf out}$ at every point in time
$t$). Clearly, at time $t$ in state $s_1$ (resp. $s_2$) 
the level is $L' = L + ({\sf in} - {\sf out})*t$ (resp.\ $L' = L +
{\sf in}*t$) where $L$ is the level at moment 0 in that state. 

\noindent Let ${\cal T}_S$ be the theory of real numbers. The problems we consider are:  
\begin{itemize}
\item[(1)] Check whether the safety condition $\Psi = L \leq L_{\sf overflow}$ is invariant 
(under jumps and flows), assuming that ${\sf in}, {\sf out}$ satisfy certain 
given properties. 
\item[(2)] Generate conditions on the parameters which guarantee that 
$\Psi$ is invariant.
\end{itemize}
$L \leq L_{\sf overflow}$ is invariant under flows iff the following
formulae are unsatisfiable w.r.t.\ ${\cal T}_S$: 
\begin{itemize}
\item[(i)] $\exists L, t (L {\leq} L_{\sf overflow} {\wedge} 0 {<} t \wedge L {<} L_{\sf alarm} {\wedge} \forall t' (0 {\leq} t' {\leq} t {\rightarrow} L {+} {\sf in}*t' < L_{\sf alarm}) \wedge L {+} {\sf in}*t > L_{\sf overflow})$,
\item[(ii)]  $\exists L, t (L {\leq} L_{\sf overflow} {\wedge} 0 {<} t
  \wedge L {\geq} L_{\sf alarm} {\wedge} \forall t'( 0 {\leq} t'
  {\leq} t {\rightarrow} L {+} {\sf in}'{*}t' \geq L_{\sf alarm})
  {\wedge} L {+} {\sf in}'{*}t > L_{\sf overflow})$,
\end{itemize}
where in (ii) ${\sf in}'$ is used as an abbreviation for ${\sf in} - {\sf
  out}$. These are formulae with alternations of quantifiers. 
Task (1) can be solved using a decision procedure for the satisfiability of the 
$\exists \forall$ fragment of the theory of reals, task (2) uses quantifier elimination. In Section~\ref{sect-ha} we will 
present this situation in detail: We will discuss the case when the 
evolution rules in a state are specified by giving bounds on the rate of 
growth of the continuous variables, then look into possibilities of
approximating parametric hybrid automata and to the verification of
systems of hybrid automata. 
}
\end{example}

\section{Decision problems in complex theories}
\label{local}
In this section we analyze a class of theories used for 
modeling reactive, real time and hybrid systems for 
which we can obtain decidability results. 

\subsection{Logic: Preliminaries}

We consider signatures of the form
$\Pi = (\Sigma, {\sf Pred})$ or many-sorted signatures of the form 
$\Pi = (S, \Sigma, {\sf Pred})$, 
where $S$ is a set of sorts, $\Sigma$ is a family of function symbols and ${\sf Pred}$
a family of predicate symbols. 
If $\Pi$ is a signature and $C$ is a set of new constants, we will denote 
by $\Pi^C$ the expansion of $\Pi$ with constants in $C$, i.e.\ 
the signature $\Pi^C = (\Sigma \cup C, {\sf Pred})$.

We assume known standard definitions from first-order logic  
such as terms, atoms, formulae, $\Pi$-structures, logical entailment,  
model, satisfiability, unsatisfiability.  
A literal is an atom or the negation of an atom; a clause is a
(finite) disjunction of literals. In this paper we refer to (finite) conjunctions of
clauses also as ``sets of clauses'', and to (finite) conjunctions of formulae 
as ``sets of formulae''. Thus, if $N_1$ and $N_2$ are finite sets of
formula then $N_1 \cup N_2$ 
will stand for the conjunction of all formulae in $N_1 \cup N_2$. 
All free variables of a clause (resp. of a set
of clauses) are considered to be implicitly universally quantified. 
We denote ``verum'' with $\top$ and ``falsum'' with $\perp$. $\perp$ is also a notation 
for the empty clause. 

\medskip
\noindent First-order theories are sets of formulae (closed under logical consequence), 
typically all consequences of a set of axioms. 
Alternatively, we may consider a set of models which defines a
theory.
Theories can be defined by specifying a set of axioms, or by specifying a
set of structures (the models of the theory). 
In this paper, (logical) theories are simply sets of sentences.

\begin{definition}[Entailment] 
 If $F, G$ are formulae and ${\mathcal T}$ is a theory we 
write: 
\begin{enumerate}
\vspace{-2mm}
\item $F \models G$ to express the fact that every model of $F$ is a
  model of $G$; 
\vspace{-2mm}
\item $F \models_{\mathcal T} G$ -- also written as ${\mathcal T} \cup F
  \models G$ and sometimes ${\mathcal T} \wedge F \models G$ -- 
to express the fact that every model of $F$ which is also a model of 
$\T$ is a model of $G$.
\end{enumerate}
\noindent If $F \models G$ we say that {\em $F$ entails $G$}. If $F \models_{\mathcal T}
 G$ we say that {\em $F$ entails $G$ w.r.t.\ ${\mathcal T}$}. 
$F \models \perp$ means that $F$ is
unsatisfiable; $F \models_{\T} \perp$ means that there is no model of
$\T$ in which $F$ is true. 
If there is a model of $\T$ which is also a
model of $F$ we say
that $F$ is satisfiable w.r.t.\ ${\mathcal T}$. 
If $F \models_{\mathcal T} G$ and $G \models_{\mathcal T} F$ we say that
 {\em $F$ and $G$ are equivalent w.r.t.\ ${\mathcal T}$}.
\end{definition}
\begin{definition}
A theory $\T$ over a signature  
$\Pi$ {\em allows quantifier elimination} if for every formula $\phi$ over  
$\Pi$ there exists a quantifier-free formula $\phi^*$ over  
$\Pi$ which is equivalent to $\phi$ modulo $\T$.  
\end{definition} 
\begin{example}
{\em Presburger arithmetic with congruence modulo $n$,  
rational linear arithmetic $LI({\mathbb Q})$ and real linear
arithmetic $LI({\mathbb R})$, the theories of real closed fields (real
numbers) and 
of algebraically closed fields, the theory of finite fields, the
theory of absolutely free algebras, and the 
theory of acyclic lists in the signature 
$\{ {\rm car}, {\rm cdr}, {\rm cons} \}$
(\cite{tarski,malcev,chang-keisler,hodges,ghilardi:model-theoretic-methods})
allow quantifier elimination.
} 
\label{examples-qe}
\end{example}
\medskip

\subsection{Theories, theory extensions}
Let ${\mathcal T}_0$ be a theory with signature 
$\Pi_0 = (S, \Sigma_0, {\sf Pred})$, where $S$ is a set of sorts, 
$\Sigma_0$ a set of function symbols, and ${\sf Pred}$ 
a set of predicate symbols. 
We consider extensions
${\mathcal T}_1 = {\cal T}_0 \cup {\cal K}$ of ${\mathcal T}_0$ 
with signature $\Pi = (S, {\Sigma}, {\sf Pred})$,
where $\Sigma = \Sigma_0 \cup \Sigma_1$
(i.e.\ the 
signature is extended by new function symbols $\Sigma_1$ whose properties are 
axiomatized by a set ${\cal K}$ of formulae). We consider two cases: 
\begin{itemize}
\item ${\mathcal K}$ consists of clauses $C(x_1, \dots, x_n)$ over the
  signature $\Pi$ containing function symbols in
  $\Sigma_1$.
\item ${\mathcal K}$ consists of {\em augmented clauses}, i.e.\ of axioms 
of the form 
$(\Phi(x_1, \dots, x_n) \vee C(x_1, \dots, x_n))$, 
where $\Phi(x_1, \dots, x_n)$ is an {\em arbitrary first-order 
formula} in the base signature $\Pi_0$ 
and $C(x_1, \dots, x_n)$ is a {\em clause} containing $\Sigma_1$-functions. 
\end{itemize}
The free variables $x_1, \dots, x_n$ are considered to be
 universally quantified. 
 
\medskip
\noindent In what follows, we will consider axiomatizations with sets of
clauses (i.e.\ conjunctions of implicitly universally quantified
clauses), 
for the sake of simplicity. However, most results hold 
for axiomatizations by means of augmented clauses -- this is the 
reason why we here provided both definitions. 

\noindent We will call a clause containing extension functions (i.e.\ functions in
$\Sigma_1$) an {\em extension clause}. 

\subsection{Locality of an extension}
The notion of locality for theory extensions was introduced in 
\cite{Sofronie-cade-05,GSW-iandc-06}.
Let $\Pi_0 {=} (\Sigma_0, {\sf Pred})$ be a signature, and ${\mathcal T}_0$ be a 
``base'' theory with signature $\Pi_0$. 
We consider 
extensions $\T := {\mathcal T}_0 \cup \K$
of ${\mathcal T}_0$ with new function symbols $\Sigma_1$
({\em extension functions}) whose properties are axiomatized using 
a set $\K$ of (universally closed) clauses 
in the extended signature $\Pi = (\Sigma, {\sf Pred})$, where $\Sigma
= \Sigma_0 \cup \Sigma_1$, 
which contain function symbols in $\Sigma_1$. 
Let $C$ be a fixed countable set of fresh constants. We denote by 
$\Pi^C$ the expansion of $\Pi$ with constants in $C$. 

If $G$ is a finite set of ground $\Pi^C$-clauses and $\K$ a set of
$\Pi$-clauses, we denote by  ${\sf st}({\mathcal K}, G)$ the set of all 
ground terms which occur in $G$ or ${\mathcal K}$. We denote by 
${\sf est}({\mathcal K}, G)$ the set of all extension ground terms (i.e.\ 
terms starting with a function in $\Sigma_1$) which occur in $G$ or
${\mathcal K}$.
In what follows, a finite set of formulae is regarded as the
conjunction of its elements; in particular 
we regard every finite set $G$
of ground clauses as the ground formula $\bigwedge_{{\cal C} \in G}
{\cal C}$. If $T$ is a set of ground terms in the signature  $\Pi^C$, 
we denote by $\K[T]$ the set of all instances of $\K$ in which the terms 
starting with a function symbol in $\Sigma_1$ are in $T$. 
Formally: 
\begin{align*} 
\K[T] := \{ \varphi\sigma \,|\; & \forall \bar{x}.\,
\varphi(\bar{x}) \in \K, 
\text{ where (i) if } f \in \Sigma_1 \text{ and } t = f(t_1,...,t_n)
\text{ occurs in } \varphi\sigma\\[-1ex]                                     
& \text{ then } t \in T \text{; (ii) if } x
\text{ is a variable that does not appear below some } \\[-1ex]
& \text{ $\Sigma_1$-function in } \varphi \text{ then } \sigma(x) = x \}.
\end{align*}
For any set $G$ of ground $\Pi^C$-clauses we write 
$\K[G]$ for $\K[{\sf est}({\mathcal K}, G)]$.

\medskip
\noindent We focus on the following types of locality of 
an extension 
${\mathcal T}_0 \subseteq {\mathcal T}_1 = {\mathcal T}_0 \cup {\mathcal K}$ with ${\mathcal K}$ 
a set of clauses (resp.\ augmented clauses for ${\sf (ELoc)}$). 

\medskip
\noindent
\begin{tabular}{@{}ll}
${\sf (Loc)}$  & For every finite set $G$ of $\Pi^C$-ground clauses 
                     ${\mathcal T}_1 {\cup} G \models \perp$ iff 
                     ${\mathcal T}_0 {\cup} {\mathcal K}[G] {\cup} G \models \perp$ \\
${\sf (ELoc)}$  & For every formula $F = F_0 \cup G$,
where $F_0$ is a finite set of $\Pi_0^C$-sentences and \\
&  $G$ is a finite set of ground $\Pi^C$-clauses, 
${\cal T}_1 \cup F_0 \cup G \models \perp$ iff 
${\cal T}_0 \cup {\cal K}[G] \cup F_0 \cup G \models \perp$
\end{tabular}

\medskip
\noindent 
\begin{definition}[\cite{Sofronie-cade-05,GSW-iandc-06}]
We say that an extension ${\mathcal T}_0 \subseteq {\mathcal T}_1$ is 
{\em local} if it satisfies condition ${\sf (Loc)}$. We refer to 
condition ${\sf (ELoc)}$ as {\em extended locality} condition.
\end{definition}

\medskip
\noindent 
The more general notions such as $\Psi$-locality and $\Psi$-extended locality 
of a theory extension were introduced in 
\cite{ihlemann-jacobs-sofronie-tacas08} to encompass situations 
in which the instances to be considered are described 
by a closure operation $\Psi$.

\begin{definition}[\cite{ihlemann-sofronie-ijcar10}]
\label{def-psi-loc}
Let $\Psi$ be a map associating with 
every set $T$ of ground $\Pi^C$-terms a set $\Psi(T)$ of ground $\Pi^C$-terms. 
For any set $G$ of ground $\Pi^C$-clauses we write 
$\K[\Psi_{\mathcal K}(G)]$ for $\K[\Psi({\sf est}({\mathcal K}, G))]$.
%
Let $\T_0 \cup \K$ be an extension of $\T_0$ 
with clauses in ${\mathcal K}$. 
We say that ${\mathcal K}$ is $\Psi$-local 
if it satisfies: 

\medskip
\noindent \begin{tabular}{@{}l@{}l}
$({\sf Loc}^{\Psi})$ & 
For every finite set $G$ of ground clauses, ${\cal T}_0 \cup {\mathcal K} \cup G \models \perp$ iff 
${\cal T}_0 \cup {\mathcal K}{[\Psi_{\mathcal K}(G)]} \cup G \models \perp$. 
\end{tabular}

\medskip
\noindent Condition $({\sf ELoc}^{\Psi})$ is defined analogously. If
${\Psi}({\sf est({\mathcal K}, G)}) = {\sf est}({\mathcal K}, G)$ 
we recover the notions $({\sf Loc})$ resp.\ $({\sf ELoc})$.
\end{definition} 

\subsection{Partial Structures} 

\noindent In \cite{Sofronie-cade-05} we showed that 
local theory extensions can be 
recognized by showing that certain partial models embed into total
ones, and in \cite{ihlemann-sofronie-ijcar10} we established similar
results for $\Psi$-local theory extensions and generalizations
thereof. 
We introduce the main  
definitions here.

Let $\Pi = (\Sigma, {\sf Pred})$ be a
first-order signature with set of function symbols $\Sigma$ 
and set of predicate symbols ${\sf Pred}$. 
A \emph{partial $\Pi$-structure} is a structure 
$\A = (A, \{f_\A\}_{f\in\Sigma}, \{P_\A\}_{P\in \Pred})$, 
where $A$ is a non-empty set, for every $n$-ary $f \in \Sigma$, 
$f_\A$ is a partial function from $A^n$ to $A$, and for every $n$-ary 
$P \in {\sf Pred}$, $P_\A \subseteq A^n$. We consider constants (0-ary functions) to be always 
defined. $\A$ is called a \emph{total structure} if the 
functions $f_\A$ are all total. 
Given a (total or partial) $\Pi$-structure $\A$ and $\Pi_0 \subseteq \Pi$ 
we denote the reduct of 
$\A$ to $\Pi_0$ by $\A{|_{\Pi_0}}$.

The notion of evaluating a term $t$ with variables $X$ w.r.t.\ 
an assignment $\beta : X \rightarrow A$
 for its variables in a partial structure
$\A$ is the same as for total algebras, except that the evaluation is
undefined if $t = f(t_1,\ldots,t_n)$ 
and at least one of
$\beta(t_i)$ is undefined, or else $(\beta(t_1),\ldots,\beta(t_n))$ is
not in the domain of $f_\A$.

\begin{definition}
A \emph{weak $\Pi$-embedding} between two partial $\Pi$-structures
$\A$ and $\B$, where 
$$\A = (A, \{f_\A\}_{f\in \Sigma}, \{P_\A\}_{P \in \Pred}) \text{ and
} \B = (B, \{f_\B\}_{f\in \Sigma}, \{P_\B\}_{P \in \Pred})$$
is a total map $\varphi : A \rightarrow B$ such that 
\begin{enumerate}
\item[(i)] $\varphi$ is
an embedding w.r.t.\ ${\sf Pred} \cup \{ = \}$, i.e.\ 
 for every $P \in {\sf Pred} \cup \{ = \}$ with arity $n$ and every 
$a_1, \dots, a_n \in \A$, 
$(a_1, \dots, a_n) \in P_\A$ if and only if $(\varphi(a_1), \dots, \varphi(a_n))\in P_\B$; 
\item[(ii)]  
whenever $f_\A(a_1, \dots, a_n)$ is defined (in $\A$),  then 
$f_\B(\varphi(a_1), \dots, \varphi(a_n))$ is defined (in $\B$) and 
$\varphi(f_\A(a_1, \dots, a_n)) = f_\B(\varphi(a_1), \dots, \varphi(a_n))$,
for all $f \in \Sigma$. 
\end{enumerate}
\end{definition} 
%

\begin{definition}[Weak validity]
Let $\A$ be a partial $\Pi$-algebra
and $\beta : X {\rightarrow} A$ a valuation for its variables.
$(\A, \beta)$ {\em weakly satisfies a clause  $C$} (notation: 
$(\A, \beta) \models_w C$) if either some of the literals in 
$\beta(C)$ are not defined or otherwise all literals are defined and
for at least one literal $L$ in $C$, $L$ is true in $\A$ w.r.t.\ $\beta$. 
$\A$ is a {\em weak partial model} 
of a set of clauses ${\mathcal K}$ if $(\A, \beta)  \models_w C$ for every 
valuation 
$\beta$ and every clause $C$ in ${\mathcal K}$. 
\end{definition}

\subsection{Recognizing $\Psi$-Local Theory Extensions} 
In \cite{Sofronie-cade-05} we proved that if 
every weak partial model of an extension 
${\mathcal T}_0 \cup {\mathcal K}$ of a base theory ${\mathcal T}_0$ 
with total base functions can be embedded into a total
model of the extension, then the extension is local. 
In \cite{ihlemann-jacobs-sofronie-tacas08} we lifted these results to $\Psi$-locality. 

Let $\alg{A} = (A, \{ f_{\A} \}_{f \in \Sigma_0 \cup \Sigma_1} \cup C, \{ P_\A
\}_{P \in {\sf Pred}})$ be a partial $\Pi^C$-structure with total
$\Sigma_0$-functions. 
Let $\Pi^A$ be the extension of the signature $\Pi$ with constants 
from $A$. We denote by $T(\A)$ the following set of ground 
$\Pi^A$-terms: 
$$T(\A) := \{ f(a_1,...,a_n) \,|\; f \in \Sigma, a_i \in A, i=1,\dots,n, f_{\A}(a_1,...,a_n) \text{ is defined }  \}. $$
Let ${\sf PMod}_{w,f}^\Psi({\Sigma_1}, {\mathcal T})$ be the class of all
weak partial models $\A$ of ${\mathcal T}_0 \cup {\mathcal K}$, such that
$\A{|_{\Pi_0}}$ is a total model of $\T_0$, the
$\Sigma_1$-functions are possibly partial, $T(\A)$ is finite and 
all terms in $\Psi({\sf est}(\K, T(\A)))$ are
defined (in the extension $\A^A$ with
constants from $A$).
We consider the following embeddability property of partial algebras:

\medskip
\begin{tabbing}
\= $({\sf Emb}_{w,f}^\Psi)$ \quad \= Every $\alg{A} \in {\sf PMod}_{w,f}^\Psi({\Sigma_1},  \T)$ weakly embeds into a total model of $\T$.
\index{$\Psi$-!embeddability}
\end{tabbing}

\medskip
\noindent 
We also consider the properties $({\sf EEmb}_{w,f}^{\Psi})$,  
which additionally requires the embedding to be {\em elementary} and 
$({\sf Comp}_f)$  which requires that every  structure $\alg{A} \in {\sf
  PMod}_{w,f}^\Psi({\Sigma_1}, \T)$ embeds
into a total model of $\T$ {\em with the same support}. 

\medskip

\noindent When establishing links between locality and embeddability we require 
that the clauses in $\K$
are \emph{flat} 
and \emph{linear} w.r.t.\ $\Sigma$-functions.
When defining these notions we distinguish between ground and 
non-ground clauses.

\begin{definition}
An {\em extension clause $D$ is flat} (resp. \emph{quasi-flat}) 
when all symbols 
below a $\Sigma_1$-function symbol in $D$ are variables 
(resp. variables or ground $\Pi_0$-terms).
$D$ is \emph{linear}  if whenever a variable occurs in two terms of $D$
starting with $\Sigma_1$-functions, the terms are equal, and 
no term contains two occurrences of a variable.

\noindent A {\em ground clause $D$ is flat} if all symbols below a $\Sigma_1$-function 
in $D$ are constants.
A {\em ground clause $D$ is linear} if whenever a constant occurs in
two terms in $D$ whose root symbol is in  $\Sigma_1$, the two terms are identical, and
if no term which starts with a $\Sigma_1$-function contains two occurrences of the same constant.
\label{flat}
\end{definition} 
 \begin{definition}[\cite{ihlemann-sofronie-ijcar10}]
With the above notations, let $\Psi$ be a map associating with 
$\K$ and a set of $\Pi^C$-ground terms $T$ 
a set $\pK(T)$ of $\Pi^C$-ground terms. 
We call $\pK$ a \emph{term closure operator} if the following
holds for all sets of ground terms $T, T'$: 
\begin{enumerate}
\item $\mathrm{est}(\K, T) \subseteq \pK(T)$,
\item $T \subseteq T' \Rightarrow \pK(T) \subseteq \pK(T')$,
\item $\pK(\pK(T)) \subseteq \pK(T)$,
\item for
  any map $h: C \rightarrow C$, $\bar{h}(\pK(T)) = \Psi_{\bar{h}\K}(\bar{h}(T))$,
  where $\bar{h}$ is the canonical extension of $h$ to extension
  ground terms.
\end{enumerate}
\end{definition}

\begin{theorem}[\cite{ihlemann-jacobs-sofronie-tacas08,
    ihlemann-sofronie-ijcar10}]
Let ${\mathcal T}_0$ be a first-order theory and $\K$ a set of universally closed flat clauses in the signature
$\Pi$. The following hold: 
\begin{enumerate}
\item If all clauses in $\K$ are linear  
and $\Psi$ is a term closure operator such 
that for every flat set of ground terms $T$, $\Psi(T)$ is
flat, then $({\sf Emb}_{w,f}^\Psi)$ implies $({\sf Loc}^{\Psi})$
and $({\sf EEmb}_{w,f}^\Psi)$ implies $({\sf ELoc}^{\Psi})$.

\item If  the extension ${\mathcal T}_0 \subseteq {\mathcal T} {=} {\mathcal T}_0 {\cup} {\mathcal K}$
satisfies $({\sf Loc}^{\Psi})$ then $({\sf Emb}_{w,f}^\Psi)$ holds; 
if it satisfies $({\sf ELoc}^{\Psi})$ then $({\sf EEmb}_{w,f}^\Psi)$ holds.
\end{enumerate}
\label{check-loc}
\end{theorem}
If we can guarantee that $({\sf Emb}_{w,f})$ holds and the support of the total model which we obtain 
is the same as the support of the partial model we start with
-- condition known as 
``completability of models without changing the support'', (${\sf
  Comp}_f$) --  then 
condition $({\sf ELoc})$ is guaranteed.
%
%

\noindent The following locality transfer
result proved in \cite{ihlemann-sofronie-ijcar10} 
is useful in this paper. For the sake of simplicity we state it for
simple locality; it can easily be extended to $\Psi$-locality.
\begin{theorem}[\cite{ihlemann-sofronie-ijcar10}]
\label{transfer}
Let $\Pi_0 = (\Sigma_0, {\sf Pred})$ be a signature, 
${\cal T}_0$ a  $\Pi_0$-theory,  
$\Sigma_1$ and $\Sigma_2$ two disjoint sets of new function symbols,
$\Pi_i =  (\Sigma_0 \cup \Sigma_i, {\sf Pred})$, $i = 1,2$, and  
${\cal K}$ a set of flat and linear $\Pi_1$-clauses. 
Assume that the extension 
${\cal T}_0 \subseteq {\cal T}_0 \cup {\cal K}$ satisfies condition 
${\sf ELoc}$ as a consequence of an embeddability condition in which the 
support of the models does not change.  
Let ${\cal T}_2$ be a $\Pi_2$-theory such that 
${\cal T}_0 \subseteq {\cal T}_2$. Then the extension 
${\cal T}_2 \subseteq {\cal T}_2 \cup {\cal K}$ satisfies 
condition ${\sf ELoc}$ as well.  
\end{theorem}

\subsection{Examples of local theory extensions}
\label{examples-local-extensions}

We present some examples of theory extensions which were proved to be $\Psi$-local 
in previous work, and which appear in a natural way in the
verification problems we consider in this paper. 

\medskip
\noindent {\bf Free and bounded functions.} 
Let ${\cal T}_0$ be a theory with signature $\Pi_0 =
(\Sigma_0, {\sf Pred})$. 
Any extension of ${\cal T}_0$ with free function symbols in a
set $\Sigma_1$ disjoint from $\Sigma_0$ is local. 
Assume ${\cal T}_0$ contains a 
binary predicate $\leq \in {\sf Pred}$, which is a partial order and $f \not\in \Sigma_0$. 
For $1 \leq i \leq m$ let $t_i(x_1, \dots, x_n)$ and 
$s_i(x_1,\dots, x_n)$ 
be terms in the signature $\Pi_0$ 
and $\phi_i(x_1, \dots, x_n)$ 
be $\Pi_0$-formulae with (free) variables among $x_1, \dots, x_n$, 
such that 
(i) ${\cal T}_0 \models \forall {\overline x} 
(\phi_i({\overline x})\rightarrow \exists y (s_i({\overline x}) \leq y \leq t_i({\overline x})))$, 
and 
(ii)
if $i \neq j$, $\phi_i \wedge \phi_j \models_{{\cal T}_0} \perp$.
The extension of ${\cal T}_0$ with the function $f$ satisfying the
axiom ${\sf GB}(f)$,  
${\cal T}_0 \subseteq {\cal T}_0 \cup {\sf GB}(f)$,  
satisfies condition ${\sf ELoc}$ \cite{sofronie-ihlemann-ismvl-07,ihlemann-jacobs-sofronie-tacas08}, if  

\smallskip
${\sf GB}(f)  =  \displaystyle{\bigwedge_{i = 1}^m} {\sf GB}^{{\phi_i}}(f), \text{ where~~ }
{\sf GB}^{{\phi_i}}(f):~~ \forall {\overline x} (\phi_i({\overline x}) \rightarrow  s_i({\overline x}) \leq f({\overline x}) \leq t_i({\overline x}))$

\smallskip
\noindent (in this last case $\Sigma_1 = \{ f \}$ and ${\cal K} = {\sf
  GB}(f)$). The locality proof uses the fact that every partially defined
function (possibly satisfying boundedness conditions) can be extended
to a total function (satisfying the same boundedness conditions).

\bigskip
\noindent {\bf Monotonicity, boundedness for monotone functions.}
Any extension of a $\Pi_0$-theory ${\cal T}_0$ where $\Pi_0 =
(\Sigma_0, {\sf Pred})$, for which $\leq \in {\sf Pred}$ is a partial order 
with functions in a set $\Sigma_1$ disjoint from $\Sigma_0$, such that 
every $f \in \Sigma_1$ satisfies\footnote{For $i \in I$, $\sigma_i  {\in} \{ -, + \}$, 
and $\leq^+ = \leq, \leq^- = \geq$.} 
conditions ${\sf Mon}^{\sigma}(f)$ and
${\sf Bound}^t(f)$, is local
\cite{sofronie-ihlemann-ismvl-07,ihlemann-jacobs-sofronie-tacas08}. Here: 

\smallskip
$\begin{array}{ll}
{\sf Mon}^{\sigma}(f) &   \displaystyle{\bigwedge_{i \in I}} x_i {\leq_i}^{\sigma_i} y_i \wedge \bigwedge_{i \not\in I} x_i = y_i \rightarrow f(x_1,.., x_n) \leq f(y_1,.., y_n)\\
{\sf Bound}^t(f) &  \forall x_1, \dots, x_n (f(x_1, \dots, x_n) \leq t(x_1, \dots, x_n)) 
\end{array}$

\smallskip
\noindent where $\sigma$ depends on $f$, $t(x_1, \dots, x_n)$ is a $\Pi_0$-term 
with variables among $x_1,$ $\dots,$ $x_n$ whose 
associated function is concanve and has the same monotonicity as $f$ in any model.  

\noindent The extensions satisfy condition $({\sf ELoc})$ if e.g.\ in 
${\cal T}_0$ all finite and empty infima (or suprema) exist.

\smallskip
\noindent The locality proof uses the fact that every monotone 
partial function on a poset can be extended
to a total function on the Dedekind-McNeille completion of the poset, 
cf.\ e.g.\ \cite{sofronie-ihlemann-ismvl-07}.

\medskip
\noindent {\bf Update rules \cite{ihlemann-jacobs-sofronie-tacas08}.}
We consider update rules in which some of the function symbols are
updated depending on a partition of their
domain of definition. 
Let ${\cal T}_0$ be a base theory with signature $\Sigma_0$ and
$\Sigma \subseteq \Sigma_0$. Let $\Sigma' = \{ f' \mid f \in \Sigma
\}$ (disjunct from $\Sigma_0$). 
Consider a family ${\sf Update}(\Sigma, \Sigma') = \bigcup_{f \in
  \Sigma}{\sf Update}(f, f')$, where for every $f \in \Sigma$, 
${\sf Update}(f, f')$ has one of the forms:

$(1) \quad  \forall {\overline x} (\phi^f_i({\overline x}) \rightarrow
    f'({\overline x}) = t_i) \quad  i =1, \dots, m,$ 

\noindent or, in case $\leq$ is a partial
      ordering in the theory $\T_0$: 
 
$(2) \quad \forall {\overline x} (\phi^f_i({\overline x}) \rightarrow
    s_i \leq f'({\overline x}) \leq t_i) \quad  i =1, \dots, m, $

\noindent describing the way $f'$ is defined  
depending on a finite set $\{ \phi^f_i \mid i \in I \}$ of
$\Sigma_0$-formulae such that
\begin{enumerate}
\item[(i)] $\phi^f_i({\overline x}) \wedge \phi^f_j({\overline x})
  \models_{{\cal T}_0} \perp $  for $i {\neq} j$, 
\vspace{-3mm}
\item[(ii)] $s_i, t_i$ are terms in the signature of $\T_0$ (in
  particular they can contain $f$ and other functions in $\Sigma$) and for
  condition (2): 
$\models_{\T_0} \forall {\overline x} (\phi^f_i({\overline x}) \rightarrow
    s_i \leq t_i)$ for every $i = 1, \dots, m$. 
\end{enumerate}
Then the extension of ${\cal T}_0$ with new function symbols $\Sigma_1
= \Sigma'$ satisfying axioms ${\sf Update}(\Sigma, \Sigma')$
is local.  The locality proof uses the fact that every partial 
function satisfying the axioms ${\sf Update}(\Sigma, \Sigma')$ can be extended
to a total function, using the definitions in the axioms in ${\sf Update}$.

\smallskip
\noindent {\bf Convexity/concavity \cite{sofronie-ki08}.}
Let $f$ be a unary function, and 
$I = [a, b]$ a subset of the domain of definition of $f$. 
We consider the axiom:

\vspace{-4mm}
\begin{eqnarray*}
{\sf Conv}^I(f) & & \forall x, y, z \left(x, y \in I \wedge x \leq z \leq y \rightarrow \frac{f(z) - f(x)}{z - x} \leq  \frac{f(y) - f(x)}{y - x}\right).
\end{eqnarray*}

\vspace{-2mm}
\noindent 
Then ${\cal T}_0 {\subseteq} {\cal T}_0 {\cup} {\sf Conv}^I_f$ 
satisfies condition $({\sf ELoc})$ if
${\cal T}_0 = {\mathbb R}$ (the theory of reals), or 
${\cal T}_0 = {\mathbb Z}$ (e.g.\ Presburger arithmetic), or 
${\cal T}_0$ is the many-sorted combination of the theories of reals (sort ${\sf real}$) and integers (sort ${\sf int}$) and 
$f$ has arity ${\sf int} \rightarrow {\sf real}$. Concavity of a
function $f$ can be defined by ${\sf Conc}^I(f) = {\sf Conv}^I(-f)$.
\noindent The locality proof \cite{sofronie-ki08} uses the fact that
every partial algebra which weakly satisfies ${\sf Conv}^I_f$ (resp.\ ${\sf Conc}^I_f$), in
which the function $f$ has a finite definition domain can be extended
to a total model of  ${\sf Conv}^I_f$ (resp.\ ${\sf Conc}^I_f$) 
as follows: Let $p_1, \dots, p_n \in {\mathbb R}$ be the points at
which $f$ is defined. 
Let $f : {\mathbb R} \rightarrow {\mathbb R}$ be obtained by linear
interpolation from $f$. 
Then $f$ is convex. All other cases are proved similarly.
As pointed out in \cite{sofronie-ki08}, the conditions above can be combined 
with continuity and sometimes also with the derivability of the functions.

\medskip
\noindent {\bf Linear combinations of functions.}
Let $f_1, \dots, f_n$ be unary function symbols.
The extension  ${\mathbb R} \subseteq 
{\mathbb R} \cup {\sf BS}$ satisfies condition $({\sf ELoc})$, 
where ${\sf BS}$ contains conjunctions of axioms of type  

$\displaystyle{\forall t (a {\leq} \sum_{i = 1}^n a_i f_i(t) {\leq}
  b)}, \quad \quad \displaystyle{\forall t (g(t) {\leq} \sum_{i = 1}^n a_i f_i(t) {\leq}
  f(t))}, \quad \quad \text{ or }$

$\displaystyle{\forall t, t' (t {<} t' \rightarrow a {\leq} \sum_{i=1}^n a_i
  \frac{f_i(t') {-}  f_i(t)}{t'{-}t} {\leq} b)},$

\noindent where $a, b \in {\mathbb R}$ and
\begin{itemize}
\item[(i)] $g$ is a function
symbol satisfying condition ${\sf Conv}(g)$, or $g(t)$ is a term over the
theory of the real numbers with $t$ as only free variable such that
the associated function $f_{g(t)} : {\mathbb R} \rightarrow {\mathbb R}$ is convex, 
\item[(ii)] $f$ is a function 
symbol satisfying condition ${\sf Conc}(f)$, or $f(t)$ is a term over the
theory of the real numbers with $t$ as only free variable such that
the associated function 
$f_{f(t)} : {\mathbb R} \rightarrow {\mathbb R}$ is  concave,  
\item[(iii)] either $g$ and $f$ are function symbols satisfying the condition 
$\forall t (g(t) \leq f(t))$ or $g(t)$ and $f(t)$ are terms in the
theory of real numbers  with $t$ as only free variable such that 
$\models_{\mathbb R} \forall t ~ g(t) \leq f(t)$.
\end{itemize}
Using arguments similar to those in \cite{sofronie-ki08} it can be 
proved that if we additionally require the functions to be   
continuous locality is still preserved. 

\smallskip
\noindent If $I$  is an interval of the form $(-\infty, a], [a, b]$ or $[a, \infty)$ then we can define 
versions of monotonicity/boundedness, convexity/concavity
and boundedness axioms for linear combinations of functions and 
of their slopes relative to the interval $I$ (then conditions 
(i) and (ii) for $f$ and $g$ are relative to  $I$).

\subsection{Hierarchical reasoning in theory extensions}
Consider a $\Psi$-local theory extension 
${\mathcal T}_0 \subseteq {\mathcal T}_0 \cup {\mathcal K}$.
Condition $({\sf Loc}^{\Psi})$ requires that, for every set $G$ of ground 
$\Pi^c$ clauses, ${\mathcal T}_0 \cup {\cal K} \cup G \models \perp$ iff 
${\mathcal T}_0 \cup {\mathcal K}[\Psi_{\cal K}(G)] \cup G \models \perp$.

\medskip
\noindent All clauses in ${\mathcal K}[\Psi_{\cal K}(G)] \cup G$ have the 
property that the function symbols in $\Sigma_1$ have as arguments only 
ground terms, so  
${\mathcal K}[\Psi_{\cal K}(G)] \cup G$ can be flattened 
and purified: The function symbols in $\Sigma_1$ are separated from 
the other symbols by introducing, in a bottom-up manner, new  
constants $c_t$ for
subterms $t {=} f(g_1, \dots, g_n)$ with $f {\in} \Sigma_1$, $g_i$ ground 
$\Pi_0^C$-terms together with corresponding definitions $c_t {=} t$ 
($C$ is a set of constants 
which contains the constants introduced by flattening, resp.\ purification).
Flattening can be performed in time linear in the size of ${\mathcal
  K}[\Psi_{\cal K}(G)] \cup G$. 

\medskip
\noindent The set of clauses thus obtained 
has the form ${\mathcal K}_0 \cup G_0 \cup {\sf Def}$, 
where ${\sf Def}$ consists of ground unit clauses of the form 
$f(g_1, \dots, g_n) = c$, where $f \in \Sigma_1$, $c$ is a 
constant, $g_1, \dots, g_n$ are ground 
terms without $\Sigma_1$-function symbols, and ${\mathcal K}_0$ and $G_0$ do not contain $\Sigma_1$-function 
symbols.
(In what follows we always 
flatten and then purify ${\mathcal K}[\Psi_{\cal K}(G)] \cup G$ to 
ensure that the ground unit clauses in 
${\sf Def}$ are in fact of the form $f(c_1, \dots, c_n) = c$, where
$c_1, \dots, c_n, c$ are {\em constants}.)
\begin{theorem}[\cite{Sofronie-cade-05,ihlemann-jacobs-sofronie-tacas08}]
Let ${\mathcal K}$ be a set of clauses. 
Assume that ${\mathcal T}_0 \subseteq {\cal T}_1 = {\mathcal T}_0 \cup {\mathcal K}$ is a 
$\Psi$-local theory extension. 
For any set $G$ of ground clauses, 
let ${\mathcal K}_0 \cup G_0 \cup {\sf Def}$ 
be obtained from ${\mathcal K}[\Psi_{\cal K}(G)] \cup G$ by flattening and purification, 
as explained above. 
Then the following are equivalent to ${\cal T}_1 \cup G \models \perp$: 

\vspace{-2mm} 
\begin{itemize}
\item[(1)] ${\mathcal T}_0 {\cup} {\mathcal K}[\Psi_{\cal K}(G)] {\cup} G \models \perp.$ 
\item[(2)] ${\mathcal T}_0 {\cup} {\mathcal K}_0 {\cup} G_0 {\cup} {\sf Def} \models \perp.$ 
\item[(3)] ${\mathcal T}_0 \cup {\mathcal K}_0 \cup G_0 \cup {\sf Con}[G]_0 \models \perp,$ where 

$\displaystyle{~~~ {\sf Con}[G]_0  = \{ \bigwedge_{i = 1}^n c_i = d_i \rightarrow c = d \mid 
f(c_1, \dots, c_n) = c, f(d_1, \dots, d_n) = d \in {\sf Def} \}}.$\\[-3ex]
\end{itemize}
\label{lemma-rel-transl}
\end{theorem} 
A similar equivalence holds for extended ${\Psi}$-local extensions, with the 
remark that in that case ${\cal K}_0$ and $G_0$ may contain arbitrary 
$\Pi_0$-sentences \cite{ihlemann-jacobs-sofronie-tacas08,ihlemann-sofronie-ijcar10}.

\begin{example}
{\em Let ${\cal T}_0$ be the theory of real numbers, 
and ${\cal T}_1$ a local extension of ${\cal T}_0$ 
with two monotone functions $f$ and $g$.
Consider the following problem: 
$${\cal T}_0 \cup {\sf Mon}_f \cup {\sf Mon}_g \models \forall x, y, z
(0 \leq x \wedge x + y \leq z \wedge f(x+y) \leq g(x+y) \rightarrow f(x) \leq g(z)).$$
The problem reduces to checking whether 
${\cal T}_0 \cup {\sf Mon}_f \cup {\sf Mon}_g \cup G \models \perp,$
where $$G = 0 \leq c_1 \wedge c_1 + c_2 \leq c_3 \wedge
f(c_1+c_2) \leq g(c_1+c_2) \wedge 
f(c_1) \not\leq g(c_3).$$ 
%
The locality of the extension ${\cal T}_0 \subseteq 
{\cal T}_1$ means that, in order to test if 
${\cal T}_0 \cup {\sf Mon}_f \cup {\sf Mon}_g \cup G \models \perp,$
it is sufficient to test whether 
${\cal T}_0 \cup {\sf Mon}_f[G] \cup {\sf Mon}_g[G] \cup G \models 
\perp,$
where ${\sf Mon}_f[G], {\sf Mon}_g[G]$ consist of those instances of 
the monotonicity axioms for $f$ and $g$ in which the terms starting with 
the function symbol $f$ and $g$ occur already in $G$:

\medskip
{\small \noindent 
$\begin{array}{@{}llll ll}
{\sf Mon}_f[G] & {=} & c_1 \leq c_1 + c_2 \rightarrow f(c_1) {\leq}
f(c_1 + c_2) &    ~~{\sf Mon}_g[G] & {=} & c_3 {\leq} c_1 + c_2
\rightarrow g(c_3) {\leq} g(c_1 + c_2) \\
& & c_1 + c_2 {\leq} c_1  \rightarrow f(c_1 + c_2) {\leq} f(c_1) &
\quad & &   c_1 + c_2 {\leq}  c_3 \rightarrow g(c_1 + c_2) {\leq} g(c_3) 
\end{array}$}

\medskip
\noindent 
In order to check the satisfiability of ${\cal T}_0 \cup {\sf
  Mon}_f[G] \cup {\sf Mon}_g[G] \cup G$  we purify it, 
introducing definitions for the term starting with extension functions
in a bottom-up fashion: 
$d_1 = c_1 + c_2, f(c_1) = e_1, f(d_1) = e_2, g(c_3) = e_3, g(d_1) = e_4$. 
The corresponding set of instances of the congruence axioms is: 
${\sf Con}_0 =  \{ c_1 = d_1  \rightarrow  e_1 = e_2, 
c_3 = d_1  \rightarrow  e_3 = e_4 \}$. 
We obtain the following set of clauses:

\medskip
{\small 
$\begin{array}{l||llllll}
\hline 
{\sf Def} & {\sf Def} & G_0& & ({\sf Mon}_f \cup {\sf Mon}_g)_0 & & {\sf Con}_0\\
\hline 
f(c_1) = e_1 & d_1 = c_1 + c_2 & 0 \leq c_2 & \quad  & c_1 \leq d_1 \rightarrow e_1
\leq e_2 & \quad & c_1 = d_1  \rightarrow  e_1 = e_2\\
f(d_1) = e_2 & & d_1 \leq c_3 & \quad & d_1 \leq c_1 \rightarrow e_2
\leq e_1 & & \\
g(c_3) = e_3 & & e_2  \leq e_4 &             & c_3 \leq d_1 \rightarrow
e_3 \leq e_4 & & c_3 = d_1 \rightarrow  e_3 = e_4\\
g(d_1) = e_4 & & e_1 > e_3     &            & d_1 \leq c_3  \rightarrow
e_4 \leq e_3 & & \\
\hline 
\end{array}$ }

\medskip
\noindent 
It is easy to see that the set of instances of the congruence axioms
corresponding to the definitions ${\sf Def}$, ${\sf Con}_0$ (last
column in the table above), is
entailed by $({\sf Mon}_f \cup {\sf Mon}_g)_0$ and therefore is redundant.

\medskip
\noindent We can use a decision procedure for the theory of real
numbers for checking the satisfiability of $({\sf
  Mon}_f[G] \cup {\sf Mon}_g[G])_0 \cup G_0$ (the theory of real
numbers is decidable \cite{tarski}). 
%
It can easily be checked that this set of formulae is unsatisfiable: 
from $0 \leq c_2$ it follows that $c_1 \leq c_1 + c_2 = d_1$ and as $c_1
\leq d_1 \rightarrow e_1 \leq e_2$ we infer that $e_1 \leq
e_2$. 
From $d_1 \leq c_3$ and $d_1 \leq c_3\rightarrow e_4 \leq e_3$ it 
follows that $e_4 \leq e_3$. Since $e_1 \leq e_2, e_2 \leq e_4$ and
$e_4 \leq e_3$ we know that 
$e_1 \leq e_3$. This yields a contradiction with $e_1 > e_3$. 
}
\end{example}

\medskip
\noindent {\bf Decidability, parameterized complexity.}
Theorem~\ref{lemma-rel-transl} allows us to show that 
if for every finite set $T$ of terms $\Psi_{\cal K}(T)$ is finite and
can be effectively constructed (and
has size computable from the size of $T$) 
then (i) decidability 
of satisfiability w.r.t.\ a $\Psi$-local extension ${\cal T}_1$ 
of a theory ${\cal T}_0$ is a consequence of the decidability of the 
satisfiability of a certain fragment of ${\cal T}_0$, and 
(ii) the complexity of such satisfiability tests in ${\cal T}_1$
can be expressed as a function of the complexity of satisfiability 
checking for a suitable fragment of ${\cal T}_0$ as explained in
Theorem~\ref{complex}. 
\begin{theorem}[\cite{Sofronie-cade-05,ihlemann-jacobs-sofronie-tacas08}]
Consider the theory extension 
${\cal T}_0 \subseteq {\cal T}_1 = {\cal T}_0 \cup {\cal K}$, where
${\cal K}$ is finite, flat and linear, with at most $k$ variables per
clause. Assume that this theory extension satisfies  
condition $({\sf Loc}^{\Psi})$ for a closure operator with the
property that for every finite set $T$ of ground terms $\Psi_{\cal K}(T)$ is finite and
can be effectively constructed (and
has size $h(|T|)$, where $h$ is a computable function and $|T|$ is the size of $T$). 
Then satisfiability of  
$G$ as in the definition of $({\sf Loc}^{\Psi})$ 
w.r.t.\ ${\cal T}_1$ is decidable 
provided ${\cal K}[\Psi_{\cal K}(G)]$ is finite and 
${\cal K}_0 \cup G_0 \cup {\sf Con}[G]_0$ belongs to 
a decidable fragment ${\cal F}$ of ${\cal T}_0$. 
If (i) the complexity of testing the satisfiability of 
a set of formulae in ${\cal F}$ of size $m$ w.r.t.\ ${\cal T}_0$ 
can be described by a function $g(m)$ and (ii) $\Psi_{\cal K}(G)$ is a set of 
terms of size $n$, then 
the complexity of checking whether $G \models_{{\cal T}_1} \perp$
is $g(n^k)$, where $k$ is the maximum number of free 
variables in a clause in ${\cal K}$ (but at least $2$). 

\noindent A similar result holds for theory extensions satisfying condition ${\sf ELoc}^{\Psi}$
and satisfiability of formulae $F = F_0 \cup G$ as in the definition
of $({\sf ELoc}^{\Psi})$. 
\label{complex}
\end{theorem}
{\em Proof:} The decision procedure based on the hierarchical
reduction method 
Theorem~\ref{lemma-rel-transl} with the complexity analysis is 
presented in Algorithm~\ref{alg-hier-red}. 

\noindent The correctness is a consequence of 
the locality of the extension and of Theorem~\ref{lemma-rel-transl}. \QED

\begin{algorithm}[t]
\caption{Checking ground satisfiability w.r.t.\
  a theory extension  $\T_0 \subseteq \T_0 \cup \K$}

{\bf Assumption:} The theory extension  $\T_0 \subseteq \T_0 \cup \K$
is $\Psi$-local and satisfies all conditions in
Thm.~\ref{complex}. 

\medskip
\begin{tabular}{ll}
{\bf Input:} & A set of ground clauses $G$. \\
{\bf Task:} & Decide whether $G$ is satisfiable w.r.t.\ $\T_0 \cup
\K$.\\
\hline 
\end{tabular}

\begin{description} 
\item[Step 1:] Compute $\Psi_{\cal K}(G)$  \hfill {\footnotesize /*
    Size: $h(|G|)$ */} 
\item[Step 2:] Compute  ${\cal K}[\Psi_{\cal K}(G)]$  
\hfill {\footnotesize /* Size: $|{\cal
    K}|*h(|G|)^k$,

\vspace{-1mm}
\hfill  where $k$ is the maximum number of free variables
  in a clause of $\K$ */ } 
\item[Step 3:] Compute ${\cal K}_0 \cup G_0$ by
  flattening and purification from ${\cal K}[\Psi_{\cal K}(G)] \cup
  G$. 

 \hfill {\footnotesize /* Size  $|{\cal
    K}|*h(|G|)^k + |G|$ */ } 

\item[Step 4:] Compute ${\sf Con}[G]_0$  \hfill {\footnotesize /* Size at most
  $h(|G|)^2$ */ } 
\item[Step 5:] Use decision procedure for ${\cal F}$  to check
  satisfiability of ${\cal K}_0 \cup G_0
  \cup {\sf Con}[G]_0$. 

\hfill {\footnotesize /* Complexity:  $g(m)$, where $m$ is the size of ${\cal K}_0 \cup G_0
\cup {\sf Con}[G]_0$, i.e.\ 

\vspace{-1mm}
\hfill ~~ $m = |{\cal
    K}|*h(|G|)^k + |G| + h(|G|)^2 \in
O(h(|G|^{{\sf max}(k, 2)}))~~~$ 

\vspace{-1mm}
\hfill ~~ (for a fixed $\K$, $|{\cal K}|$ and $k$ can be
considered to be constants) */ }  
\end{description}
\label{alg-hier-red}
\end{algorithm}

\medskip
\noindent {\bf Chains of Theory Extensions.} 
We can also consider chains of theory extensions of the form: 
\begin{eqnarray*}
{\mathcal T}_0 & \subseteq &  {\mathcal T}_1  =  {\mathcal T}_0 \cup
{\mathcal K}_1  ~~~\subseteq~~~  {\mathcal T}_2 = {\mathcal T}_0  \cup
{\mathcal K}_1 \cup {\mathcal K}_2 ~~~\subseteq~~~ \dots ~~~ \subseteq~~~  {\mathcal
  T}_n = {\mathcal T}_0 \cup {\mathcal K}_1 \cup...\cup {\mathcal K}_n
\end{eqnarray*}
where for all $1 \leq i
\leq n$, $\T_i$ is a local extension of $\T_{i-1}$. 

\noindent 
For a chain of local extensions  a satisfiability check w.r.t.\ the last extension can 
be reduced (in $n$ steps) to a satisfiability check w.r.t.\  
${\mathcal T}_0$. The only restriction we need to impose in order to
ensure that such a reduction is possible is that at each step the
clauses reduced so far need to be ground. 
Groundness is assured if each variable in a clause appears at least
once under 
an extension function.
This iterated instantiation procedure for chains of local theory
extensions has been implemented in H-PILoT \cite{hpilot}.\footnote{H-PILoT  ~allows the user to specify a chain of extensions by 
tagging the extension functions with their place in the chain 
(e.g., if $f$ occurs in ${\mathcal K}_3$ but not in ${\mathcal K}_1
\cup {\mathcal K}_2$ it is declared as level 3). }

\subsection{Symbol elimination in local theory extensions}
\label{symbol-elimination}

In \cite{sofronie-ijcar2016, sofronie-lmcs-2018} we identified
situations in which hierarchical symbol elimination is possible. 

Let $\Pi_0 = (\Sigma_0, {\sf Pred})$. 
Let ${\mathcal T}_0$ be a base theory with signature $\Pi_0$.  
We consider theory extensions $\T_0
\subseteq \T = \T_0 \cup \K$, in which among the extension functions
we identify a set of {\em parameters} $\Sigma_P$ (function and constant symbols). 
Let $\Sigma$ be a signature consisting of extension symbols which are not
parameters (i.e.\ such that $\Sigma \cap (\Sigma_0 \cup \Sigma_P) =
\emptyset$).
%
We assume that ${\mathcal K}$ is a set of clauses 
in the signature $\Pi_0 {\cup} \Sigma_P {\cup} \Sigma$ in which all 
variables occur also below functions in $\Sigma_1 = \Sigma_P \cup
\Sigma$.

\medskip
\begin{algorithm}[h!]
\caption{Algorithm for Symbol Elimination in Theory Extensions}

\begin{tabular}{ll}
{\bf Input:} & $G$, a finite set of ground clauses in the signature $\Pi^C$; \\
& $T$, a finite set of ground $\Pi^C$-terms  with ${\sf est}({\cal K}, G) \subseteq T$. \\

{\bf Output:} & Universal formula $\forall y_1 \dots y_n \Gamma_T(y_1,
\dots, y_n)$ over $\Pi_0 {\cup} \Sigma_P$\\
& such that ${\cal T}_0 \cup {\cal K} \cup \forall y_1 \dots y_n \Gamma_T(y_1,
\dots, y_n) \cup G$ unsatisfiable. \\
\hline 
\end{tabular}

\begin{description}
\item[Step 1] Compute the set of $\Pi_0^C$ clauses $\K_0 \cup G_0 \cup
  {\sf Con}_0$  from $\K[T] \cup G$  using the purification step
  described in  Theorem~\ref{lemma-rel-transl} (with set of extension
  symbols $\Sigma_1$). 


\item[Step 2]  $G_1 := {\mathcal K}_0 \cup G_0\cup {\sf Con}_0$. 
Among the constants in $G_1$, identify 
\begin{enumerate}
\vspace{-1mm}
\item[(i)] the constants
${\overline c_f}$, $f \in \Sigma_P$, where either $c_f = f \in
\Sigma_P$ is a constant
parameter, or $c_f$ is 
introduced by a definition $c_f := f(c_1, \dots, c_k)$ in the hierarchical
reasoning method, 
\vspace{-1mm}
\item[(ii)] all constants  ${\overline c_p}$ 
occurring as arguments of functions in $\Sigma_P$ in such definitions. 
\vspace{-1mm}
\end{enumerate}
Let ${\overline  c}$ be the remaining constants.\\
Replace the constants in ${\overline  c}$
with existentially quantified variables ${\overline x}$ in $G_1$,
i.e.\ 
replace $G_1({\overline c_p}, {\overline c_f}, {\overline c})$ 
with $G_1({\overline c_p}, {\overline c_f}, {\overline x})$, and
consider the formula
$\exists {\overline x} G_1({\overline c_p},{\overline c_f}, {\overline x})$.


\item[Step 3] Compute a quantifier-free formula 
$\Gamma_1({\overline c_p}, {\overline c_f})$  equivalent to 
$\exists {\overline x} G_1({\overline c_p}, {\overline c_f},{\overline
  x})$ w.r.t.\ $\T_0$ using a  method for quantifier elimination in 
${\mathcal T}_0$.  


\item[Step 4] Let $\Gamma_2({\overline c_p})$ be the formula 
obtained by replacing back in $\Gamma_1({\overline c_p}, {\overline c_f})$ 
the constants $c_f$ introduced by definitions $c_f := f(c_1, \dots,
c_k)$ with the terms $f(c_1, \dots,c_k)$. 

Replace ${\overline c_p}$ with existentially quantified variables ${\overline y}$. 


\item[Step 5] Let $\forall {\overline y} \Gamma_T({\overline y})$ be
  $\forall {\overline y} \neg \Gamma_2({\overline y})$. 
\end{description}
\label{alg-symb-elim}
\end{algorithm}

\medskip

\noindent 
We identify situations in which we can
generate, for every ground formula $G$,   
a (universal) formula $\Gamma$ representing a family of
constraints on the parameters of $G$ 
such that $\T \cup \Gamma \cup G \models \perp$. 
We consider base theories which allow quantifier elimination, and use quantifier 
elimination to generate the formula $\Gamma$.

\begin{theorem}[\cite{sofronie-ijcar2016, sofronie-lmcs-2018}] 
Assume that ${\mathcal T}_0$ allows quantifier elimination.  
For every finite set of ground clauses $G$, and every finite set $T$ of terms over the 
signature $\Pi^C$, where $\Pi = (\Sigma_0 \cup \Sigma \cup \Sigma_P,
{\sf Pred})$, with 
${\sf est}(\K, G)
\subseteq T$, we can construct a universally quantified 
$\Pi_0 \cup \Sigma_P$-formula 
$\forall {\overline y} \Gamma_T({\overline y})$ 
with the property that  for every
structure ${\mathcal A}$ with signature 
$\Pi_0 \cup \Sigma \cup \Sigma_P \cup C$ 
which is a model of ${\mathcal T}_0 \cup {\mathcal K}$, if 
${\mathcal A} \models \forall {\overline y}
\Gamma_T({\overline y})$ then ${\mathcal A} \models \neg G$, 
i.e.\ such that ${\mathcal T}_0 \cup {\mathcal K} \cup \forall {\overline y}
\Gamma_T({\overline y}) \cup 
  G$ is unsatisfiable. 
\label{inv-trans-qe}
\end{theorem}
{\em Proof:} 
We construct a universal 
formula $\forall y_1 \dots y_n \Gamma_T(y_1, \dots, y_n)$ over the
signature $\Pi_0 {\cup} \Sigma_P$ with the desired properties by following
Steps 1--5 in Algorithm~\ref{alg-symb-elim}. \QED

\medskip
\noindent A similar approach is used in \cite{sofronie-ijcar2010} 
for generating constraints on parameters 
which guarantee safety of parametric systems. 

\

\noindent We denote by $\forall {\overline y} \Gamma_G({\overline y})$
the formula 
obtained when $T = {\sf est}(\K, G)$. 

\begin{theorem}[\cite{sofronie-ijcar2016, sofronie-lmcs-2018}]  
   If the extension ${\mathcal T}_0 \subseteq {\mathcal T}_0 \cup
  {\mathcal K}$ satisfies condition $({\sf Comp}_{f})$ and $\K$ is flat
  and linear and every variable in $\K$ occurs at least once below an
  extension term then 
$\forall y \Gamma_G(y)$ is entailed by every
  conjunction $\Gamma$ of 
  clauses with the property that ${\mathcal T}_0 \cup \Gamma \cup
  {\mathcal K} \cup G$ is unsatisfiable (i.e.\ it is a weakest such constraint). 
\label{symb-elim-weakest}
\end{theorem}
\noindent A similar result can be established for $\Psi$-locality and 
for chains of local theory extensions. 

\begin{theorem}[\cite{sofronie-lmcs-2018}]  
 Assume that we have the following chain of theory extensions:
\[ {\mathcal T}_0 ~~~\subseteq~~~ {\mathcal T}_0 \cup {\mathcal K}_1
~~~\subseteq~~~  
{\mathcal T}_0 \cup \K_1 \cup \K_2 ~~~\subseteq~~~ \dots ~~~\subseteq~~~ 
{\mathcal  T}_0 \cup \K_1 \cup \K_2 \cup \dots \cup \K_n \] 
where every extension in the chain satisfies condition 
$({\sf Comp}_{f})$, $\K_i$ are all flat
  and linear, and in all $\K_i$ all variables occur below the 
extension terms on level $i$. 

Let $G$ be a set of ground clauses, and let $G_1$ be 
the result of the hierarchical reduction of satisfiability 
of $G$ to a satisfiability test w.r.t.\ $\T_0$. Let $T(G)$ be the set
of all instances used in the chain of hierarchical reductions and let 
$\forall y \Gamma_{T(G)}(y)$ be the formula obtained by applying 
Steps 2--5 to $G_1$ with set of ground terms $T(G)$.
Then $\forall y \Gamma_{T(G)}(y)$ 
is entailed by every
  conjunction $\Gamma$ of 
  clauses with the property that ${\mathcal T}_0 \cup \Gamma \cup
  {\mathcal K}_1 \cup \dots \cup \K_n  \cup G$ is unsatisfiable (i.e.\ it is a weakest such constraint). 
\label{symb-elim-weakest-chains}
\end{theorem}
Examples illustrating the way Algorithm~\ref{alg-symb-elim} can be
used for symbol elimination (both for theory extensions and for chains
of theory extensions) are presented in detail in
\cite{sofronie-lmcs-2018}.

\section{Verification problems for parametric systems}
\label{verif}
We identify situations in which 
decision procedures 
exist for invariant checking and bounded model checking, as well 
as methods for obtaining constraints between the parameters which guarantee 
that certain properties are invariants. 
\subsection{Systems modeled using transition constraints}
\label{trans}
We specify reactive 
systems using tuples 
$(\Pi_S, {\cal T}_S, T_S)$ where $\Pi_S$ is a signature, ${\cal T}_S$ is a 
$\Pi_S$-theory (describing the data types used in the specification
and their properties), and 
$T_S = (V, \Sigma, {\sf Init}, {\sf Update})$ is a transition constraint 
system which specifies:
the variables ($V$) and function symbols ($\Sigma$) 
whose values change over time, where $V \cup \Sigma \subseteq \Sigma_S$; 
a formula ${\sf Init}$ specifying the properties of initial states; 
a formula ${\sf Update}$
with variables in $V {\cup} V'$ and 
function symbols in $\Sigma {\cup} \Sigma'$ (where $V'$ and $\Sigma'$ are 
new copies of $V$ resp.\ $\Sigma$, denoting the variables resp.\ functions 
after the transition) which specifies the relationship between the values 
of variables $x$ (functions $f$) before a transition and their values
$x'$ ($f'$) after the transition. 
 
\noindent We consider   
{\em invariant checking} and {\em bounded model checking} problems\footnote{In what follows we only address 
invariant checking; the problems which 
occur in bounded model checking are similar.
}, cf.\ \cite{MannaPnueli}:

\smallskip
\noindent {\bf Invariant checking.} 
A formula $\Phi$ is an inductive invariant of a system $S$ with 
theory ${\cal T}_S$ and  
transition constraint system 
$T_S {=} (V, \Sigma, {\sf Init}, {\sf Update})$ if:
\begin{itemize}
\item[(1)] $\mathcal{T}_S \wedge {\sf Init} \models \Phi$ and 
\vspace{-3mm}
\item[(2)] $\mathcal{T}_S \wedge \Phi \wedge {\sf Update} \models \Phi'$,  
where $\Phi'$ results from $\Phi$ by replacing each $x \in
\mathcal{V}$ by $x'$ and each $f \in \Sigma$ by $f'$.
\end{itemize}
\smallskip
\noindent {\bf Bounded model checking.}
We check whether, for a fixed $k$, unsafe states are 
reachable in at most $k$ steps. Formally, we check whether: 

\noindent 
$\displaystyle{~~~ {\cal T}_S \wedge {\sf Init}_0 \wedge 
\bigwedge_{i = 0}^{j-1} {\sf Update}_i \wedge \neg \Phi_j 
\models \perp \quad \text{ for all } 0 \leq j \leq k}$, 

\noindent 
where ${\sf Update}_i$ is obtained from ${\sf Update}$ by 
replacing every $x {\in} V$  by $x_i$, every  
$f {\in} \Sigma$ by $f_i$, and each 
$x' {\in} V'$, $f' {\in} \Sigma'$ by 
$x_{i+1}, f_{i+1}$; 
${\sf Init}_0$ is ${\sf Init}$ with $x_0$ replacing $x \in V$ and 
$f_0$ replacing $f {\in} \Sigma$; 
$\Phi_i$ is obtained from $\Phi$ similarly. 

\medskip
\noindent Situations in which invariant checking 
{\em under given constraints on parameters} is decidable and
possibilities of 
{\em deriving constraints between parameters} which guarantee that 
a certain safety property is an invariant of the system are discussed
in Section~\ref{sect-tcs}. 

\subsection{Systems modeled using  hybrid automata} 

Hybrid automata were introduced in \cite{Henzinger} to describe
systems with discrete control (represented by a finite set of 
control modes) such that in every control mode certain variables can
evolve continuously in time according to precisely specified rules.

\begin{definition}[\cite{Henzinger}]
A hybrid automaton is a tuple 
$S = (X, Q, {\sf flow}, {\sf Inv}, {\sf Init}, E, {\sf guard}, {\sf jump})$
consisting of: 
\begin{itemize}
\vspace{-2mm}
\item[(1)] 
A finite set $X = \{ x_1, \dots, x_n \}$ of real valued variables
(which can change in time, and are therefore regarded as functions
$x_i : {\mathbb R} \rightarrow {\mathbb R}$) and 
a finite set $Q$ of control modes; 
\item[(2)] A family $\{ {\sf flow}_q \mid q \in Q \}$ of 
predicates over the variables in
$X \cup {\dot X}$ (where ${\dot X} =\{ {\dot x_1}, \dots, {\dot x_n} \}$, 
where ${\dot x_i}$ is the derivative of $x_i$)  
specifying the 
continuous dynamics in each control mode\footnote{This means that we assume that the functions
$x_i : {\mathbb R} \rightarrow {\mathbb R}$ are differentiable during flows.}; 
a family $\{ {\sf Inv}_q \mid q \in Q \}$ of 
predicates over the variables in $X$ defining the  
invariant conditions for each control mode; 
and a family $\{ {\sf Init}_q \mid q \in Q \}$ of 
predicates over the variables in $X$, defining the initial 
states for each control mode. 
\item[(3)] A finite multiset $E$ 
with elements in $Q {\times} Q$ (the control switches). 
Every $(q, q') \in E$ is a 
directed edge between $q$ (source mode) and $q'$ (target mode); 
a family of guards 
$\{ {\sf guard}_e \mid e \in E \}$ (predicates over $X$); and a family of 
jump conditions $\{ {\sf jump}_e \mid e \in E \}$ 
(predicates over $X \cup X'$, 
where $X' =\{ x'_1, \dots, x'_n \}$ is a copy of $X$ consisting of ``primed'' 
variables). 
\end{itemize}
\end{definition}
A {\em state} of $S$ is a pair $(q, a)$ consisting
of a control mode $q \in Q$ and a vector $a = (a_1, \dots, a_n)$
that represents a value $a_i \in {\mathbb R}$ for each variable $x_i \in X$.
A state $(q, a)$ is {\em admissible} if ${\sf Inv}_q$ 
is true when each $x_i$ is replaced by $a_i$.
There are two types of {\em state change}: 
(i) A {\em jump} is an instantaneous transition 
that changes the control location and the values of variables in $X$ according
to the jump conditions;   
(ii) In a {\em flow}, the state can change due to the evolution in a 
given control mode over an interval of time: the values of the
variables in $X$  
change continuously according to the flow rules of the current
control location; all intermediate states are admissible.
A {\em run} of $S$ is a finite
sequence $s_0 s_1 \dots s_k$ of  admissible states such that
(i) the first state $s_0$ is an initial state of $S$ (the values of 
the variables satisfy ${\sf Init}_q$ for some $q \in Q$),
(ii) each pair $(s_j, s_{j+1})$ is either a jump of $S$ or the endpoints of a 
flow of $S$.

\noindent {\em Notation.} In what follows we use the following notation.
If $x_1, \dots, x_n \in X$ we denote 
the sequence $x_1, \dots, x_n$ with ${\overline x}$, the sequence 
$\dot{x}_1, \dots, \dot{x}_n$ with $\overline{\dot{x}}$, and  
the sequence of values $x_1(t), \dots, x_n(t)$ of these variables 
at a time $t$ with ${\overline x}(t)$.

\medskip
\noindent A formula is an inductive invariant of a hybrid automaton if
it is true in all initial states and it is preserved under all jumps and
flows. Situations in which invariant checking 
{\em under given constraints on parameters} is decidable, and
possibilities of 
{\em deriving constraints between parameters} which guarantee that 
a certain safety property is an invariant are described in
Section~\ref{sect-ha}. 

\section{Systems modeled using transition constraints}
\label{sect-tcs}

Consider a system $S$ specified by a tuple $(\Pi_S, {\cal T}_S, T_S)$ where $\Pi_S$ is a signature, ${\cal T}_S$ is a 
$\Pi_S$-theory (describing the data types used in the specification
and their properties), and 
$T_S = (V, \Sigma, {\sf Init}, {\sf Update})$ is a transition constraint 
system. 
Assume that the signature $\Pi_S$
extends a ``base signature'' $\Pi_0 = (\Sigma_0, {\sf Pred})$ of interpreted function and predicate
symbols with new function symbols in a set $\Sigma_1$ with $V \cup \Sigma
\subseteq \Sigma_1$,  and that some of these additional constants
and functions used in the description of $T_S$ can
be considered to be parametric. We denote the set of these constants
and functions with $\Sigma_P \subseteq \Sigma_1$, and the extension of $\Pi_0$ with these
parametric symbols with $\Pi_P$.

\noindent Let $\Phi$ be a formula over the signature $\Pi_S$ specifying a property of the system $S$. 
\smallskip
\noindent  
Let $\Gamma$ be a formula over the signature $\Pi_P$ describing additional constraints on the parameters.
To check whether a formula $\Phi$ is an inductive invariant 
under the constraints  $\Gamma$ we need to analyze whether the
following holds:  
\label{label-problems} 
$$\begin{array}{lrcl}
(1) \quad \quad \quad & \T_S \wedge \Gamma \wedge {\sf
  Init} \wedge \neg \Phi   & \models  & \perp \\
(2) & \T_S \wedge \Gamma \wedge \Phi \wedge {\sf Update} \wedge \neg
\Phi' & \models & \perp, 
\end{array}$$
where $\Phi'$ is obtained from $\Phi$ by replacing every symbol in $V
\cup \Sigma$ with its primed version. 
These are satisfiability problems for possibly 
quantified formulae. In general satisfiability of such types of
formulae is not decidable.  
We are interested in 
identifying situations in which the problems above are decidable, 
and also in the dual problem 
of inferring a set ${\overline \Gamma}$ of {\em most general constraints} 
on the parameters which guarantee 
that $\Phi$ is an invariant. 

\begin{definition}  
Let ${\overline \Gamma}$ be a formula over the signature $\Pi_P$ expressing constraints 
on the parameters in $\Sigma_P$. 
We say that ${\overline \Gamma}$ is a {\em weakest condition under which 
$\Phi$ is an inductive invariant} iff for every formula $\Gamma$ expressing 
constraints on the parameters, $\Phi$ is an inductive invariant under $\Gamma$ iff  ${\cal T}_S \cup \Gamma \models {\overline \Gamma}$. 
\end{definition}
We distinguish two types of situations. 
\begin{description}
\item[(1) Simple transition systems; only variables are updated:] 
If the description of the system does not use complicated
data structures, $\T_S$ is the extension of a theory $\T_0$ with
additional (free) constants, $\Sigma = \emptyset$,
i.e.\ only variables are updated,  the updates are 
quantifier-free formulae (typically assignments for the variables), 
and $\Gamma, {\sf Init}$ and
$\Phi$ are in a fragment of $\T_S$ for which satisfiability is
decidable, then invariant
checking is decidable. 

If $\T_0$ allows quantifier elimination then $\T_0$ is decidable, so 
satisfiability of arbitrary $\Pi_S$-formulae w.r.t. $\T_S$ is 
decidable. 
Therefore, $\Gamma, {\sf Init}$ and 
$\Phi$ can be arbitrary $\Pi_S$-formulae. 
In this case, methods for quantifier elimination in $\T_0$ can be used for
synthetising constraints on the parameters.

This simple situation is discussed in
Section~\ref{sect:case1}. 

\item[(2) Complex systems, complex updates:] If we allow for
  functional parameters that are assumed to satisfy certain certain
  properties  and 
consider more complicated system descriptions, using various types of
data structures, we can still guarantee that invariant checking is
decidable if the formulae $\Gamma, {\sf Init}$ and
$\Phi$ and $\neg \Phi$ are in a decidable fragment of $\T_S$. 

However, in such situations $\T_S$ will not necessarily allow quantifier elimination. 
In Section~\ref{sect:case2} we analyze possibilities of using
hierarchical symbol elimination for constraint synthesis. 

The general case, in which also general (possibly global) updates for 
functions symbols are allowed, is discussed in
Section~\ref{sect:case3}. 
\end{description} 
To distinguish between the conditions mentioned above, in what follows 
we will use some of the following conditions for formulae and theories: \label{page-assumptions}  
\begin{description}
\item[Assumption QE($\T_S$):] ${\cal T}_S$ allows quantifier
  elimination. 

\vspace{-3mm}
\item[Assumption ELoc($\T_S$):] ${\cal T}_S$ is an extension of a  $\Pi_0$-theory 
${\cal T}_0$ with a set ${\cal K}$ of flat and linear clauses 
satisfying  ${\sf Comp}_{\sf f}$ -- thus also ${\sf ELoc}$ 
(and the additional requirement in Thm.\ \ref{transfer}), 
and s.t.\ all variables occurring in 
clauses in ${\cal K}$ occur below an extension function.

\item[Assumption Ground(${\sf Init}$):] ${\sf Init}({\overline x})$ is
  a quantifier-free $\Pi_S$-formula.

\vspace{-3mm}
\item[Assumption Ground($\Phi$):]  $\Phi({\overline x})$ and  $\neg
  \Phi({\overline x})$ are quantifier-free $\Pi_S$-formulae.

\vspace{-3mm}
\item[Assumption Ground(${\sf Update}$):] ${\sf Update} = {\sf Update}_{\Sigma, \Sigma'}({\overline x},
{\overline x'})$ is a
  quantifier-free $\Pi_S \cup V \cup \Sigma \cup V' \cup \Sigma'$-formula describing
  the updates of the variables $v \in V$ and of the function symbols
  $f \in \Sigma$. 
\end{description} 
\noindent {\bf Remark:} Although in assumptions {\bf Ground}({\sf
  Init}), {\bf Ground}($\Phi$) and {\bf Ground}({\sf Update}) the
involved formulae are not strictly-speaking ground, they only contain
free variables. Since we are testing satisfiability, the free
variables are replaced with Skolem constants, so in this case we need
to check satisfiability of ground formulae. This justifies the name we
used for these assumptions.   
\begin{example}
{\em 
We illustrate this type of assumptions on an example. 
Consider the system $S$ specified by the triple 
$(\Pi_S, {\cal T}_S, T_S)$, where ${\cal T}_S$ is the extension of 
the theory of linear arithmetic with an uninterpreted function 
symbol $f$ (modeling an array) and constants $n, l$ and $u$, and 
$T_S = (V, \Sigma, {\sf Init}, {\sf Update})$, where 
$V = \{ x, y \}, \Sigma = \{ f \}$, and: 
\begin{itemize}
\item ${\sf Init} :=  (x \geq 0 \wedge y \leq 2 \wedge f(x) \geq 4)$, stating that in
  the initial states the value of the variable $x$ is positive, the
  value of $y$ is at most 2 and the
  value of $f(x)$ is at least 4.
\item ${\sf Update} :=  x' := x + 2 \wedge y' = y - 3$, stating that the value of the
  variable $x$ is increased with 2 and the value of $y$ is decreased
  with 3.
\end{itemize}
Let $\Phi$ be the formula $(0 \leq x \leq n \wedge l \leq f(x) \leq
u)$, stating that the value of the variable $x$ is between 0 and $n$
and  $f(x)$ is between $l$ and $u$. 
\noindent 
Then Assumptions {\bf  Ground}({\sf Init}) , {\bf  Ground}({\sf Update}) and 
{\bf  Ground}($\Phi$) hold. 
}
\end{example}

\begin{description}
\item[Assumption Loc(${\sf Init}$):] ${\sf Init}$ is a set of clauses s.t.\
$\T_0 \subseteq \T_0 \cup \K \cup {\sf Init}$ is a local theory
extension; 
\vspace{-3mm}
\item[Assumption Loc($\Phi$):]  $\Phi$ is a set of clauses
  s.t.\ $\T_0 \subseteq \T_0 \cup \K \cup \Phi$ is a
local theory extension;  
\vspace{-3mm}
\item[Assumption Loc(${\sf Update}$):] ${\sf Update}$ is a set of
  clauses over the signature $\Pi_S$, 
primed and unprimed versions of the symbols
  in $V$ and $\Sigma$ such that
$\T_0 \cup \K \cup \Phi \subseteq \T_0 \cup \K \cup \Phi
\cup {\sf Update}$ is a local theory
extension.

\item[Assumption ELoc(${\sf Init}$):] ${\sf Init}$ is a set of  $\Pi_S$-clauses such that 
$\T_0 \subseteq  T_0 \cup \K \cup {\sf Init}$ is a theory
extension satisfying  ${\sf Comp}_{\sf f}$. 
\vspace{-3mm}
\item[Assumption ELoc($\Phi$):]  $\Phi$ is a set of  $\Pi_S$-clauses
  such that $\T_0 \subseteq \T_0 \cup \K \cup \Phi$ is a
theory extension satisfying  ${\sf Comp}_{\sf f}$.
\vspace{-3mm}
\item[Assumption ELoc(${\sf Update}$):] ${\sf Update}$ is a set of
  clauses over the signature $\Pi_S \cup V' \cup \Sigma'$ such that
$\T_0 \cup \K \cup \Phi \subseteq \T_0 \cup \K \cup \Phi
\cup {\sf Update}$ is a theory
extension satisfying  ${\sf Comp}_{\sf f}$. 
\end{description}

\begin{example}
{\em 
We illustrate the last two types of assumptions with examples: 
Consider the system $S$ specified by the triple 
$(\Pi_S, {\cal T}_S, T_S)$, where ${\cal T}_S$ is the extension of 
the theory $\T_0$ of linear integer arithmetic with an uninterpreted function 
symbol $f$ (modeling an array) and constants $n, l$ and $u$, and 
$T_S = (V, \Sigma, {\sf Init}, {\sf Update})$, where 
$V = \{ x \}, \Sigma = \{ f \}$, and: 
\begin{itemize}
\item ${\sf Init}:= \forall x ( 1 \leq x \leq n \rightarrow f(x) \geq
  0 )$, stating that in the initial state the values stored at
  all positions $x$ with $1 \leq x \leq n$ in the
  array modeled by $f$ are positive.
\end{itemize}
$~~~~~~ \begin{array}{@{}lllrcll}
\bullet  & {\sf Update}:= \{ & \forall x & ( 1 \leq x \leq n &
\rightarrow & 
  f'(x) = f(x) + 1) & \wedge \\
& & \forall x & (x > n & \rightarrow & f'(x) = f(x)) & \wedge \\
& & \forall x & (x < 1 & \rightarrow & f'(x) = f(x)) & \} 
\end{array}$ 
\begin{quote}
is a formula expressing the fact that
the array is updated such that for all 
positions from 1 to $n$ the value stored in $f$ is increased by one 
and the value stored in $f$ is not changed elsewhere.
\end{quote}

\noindent Let $\Phi$ be the formula $\forall x (1 \leq x \leq n
\rightarrow  l \leq f(x) \leq u)$, expressing the fact that for every
index in the range $[1, n]$, the value of $f$ is between $l$ and $u$.
\noindent Then we have the following local theory extensions: 
$$\begin{array}{rcl}
 \T_0 & \subseteq & \T_0 \cup {\sf Init} \\
 \T_0 & \subseteq & \T_0 \cup \Phi \\
 \T_0 \cup \Phi & \subseteq & \T_0 \cup \Phi \cup {\sf Update}
\end{array}$$
thus, Assumptions {\bf  Loc}({\sf Init}), {\bf  Loc}({\sf Update}) 
and {\bf  Loc}($\Phi$) hold. Since all theory extensions above satisfy 
in fact condition ${\sf Comp}_{\sf f}$, Assumptions {\bf  ELoc}({\sf Init}), {\bf  ELoc}({\sf Update}) 
and {\bf  ELoc}($\Phi$) hold.
}
\end{example}

\subsection{Case 1: Simple transition systems; only variables are updated } 
\label{sect:case1}

Assume first that in the description of the transition system $T_S$ only
constants are used, i.e.\ $\T_S$ is the extension of a theory $\T_0$ with
additional (free) constants and $\Sigma = \emptyset$, 
i.e. only variables are updated,  and the updates are 
quantifier-free formulae (typically assignments for the variables).
Theorems~\ref{case1:inv-checking-dec} and~\ref{thm:case1} identify situations in which invariant
checking is decidable and, if a formula is not an invariant, a
constraint on the parameters can be obtained under which invariance is guaranteed.  
\begin{theorem}
Let $(\Pi_S, \T_S, T_S)$ be a specification of a system $S$, where 
$T_S = (V, \Sigma, {\sf Init}, {\sf Update})$ is a transition system. 
Assume that $\Sigma = \emptyset$, i.e.\ 
the formulae ${\sf Update}({\overline x}, {\overline x}')$ describe
updates of the variables  and are quantifier-free and  the formulae 
${\sf Init}({\overline x}), \Phi({\overline x}),$ and $\neg \Phi({\overline x})$, 
all belong to a class  ${\cal F}$ of $\Pi_S$-formulae (closed under
conjunctions) for which
satisfiability is decidable. Let $\Gamma({\overline p})$ be a set
(i.e. conjunction) of constraints on the parameters ${\overline p}$ in $\Sigma_P$ 
belonging also to the fragment ${\cal F}$. 
Then   checking whether the formula $\Phi$ is an invariant (under conditions 
$\Gamma({\overline p})$) 
is decidable. 
\label{case1:inv-checking-dec}
\end{theorem} 
{\em Proof:} $\Phi$ is an inductive invariant under conditions 
$\Gamma({\overline p})$ iff 
(1) $\Gamma({\overline p}) \wedge {\sf Init}({\overline x}) \wedge \neg
\Phi({\overline x})$ is unsatisfiable w.r.t.\ $\T_S$, and 
(2) $\Gamma({\overline p}) \wedge \Phi({\overline x})
\wedge {\sf Update}({\overline x}, {\overline x'}) \wedge \neg
\Phi({\overline x'})$ is unsatisfiable w.r.t.\ $\T_S$. 
\noindent The formulae in (1) and (2) are in ${\cal F}$, thus a decision
procedure for satisfiability for ${\cal F}$ can be used to check their 
satisfiability. 
\QED

\medskip
\noindent If $\T_S$ is the extension with free constants of a theory
$\T_0$ that allows quantifier elimination we can establish more
general results. 

\begin{lemma}
Let $\T_0$ be a theory allowing quantifier elimination. 
Then checking satisfiability of arbitrary formulae in any extension $\T_S$
of $\T_0$ with free constants in a set $C$  is decidable. 
\label{lemma-qe-dec}
\end{lemma} 
{\em Proof:} Let $F(x_1, \dots, x_n, c_1, \dots, c_m)$ 
be an arbitrary $\Pi_0^C$-formula containing free
variables ${x_1, \dots, x_n}$ and constants $c_1, \dots, c_m \in C$. 
Then 
$F(x_1, \dots, x_n, c_1, \dots, c_m)$ is satisfiable in the extension
of  $\T_0$ with free constants in $C$
iff the formula $\exists x_1, \dots, x_n \exists x_{c_1}, \dots,
x_{c_m} F(x_1, \dots, x_n, x_{c_1}, \dots, x_{c_m})$ is valid w.r.t.\ $\T_0$, 
where $x_{c_1}, \dots, x_{c_m}$ are variables replacing the constants 
$c_1, \dots, c_m$. 
Since $\T_0$ allows quantifier elimination, $\exists x_1, \dots, x_n \exists x_{c_1}, \dots,
x_{c_m} F(x_1, \dots, x_n, c_1, \dots, c_m)$ is valid iff it is
satisfiable iff it is equivalent to $\top$ w.r.t. $\T_0$. This can be
checked  algorithmically using a method for eliminating the quantifiers
in $\T_0$. \QED

\begin{theorem}
Let $(\Pi_S, \T_S, T_S)$ be a specification of a system $S$, where 
$\Pi_S = (\Sigma_S, {\sf Pred})$, 
$T_S = (V, \Sigma, {\sf Init}, {\sf Update})$ is a transition system. 
Assume that $\Sigma = \emptyset$, i.e.\ 
the formulae ${\sf Update}({\overline x}, {\overline x}')$ describe
updates of the variables in $V$. Let ${\sf Init}$ and $\Phi$ be 
arbitrary $\Pi_S$-formulae and let ${\sf Update}$ be an arbitrary 
formula over $\Pi_S^{V'}$. 
Assume that $\T_S$ is the extension with
additional free constants of a theory $\T_0$ with signature
$(\Sigma_0, {\sf Pred})$ which allows 
quantifier elimination, i.e. 
$\Sigma_S$ is the disjoint union of $\Sigma_0$, $\Sigma_P$ (a set of
parametric constants) and $\Sigma_n$ (the remaining, non-parametric,
additional constants).  
%
%
Then the following hold: 
\begin{itemize}
\item[(a)] Checking whether a formula $\Phi$ is an inductive invariant
  is decidable. 
\item[(b)] We can use a method for quantifier elimination for $\T_0$ to effectively  
construct a weakest condition ${\overline \Gamma}$ 
on the parameters $\Sigma_P$ under which 
$\Phi$ is an invariant.
\end{itemize}
\label{thm:case1}
\end{theorem}
\noindent {\em Proof:} (a) is a direct consequence of
Lemma~\ref{lemma-qe-dec}. The complexity of the problem of checking
whether a formula $\Psi$ is an inductive invariant depends on the 
complexity of the method for quantifier elimination in $\T_0$ which is
used. 

\smallskip
\noindent (b) As in the proof of Lemma~\ref{lemma-qe-dec}, we will 
think of all the constants in $\Sigma_S \backslash \Sigma_0$ as 
variables. We will denote by  ${\overline p}$ the sequence of
variables corresponding to the parameters in the formulae we consider
and with ${\overline x_r}$ the remaining variables associated with
symbols in $\Sigma_S \backslash \Sigma_0$. With ${\overline x}$ we
denote sequences of variables in $V$, ${\overline x'}$ represent the
corresponding variables in $V'$.  

\smallskip
\noindent (1) Clearly,  ${\sf Init}({\overline x}) \wedge \neg
\Phi({\overline x})$ is satisfiable w.r.t.\ $\T_S$, iff 
$\exists {\overline x_r} ({\sf Init}({\overline x}) \wedge \neg
\Phi({\overline x}))$ is satisfiable w.r.t.\ $\T_0$. 
Since $\T_0$ allows quantifier elimination, there exists a 
quantifier-free formula $F_i({\overline p})$ equivalent w.r.t.\ $\T_0$
with $\exists {\overline x_r} ({\sf Init}({\overline x}) \wedge \neg
\Phi({\overline x}))$  (which is a formula with free variables
${\overline p}$). Let $\Gamma_i := \neg F_i({\overline p})$. 
It is easy to see that $\Gamma_i \wedge {\sf Init}({\overline x}) \wedge \neg
\Phi({\overline x})$ is unsatisfiable w.r.t.\ $\T_S$, and that 
$\Gamma_i$ is a weakest formula with this property: For every other
formula $\Gamma'_i$ over the parameters ${\overline p}$ such that 
$\Gamma'_i \wedge {\sf Init}({\overline x}) \wedge \neg
\Phi({\overline x})$ is unsatisfiable w.r.t.\ $\T_S$ we also have 
$\Gamma'_i \wedge \exists {\overline x_r} ({\sf Init}({\overline x}) \wedge \neg
\Phi({\overline x}))$ unsatisfiable w.r.t.\ $\T_0$, hence 
$\Gamma'_i \models_{\T_0} \neg \exists {\overline x_r} ({\sf Init}({\overline x}) \wedge \neg
\Phi({\overline x}))$, so $\Gamma'_i \models_{\T_0} \Gamma_i$. \QED

\smallskip
\noindent (2) Clearly,  $\Phi({\overline x})
\wedge {\sf Update}({\overline x}, {\overline x'}) \wedge \neg
\Phi({\overline x'})$ is satisfiable w.r.t.\ $\T_S$ iff the formula 
 $\exists {\overline x_r} \exists {\overline x'} (\Phi({\overline x})
\wedge {\sf Update}({\overline x}, {\overline x'}) \wedge \neg
\Phi({\overline x'}))$ is satisfiable w.r.t.\ $\T_0$. 
Since $\T_0$ allows quantifier elimination, there exists a 
quantifier-free formula $F_u({\overline p})$ equivalent w.r.t.\ $\T_0$
with $\exists {\overline x_r} \exists {\overline x'} (\Phi({\overline x})
\wedge {\sf Update}({\overline x}, {\overline x'}) \wedge \neg
\Phi({\overline x'}))$  (which is a formula with free variables
${\overline p}$). Let $\Gamma_u := \neg F_u({\overline p})$. 
It can be seen as before that $\Gamma_u \wedge  (\Phi({\overline x})
\wedge {\sf Update}({\overline x}, {\overline x'}) \wedge \neg
\Phi({\overline x'}))$ is unsatisfiable w.r.t.\ $\T_S$, and that
$\Gamma_u$ is a weakest formula with this property. 

\smallskip
\noindent We can now set $\overline{\Gamma} := \Gamma_i \wedge \Gamma_u$. 
\QED

\begin{example} 
{\em 
Consider the system in Example~\ref{ex1} in
Section~\ref{illustration}. Let ${\cal T}_S$ be ${\mathbb R}$, the
theory of real numbers, and $T_S$ be the transition constraint system
describing the following transition system, with parameters ${\sf in}, 
{\sf out},  L_{\sf overflow}$ and $L_{\sf alarm}$ (all
constants), where ${\sf Init}(L) = L \leq L_{\sf overflow}$ and

\medskip
{\footnotesize {\noindent

\

\

\smallskip
\hspace{1.6cm}\leavevmode
\epsfverbosetrue %
\def\epsfsize#1#2{0.35#1}              
~~~~~~~~~ \epsffile{water-tank1.eps} 
}

\medskip

\vspace{-2.2cm}
{\footnotesize \noindent $\begin{array}{lcl}
~~~& \!\!L' {:=} L {+} {\sf in} & \\[5ex]
~~~L' {:=} L {+} {\sf in} {-} {\sf out} ~~~& ~~~~~~~~~~~~~~~~~~~~~~~~~~~~~~~~~~& L' {:=} L {+} {\sf in} \\[6ex]
~~~& \!\!\!\!\!\!\!\!L' {:=}  L {+} {\sf in} {-} {\sf out} & \\
\end{array}$ 
}}

\

\

\noindent \begin{tabular}{@{}l@{}l} 
${\sf Update}(L, L') := $ & $\{ L \leq L_{\sf alarm} \rightarrow L' = L
  + {\sf in}, ~~~ L > L_{\sf alarm} \rightarrow L' = L
  + {\sf in} - {\sf out} \}$. 
\end{tabular}

\medskip
\noindent Assume that a set $\Gamma$ of  constraints 
on the parameters is given, e.g.\  
$$\Gamma = \{ {\sf in} = {\sf out} {-} 10, {\sf in}  = L_{\sf
  overflow} {-} L_{\sf alarm} {-} 10, {\sf in} > 0, {\sf out} > 0,  L_{\sf alarm} < L_{\sf overflow} \}.$$ 
The formula $L {\leq} L_{\sf overflow}$ is an inductive invariant iff
it holds in the initial states and the
formulae (i), (ii) are unsatisfiable w.r.t.\ ${\cal T}_S {\cup}
\Gamma$:
\begin{itemize}
\item[(i)] $\exists L, L' ( L \leq L_{\sf overflow} \wedge L > L_{\sf alarm} \wedge L' = L + {\sf in} - {\sf out} \wedge L' > L_{\sf overflow})$;
\item[(ii)] $\exists L, L' (L \leq L_{\sf overflow} \wedge L \leq L_{\sf alarm} \wedge L' = L + {\sf in} \wedge L' > L_{\sf overflow})$.
\end{itemize}
To check this we can use any decision procedure for real numbers (or real
closed fields), or for linear arithmetic over the reals. 
We can start with a smaller set of constraints on the parameters, for instance $\Gamma =
\{ L_{\sf alarm} < L_{\sf overflow} \}$. In this case, 
Theorem~\ref{thm:case1}(b) can be used in two different ways. 
\begin{itemize}
\vspace{-2mm}
\item On the one hand, we can use a quantifier elimination method -- for real
closed fields or for linear arithmetic over the reals -- for
eliminating the existentially quantified variables $\exists L, L'$ and
obtain a constraint on the parameters ${\sf in}, {\sf out}, L_{\sf
  overflow}$ and $L_{\sf alarm}$. 

For instance, the Fourier-Motzkin quantifier elimination method
applied on formula (i) yields the equivalent formula $L_{\sf overflow}
> L_{\sf alarm} \wedge {\sf in} > {\sf out}$. The negation,
together with the assumption that $L_{\sf overflow}
> L_{\sf alarm} $ yields condition $\Gamma_i = ({\sf in} \leq {\sf
  out})$.  
Using  the Fourier-Motzkin quantifier elimination method
applied on formula (ii) yields $0 < {\sf in} \wedge L_{\sf overflow} - L_{\sf alarm} <
{\sf in}$. Negating this formula we obtain condition $\Gamma_u = ({\sf
  in} \leq 0 \vee {\sf
  in} \leq L_{\sf overflow} - L_{\sf alarm})$; under the 
additional assumption that ${\sf in} > 0$ we would obtain
condition  $\Gamma_u = ({\sf in} \leq L_{\sf overflow} - L_{\sf  alarm})$. 
The weakest constraint on the parameters which guarantees that 
$L \leq L_{\sf overflow}$ is an inductive invariant is 
$\overline{\Gamma} = \Gamma_i \wedge \Gamma_u$.

\vspace{-3mm}
\item On the other hand, it can be used for generating 
invariants with a given shape, expressed using undetermined constants 
which can also be considered to be parameters.
\end{itemize}
}
\label{ex-case1-1}
\end{example}

\subsection{Case 2: Only variables are updated; some parameters are functions} 
\label{sect:case2}

Assume now that only variables change their 
value in updates, but some parameters of the system 
are functions. We assume that the theory $\T_S$ might contain
additional data structures, possibly with their axiomatization. Thus,
we assume that ${\cal T}_S = {\cal T}_0 \cup {\cal K}$, where $\T_0$
is a base theory and ${\cal K}$ is a set of axioms axiomatizing the
properties of the extension functions, possibly including the
parameters. 

\smallskip 
\noindent 
In this case the problem of checking whether a formula $\Psi$ is an
inductive invariant can be proved to be decidable under more
restrictive conditions than in Case 1.

\begin{theorem}
Let $(\Pi_S, \T_S, T_S)$ be a specification of a system $S$, where 
$\Pi_S = (\Sigma_S, {\sf Pred})$, 
$T_S = (V, \Sigma, {\sf Init}, {\sf Update})$ is a transition system. 
Assume that $\Sigma = \emptyset$, i.e.\ 
the formulae ${\sf Update}({\overline x}, {\overline x}')$ describe
updates of the variables in $V$.
Let $\Gamma({\overline p}) = \Gamma_0 \cup \Gamma_{\Sigma}$ be a set of 
additional constraints on the parameters of $T_S$ (not included in
$\T_0 \cup \K$)  
where:  
(i) $\Gamma_0$ is a set of constraints on the non-functional parameters 
and (ii) $\Gamma_{\Sigma}$ is a set of axioms containing functional
parameters. 
Assume that the formulae 
${\sf Init}({\overline x}), \Phi({\overline x}), \neg \Phi({\overline
  x}), {\sf Update}({\overline x}, {\overline x}'), \Gamma_0$, 
belong to a fragment  ${\cal F}$  (closed under conjunction) of the theory ${\cal T}_S\cup \Gamma_{\Sigma}$ extended
with new free constants in $V'$, for which 
checking satisfiability is decidable. 
Then  checking whether the formula $\Phi$ is an invariant 
(under conditions $\Gamma({\overline p})$ on the parameters) is decidable.
\label{dec-case2} 
\end{theorem}
\noindent {\em Proof:}  (1) Decidability of problems of type (1) on
page~\pageref{label-problems} is a consequence 
of the fact that ${\sf Init}({\overline x}) \wedge \neg \Phi({\overline x}) 
\models_{{\cal T}_S \cup \Gamma} \perp$ iff 
$\Gamma_0 \wedge {\sf Init}({\overline x}) \wedge \neg \Phi({\overline x})$ 
is unsatisfiable 
w.r.t.\ ${\cal T}_S \cup \Gamma_{\Sigma}$. Since, by assumption, 
$\Gamma_0 \wedge {\sf Init}({\overline x}) \wedge \neg \Phi({\overline
  x})$  is in ${\cal F}$, a decision procedure for satisfiability of
formulae in ${\cal F}$ can be used. 

\smallskip
(2) To prove decidability of problem (2) on
page~\pageref{label-problems} note that 
$\Phi({\overline x}) \wedge {\sf Update}({\overline x},{\overline x}') 
\wedge \neg \Phi({\overline x}')\models_{{\cal T}_S \cup \Gamma} \perp$  iff 
$\Gamma_0 \wedge \Phi({\overline x}) \wedge 
{\sf Update}({\overline x},{\overline x}') \wedge \neg \Phi({\overline x}')$ 
is unsatisfiable 
w.r.t.\ $ {\cal T}_S \cup \Gamma_{\Sigma}$. Since $\Gamma_0 \wedge \Phi({\overline x}) \wedge 
{\sf Update}({\overline x},{\overline x}') \wedge \neg \Phi({\overline
  x}')$ is in ${\cal F}$ also in this case, a decision procedure for
satisfiability of formulae in ${\cal F}$ w.r.t.\ $\T_S \cup
\Gamma_{\Sigma}$ can be used. \QED

\begin{lemma}
The conditions of Theorem~\ref{dec-case2} hold e.g.\ 
if $\Gamma_0$ is quantifier-free and the decidability of the problems 
above is a consequence of locality 
properties of certain theory extensions, i.e.\ under the following
assumptions. 
\begin{description}
\item[Loc($\T_S$) resp. Loc($\T_S \cup \Gamma_{\Sigma}$):] ${\cal
    T}_S$ (resp.\ $\T_S \cup \Gamma_{\Sigma}$) is an extension of a  $\Pi_0$-theory 
${\cal T}_0$ 
with a set ${\cal K}$ (resp.\ $\K \cup \Gamma_{\Sigma}$) of flat and linear clauses 
satisfying condition ${\sf Loc}$ s.t.\ all variables occurring in 
clauses in ${\cal K}$ (resp.\ $\K \cup \Gamma_{\Sigma}$) occur below
an extension function symbol. 

\vspace{-3mm}
\item[Ground($F$):] Assume that each of the formulae $F \in 
\{ {\sf Init}({\overline x}),~~ {\sf Update}({\overline x}, {\overline
  x}'),~~  \Phi({\overline x}) \}$  satisfy assumption 
{\bf Ground}$(F)$
(cf.\ notation introduced on page~\pageref{page-assumptions}).

\vspace{-3mm}
\item[Decidability($\T_0$):] Ground satisfiability w.r.t.\ $\T_0$ is decidable.  
\end{description}

\end{lemma}
{\em Proof:} Assumption {\bf Ground(}$F${)} for $F \in 
\{ {\sf Init}({\overline x}),~ {\sf Update}({\overline x}, {\overline
  x}'), ~ \Phi({\overline x}) \}$ ensures that these formulae are all
 quantifier free formulae (i.e.\ for satisfiability tests can be
 regarded as Skolemized ground
 formulae). Assumption {\bf Loc($\T_S$)}  ensures that the
 satisfiability of ground formulae w.r.t.\ $\T_S$ is decidable. 
Assumption {\bf Loc($\T_S \cup \Gamma_{\Sigma}$)}  ensures that the
 satisfiability of ground formulae w.r.t.\ $\T_S \cup \Gamma_{\Sigma}$
 is decidable, hence that condition (ii) holds. \QED

\medskip
\noindent For clarity, in the conditions above and Theorem~\ref{case2} we consider the 
particular case in which locality allows us to reduce all proof tasks to 
satisfiability checks for {\em ground} formulae w.r.t.\ ${\cal T}_0$. 
We will then briefly discuss the way the results extend in the presence of 
extended locality conditions.
\begin{theorem}
\label{case2}
Assume that 
{\bf Ground}($F$) holds for every formula $F \in \{ {\sf Init},~
\Phi, ~{\sf Update} \}$).  
\begin{itemize}
\vspace{-2mm}
\item[(a)] Assume that ground satisfiability of formulae in ${\cal
    T}_0$ is decidable.  
Let $\Gamma = \Gamma_0 \cup \Gamma_{\Sigma}$ be a set of additional constraints 
on $\Sigma_P {\subseteq} \Sigma_S$, 
s.t.\ 
$\Gamma_0$ is a quantifier-free $\Pi_0$-formula with no variables (only with 
parameters) and {\bf Loc}($\T_S \cup \Gamma_{\Sigma})$
holds. 
Then checking whether $\Phi$ is an invariant 
(under conditions $\Gamma$) is decidable. 

\vspace{-3mm}
\item[(b)] If the theory ${\cal T}_0$ 
has quantifier elimination, this can be used 
to construct a condition ${\overline \Gamma}$ 
on the parameters under which  
$\Phi$ is an invariant. 

If in addition the theory extension $\T_0 \subseteq \T_S = \T_0 \cup \K$ satisfies the locality condition ${\sf
  Comp}_{\sf f}$ then we can construct the weakest condition
${\overline \Gamma}$  
on the parameters under which  
$\Phi$ is an invariant.
\end{itemize}
\end{theorem}
{\em Proof:} 
Both checking whether $\Phi$ is true in the initial states and
checking invariance of $\Phi$ under updates can be formulated as 
ground satisfiability problems w.r.t.\ $\T_0 \cup \K \cup
\Gamma_{\Sigma}$.  

\smallskip
\noindent (a) Since 
${\cal T}_0  {\subseteq} {\cal T}_0 {\cup} {\cal K} 
{\cup} \Gamma_{\Sigma}$ satisfies condition ${\sf Loc}$, by
Theorem~\ref{lemma-rel-transl} the following are equivalent: 
\begin{itemize}
\item ${\sf Init} \wedge \neg \Phi \models_{\T_S \cup \Gamma} \perp$.
\item $\T_0 \cup \K \cup \Gamma_{\Sigma} \cup \Gamma_0 \cup {\sf Init} \cup \neg
  \Phi \models \perp$.
\item $\T_0 \cup \K \cup \Gamma_{\Sigma} \cup G \models \perp$, where
  $G$ is a set of ground clauses obtained from $\Gamma_0 \cup {\sf Init} \cup \neg
  \Phi$ by replacing the free variables with Skolem constants and
  transformation to clause form. 
\item $\T_0 \cup (\K \cup \Gamma_{\Sigma})[G] \cup G \models \perp$. 
\item $\T_0 \cup (\K \cup \Gamma_{\Sigma})_0 \cup G_0 \cup {\sf Def}
  \models \perp$  (with the notations used in Theorem~\ref{lemma-rel-transl}). 
\item $\T_0 \cup (\K \cup \Gamma_{\Sigma})_0 \cup G_0 \cup {\sf Con}[G]_0
  \models \perp$, where 

$\displaystyle{~~~ {\sf Con}[G]_0  = \{ \bigwedge_{i = 1}^n c_i = d_i \rightarrow c = d \mid 
f(c_1, \dots, c_n) = c, f(d_1, \dots, d_n) = d \in {\sf Def} \}}.$
\end{itemize}
The last test is a satisfiability test for ground formulae w.r.t.\
$\T_0$, a problem which we assumed to be decidable. 
Similarly for checking invariance under updates. The following are
equivalent: 
\begin{itemize}
\item $\Phi \wedge {\sf Update} \wedge \neg \Phi' \models_{\T_S \cup \Gamma} \perp$.
\item $\T_0 \cup \K \cup \Gamma_{\Sigma} \cup (\Gamma_0 \cup \Phi
  \wedge {\sf Update} \cup \neg
  \Phi') \models \perp$.
\item $\T_0 \cup \K \cup \Gamma_{\Sigma} \cup G \models \perp$, where
  $G$ is a set of ground clauses obtained from $\Gamma_0 \cup \Phi
  \cup {\sf Update} \cup \neg \Phi'$ by replacing the free variables with Skolem constants and
  transformation to clause form. 
\item $\T_0 \cup (\K \cup \Gamma_{\Sigma})[G] \cup G \models \perp$. 
\item $\T_0 \cup (\K \cup \Gamma_{\Sigma})_0 \cup G_0 \cup {\sf Def}
  \models \perp$  (with the notations used in Theorem~\ref{lemma-rel-transl}). 
\item $\T_0 \cup (\K \cup \Gamma_{\Sigma})_0 \cup G_0 \cup {\sf Con}[G]_0
  \models \perp$, where 

$\displaystyle{~~~ {\sf Con}[G]_0  = \{ \bigwedge_{i = 1}^n c_i = d_i \rightarrow c = d \mid 
f(c_1, \dots, c_n) = c, f(d_1, \dots, d_n) = d \in {\sf Def} \}}.$
\end{itemize}

\smallskip
\noindent 
(b) As explained before, the fact that the initial states satisfy $\Phi$ can clearly be 
expressed as a satisfiability problem w.r.t.\ ${\cal T}_S$, and 
can be reduced to a satisfiability problem w.r.t.\ $\T_0$ using 
hierarchic reasoning in local theory extensions. 
After purification we can use the 
symbol elimination method in Section~\ref{symbol-elimination}
to obtain a constraint $\overline{\Gamma}_i$ under
which $\Phi$ holds in the initial states.  

\smallskip
\noindent $\Phi$ is invariant under updates\footnote{We here indicate
  explicitly the variables that appear in $\Phi$ and are updated in
  order to give a better intuition about the form of the
  satisfiability   problems we have to consider.}  iff 
$\exists {\overline x} \exists {\overline x'} \, (\Phi({\overline x}) \wedge {\sf Update}({\overline x}, {\overline x'}) \wedge \neg \Phi({\overline x}')) \models_{{\cal T}_0 \cup {\cal K}}  \bot.$ 
As ${\cal T}_0 \subseteq {\cal T}_0 \cup {\cal K}$ is 
local,  we can use Steps 1-5 in the symbol elimination method in Section~\ref{symbol-elimination}
to obtain a constraint $\overline{\Gamma}_u$ under
which $\Phi$ is invariant under updates. 
Then under additional conditions on the parameters 
$\overline{\Gamma} = \overline{\Gamma}_i \wedge \overline{\Gamma}_u$
$\Phi$ is guaranteed to be an inductive invariant. 

\smallskip
\noindent 
If in addition we assume that the local extension $\T_0 \subseteq \T_0
\cup \K$ satisfies condition ${\sf Comp}_f$ then, by Theorem~\ref{symb-elim-weakest},
${\overline \Gamma}_i$ is the weakest condition under
which all initial states satisfy $\Phi$ and $\overline{\Gamma}_u$ is
the weakest condition under which $\Phi$ is invariant under updates. 
\QED
\begin{example}
{\em 
Consider the water controller in Example~\ref{ex2} in Section~\ref{illustration}
where the inflow ${\sf in}$ depends 
on time. We assume that time is discrete (modeled by the integers), 
and that the values of the water level are real numbers. 
Let ${\cal T}_S = {\cal T}_0 \cup {\sf Free}({\sf in})$ be the 
extension of the many-sorted combination ${\cal T}_0$ 
of  ${\mathbb Z}$ 
(Presburger arithmetic) and ${\mathbb R}$ (the theory of reals) with the  
free function ${\sf in}$. 
We determine a set $\Gamma$ of constraints, such that $L \leq L_{\sf overflow}$ is an 
invariant under updates under assumptions ${\Gamma}$ as follows: 
$L \leq L_{\sf overflow}$ is invariant under updates iff 
\begin{itemize}
\item[(i)] $(L {\leq} L_{\sf overflow} \wedge L {>} L_{\sf alarm} \wedge L' {=} L {+}
{\sf in}(t) {-} {\sf out} \wedge t' {=} t {+} 1 \wedge L' {>} L_{\sf
  overflow})$ is unsatisfiable and  
\item[(ii)] $(L {\leq} L_{\sf overflow} \wedge L {\leq} L_{\sf alarm}
  \wedge L' {=} L {+} {\sf in}(t) \wedge t' {=} t {+} 1 \wedge L' {>}
  L_{\sf overflow})$ is unsatisfiable.
\end{itemize}
Consider formula (i). We apply Algorithm~\ref{alg-symb-elim}
for symbol elimination described in Section~\ref{symbol-elimination},
where $\Sigma_P = \{ {\sf in}, {\sf out}, L_{\sf overflow}, L_{\sf
  alarm} \}$. 
\begin{description}
\item[Step 1:] After purification we obtain:  
$$(L {\leq}  L_{\sf overflow} \wedge L > L_{\sf alarm} \wedge L' {=} L {+}
{\sf in}_0 {-} {\sf out} \wedge t' {=} t {+} 1 \wedge L' {>} L_{\sf
  overflow}).$$
As ${\sf Def} = \{ {\sf in}_0 = {\sf
  in}(t) \}$, no instances of the congruence axioms are needed. 
\item[Step 2:] The constants corresponding to terms  starting with
  parameters are ${\sf in}_0, {\sf out}, L_{\sf overflow}, L_{\sf alarm}$. 
The constant occurring as an argument of a parameter in
    ${\sf Def}$ is $t$. The other constants are $t', L, L'$; they are
    regarded as existentially quantified variables.  
We obtain: 

$\exists L \exists L' \exists t' (L {\leq}  L_{\sf overflow} \wedge L > L_{\sf alarm} \wedge L' {=} L {+}
{\sf in}_0 {-} {\sf out} \wedge t' {=} t {+} 1 \wedge L' {>} L_{\sf
  overflow}).$

\item[Step 3:] We now eliminate the existential variables
and obtain:  
$L_{\sf overflow} >  L_{\sf alarm} \wedge {\sf in}_0 {-} {\sf
  out} > 0$. 

\item[Step 4:] We replace ${\sf in}_0$ with ${\sf in}(t)$, where $t$
  is existentially quantified and obtain:

 $\exists t (L_{\sf overflow} >  L_{\sf alarm} \wedge {\sf
   in}(t) {-} {\sf out} > 0).$
 
\item[Step 5:] We negate the formula and obtain: 
$ L_{\sf overflow} \leq L_{\sf alarm} \vee \forall t ({\sf in}(t) {-} {\sf out} \leq 0).$
\end{description}
If we assume that $L_{\sf alarm} < L_{\sf overflow}$, we obtain $\forall t ({\sf in}(t) {-} {\sf out} \leq 0).$
Thus, $\overline{\Gamma}_1 = \forall t ({\sf in}(t) - {\sf out} \leq 0)$. 
For formula (ii) we can similarly construct (under the assumption that
${\sf in} > 0$) 
$\overline{\Gamma}_2 = \forall t ({\sf in}(t) \leq 
L_{\sf overflow} {-} L_{\sf alarm})$. By the locality of the extension with the free function 
${\sf in}$, $\overline{\Gamma}_1$ and $\overline{\Gamma}_2$ are the
weakest constraints under which (i) resp. (ii) hold. 
Thus, ${\overline \Gamma} = \overline{\Gamma}_1 \wedge \overline{\Gamma}_2$ 
is the weakest constraint under which $L \leq L_{\sf overflow}$ is
invariant under updates (under the initial assumptions ${\sf in} > 0$
and $L_{\sf alarm} < L_{\sf overflow}$). 
}
\label{ex-case2}
\end{example}
The assumptions of Theorem~\ref{case2} can be further relaxed:
Theorem~\ref{thm:case1a} identifies situations in which 
possibilities for hierarchical reasoning and hierarchical symbol
elimination can be used as decision procedures for invariant checking
resp. for inferring constraints on the parameters.

\begin{theorem}
Assume that the theory ${\cal T}_S$ is the extension of a ``base
  theory'' ${\cal T}_0$ (allowing quantifier elimination) 
with additional constants and function symbols whose properties are 
axiomatized by a set $\K$ of clauses, such that $\T_0 \subseteq \T_S
\cup \Gamma_{\Sigma} =
\T_0 \cup \K \cup \Gamma_{\Sigma}$ is a local extension and all 
conditions of assumption
{\bf Loc}$(\T_S \cup \Gamma_{\Sigma})$ hold. 
 Assume that {\bf  Ground}$(\Phi)$ holds, and either {\bf
   Ground}$({\sf Init})$ holds or 
${\sf Init}$ is a set of
  clauses such that $\T_0 \subseteq \T_0
  \cup \K \cup \Gamma_{\Sigma} \cup {\sf Init}$ is a
  local theory extension and all variables in $\K \cup \Gamma_{\Sigma}\cup {\sf Init}$ occur below an
  extension function; 
%
and that ${\sf Update}({\overline x}, {\overline x}')$ and $\Gamma_0$
are quantifier-free formulae. 
If the fragment of the theory $\T_0$ to which hierarchical
  reduction in Theorem~\ref{lemma-rel-transl} leads is decidable in all cases 
then the following hold: 
\begin{itemize}
\vspace{-3mm}
\item[(a)] Hierarchical
reasoning can be used as a decision procedure for checking whether
$\Phi$ is an inductive invariant of $S$,
\vspace{-3mm}
\item[(b)] 
If the theory $\T_0$ has quantifier elimination, then the method for 
symbol elimination in theory extensions described in Section~\ref{symbol-elimination} 
can be used to effectively  
strengthen condition $\Gamma_{\Sigma} \cup \Gamma_0$ to a condition ${\overline \Gamma}$ 
on the parameters under which 
$\Phi$ is an invariant.

If in addition all theory extensions satisfy the conditions in
Theorem~\ref{symb-elim-weakest}
(all formulae in $\K, {\sf Init}, \Gamma_{\Sigma}$ and $\Phi$ are flat and linear
clauses, all local extensions satisfy condition $({\sf Comp}_{f})$, 
and  all variables occur below an extension function)  
then the method can be used to effectively  
construct {\em the weakest condition} ${\overline \Gamma}$ 
on the parameters under which 
$\Phi$ is an invariant.
\end{itemize}
\vspace{-4mm}
\label{thm:case1a}
\end{theorem}
{\em Proof:} Consequence of  Theorems~\ref{lemma-rel-transl},~\ref{complex},~\ref{inv-trans-qe} and \ref{symb-elim-weakest}. \QED

\begin{example}
{\em 
Consider the program computing the maximum of the
first $n$ elements of an array of real numbers $a$ in 
the range $[v_{\sf min}, v_{\sf max}]$.  Consider the formula 
$\Phi = \forall k (1 \leq k \leq i \rightarrow a(k) \leq {\sf max})$. 

\smallskip
\begin{minipage}[t]{.35\textwidth}
\raggedleft

$\begin{array}{ll}
1 & i := 1; {\sf max} := a[1]; \\
2 & {\sf while} ~ i < n: \\
3 & ~~~~ i := i + 1;\\
4 & ~~~~ {\sf if } ~ {\sf max} \leq a[i] ~ {\sf then } ~ {\sf max} := a[i] \\
\end{array}$

\end{minipage} \hfill
\begin{minipage}{.55\textwidth}
\raggedright


Let ${\sf Init} = \{ i = 1, {\sf max} = m \}$, where $m$ is a parameter. 
Let $\Gamma$ be the following constraint on the parameters $n, m,
v_{\sf min}, v_{\sf max}$: $\Gamma = \{ n \geq 1,  v_{\sf min} \leq
v_{\sf max}, a(1) \leq m \}$. \\
We show that if $\Gamma$ holds, $\Phi$ is an inductive invariant. 
\end{minipage}

\medskip
\noindent This task can be reduced to checking
satisfiability in the theory
${\cal T}_S$ which is the extension of the 
combination of $LI({\mathbb Z})$ (sort {\sf int}), the theory
of real numbers (sort {\sf real}) with 
a function symbol $a$ satisfying the axiom ${\cal K} = \forall k (1
\leq k \leq n \rightarrow v_{\sf min} \leq a(k) \leq v_{\sf max})$. 
The hierarchical reasoning method in
Theorem~\ref{lemma-rel-transl} can be used for this. Below we 
informally explain why the formulae are unsatisfiable.  
\begin{itemize}
\item To check that $\Phi$ is true in the initial states we have to
prove that $\Gamma\cup {\cal K} \cup {\sf Init} \cup \neg \Phi$
is unsatisfiable. 
This is so since the corresponding Skolemized formula is
unsatisfiable: 

$\begin{array}{l}
n \geq 1 \wedge v_{\sf min} \leq v_{\sf max} \wedge a(1) \leq m ~~\wedge~~ \forall k (1
\leq k \leq i \rightarrow v_{\sf min} \leq a(k) \leq v_{\sf max})
\wedge \\
i = 1 \wedge {\sf max} = m ~~\wedge~~  1 \leq k_0 \leq 1 \wedge
a(k_0) > {\sf max}
\end{array}$ 

and since $i = 1$ it follows that $k_0 = 1$, so on the one hand $a(1) \leq
m = {\sf max}$ and on the other hand $a(1) > {\sf max}$.

\item To check that $\Phi$ is invariant under the update in line 1 we have
  to prove the unsatisfiability of: 

$ \begin{array}{l} 
n \geq 1 \wedge v_{\sf min} \leq
v_{\sf max} \wedge a(1) \leq m ~~\wedge~~ \forall k (1 \leq k \leq n \rightarrow v_{\sf min}
\leq a(k) \leq v_{\sf max}) \wedge \\
\forall k (1 \leq k \leq 1 \rightarrow a(k) \leq
{\sf max}) ~~\wedge ~~ i' = 1 \wedge {\sf max}' = a(1)  ~~\wedge~~ (1 \leq k_0 \leq i') \wedge a(k_0)
> {\sf max}'. 
\end{array}$

\smallskip
\noindent 
The formula is unsatisfiable: it entails on the one hand ${\sf max}' = a(1)$ and 
on the other hand $k_0 = 1$ and hence $a(1) > {\sf max}'$. 


\item To prove that $\Phi$ is a loop invariant we prove that: 
$\Gamma \wedge \K \wedge  {\sf Update} \wedge \Phi \wedge \neg \Phi'$ is
unsatisfiable, where ${\sf Update}$  is the formula 

$i' = i + 1 \wedge ({\sf max} \leq a(i') \rightarrow {\sf
  max}' = a(i')) \wedge ({\sf max} > a(i) \rightarrow {\sf max}'
= 
{\sf max}), $

i.e. that the formula obtained after Skolemization is unsatisfiable: 

$\begin{array}{l} 
n \geq 1 \wedge v_{\sf min} \leq
v_{\sf max} \wedge  a(1) \leq m \wedge \forall k (1 \leq k \leq n \rightarrow v_{\sf min}
\leq a(k) \leq v_{\sf max}) \wedge \\
i' =  i + 1 \wedge  ({\sf max} \leq a(i') \rightarrow {\sf max}' = a(i')) \wedge
 ({\sf max} > a(i') \rightarrow {\sf max}' = {\sf max}) \wedge \\
 \forall k  (1 \leq k \leq i \rightarrow a(k) \leq {\sf
  max}) ~~\wedge~~ (1 \leq k_0 \leq i' \wedge a(k_0) > {\sf max}'). 
\end{array}$

Also this formula is unsatisfiable: Since $a(k_0) > {\sf max}'$, we
must have $k_0= i+1$ (otherwise we obtain a contradiction with
$\Phi$). We have two cases (1) ${\sf max} \leq a(i+1)$
and ${\max}' = a(i+1)$ or (2) ${\sf max} > a(i+1)$ and ${\sf max}' =
{\sf max}$. In both cases $a(i+1) \leq {\sf max}'$, which leads to a
contradiction.  
  
\end{itemize}
These are all ground satisfiability problems with respect to the extension
of linear arithmetic with a new function symbol $a$ satisfying
boundedness axioms $\forall k  (1 \leq k \leq i \rightarrow a(k) \leq {\sf
  max}) \wedge {\cal K}$. 
By the results presented in Section~\ref{examples-local-extensions},
this is a local extension, thus ground satisfiability is decidable;
for this we can use the hierarchical reduction in
Theorem~\ref{lemma-rel-transl}. 

\smallskip
\noindent The formulae $\Phi, \neg \Phi'$ and ${\sf Update}$ are in 
the fragment ${\cal F}$ consisting
of all conjunctions of ground formulae and formulae of the form 
$\forall k  (1 \leq k \leq i \rightarrow a(k) \leq {\sf max})$,  
$\forall k  (1 \leq k \leq n \rightarrow v_{\sf min} \leq a(k))$ and 
$\forall k  (1 \leq k \leq n \rightarrow a(k) \leq v_{\sf max})$ for which satisfiability
is decidable due to locality. 

\medskip
\noindent Assume now that we do not assume that condition $a(1) \leq m$
in $\Gamma$ holds, but
want to derive a constraint on the parameter $m$ which would guarantee
that $\Phi$ is an invariant of the program. It can be checked as
explained before that $\Phi$ is invariant under updates also without the additional
conditions $\Gamma$. $\Phi$ holds in the initial states iff the following formula is
unsatisfiable:  

$ \begin{array}{l} 
\forall k (1 \leq k \leq n \rightarrow v_{\sf min} \leq a(k) \leq v_{\sf max})
\wedge (i = 1) \wedge {\sf max} = m \wedge 1 \leq k_0 \leq i \wedge a(k_0) > {\sf max}
\end{array}$

\smallskip
\noindent The formula can be proved to be satisfiable; to obtain a constraint on
the parameters $n, m, v_{\sf min}, v_{\sf max}$ which would guarantee
that the formula is an invariant we follow the steps of the symbol
elimination method described in Section~\ref{symbol-elimination}. 
\begin{description}
\item[Step 1:] After instantiating and purifying the formula
  (replacing $a(k_0)$ with $a_0$) we obtain: 

$ (1 {\leq} k_0 {\leq} n \rightarrow v_{\sf min} {\leq} a_0 {\leq} v_{\sf
  max})  \wedge  i {=} 1 \wedge {\sf max} = m \wedge 
\wedge 1 {\leq} k_0 {\leq} i \wedge a_0 {>} {\sf max}.$

\item[Step 2:] Identify the constants corresponding to parameters:
  $n, v_{\sf min}, v_{\sf max}, m$.  The other symbols $i, k_0, a_0, {\sf
    max}$ are regarded as existentially quantified variables. 
\item[Step 3:] Eliminate the existentially quantified variables $i, k_0$ by replacing
  them with 1 and ${\sf max}$ by replacing it with $m$ and obtain: 

$(1 {\leq}  n \rightarrow v_{\sf min} {\leq} a_0 {\leq} v_{\sf max})  \wedge  a_0 {>} m$

then eliminate $a_0$ using a quantifier elimination
  method for linear real arithmetic: 

$n < 1 \vee (m < v_{\sf max} \wedge v_{\sf min} \leq v_{\sf max}).$

\item[Step 4:] The negation of this formula is: 
$1 \leq n \wedge (v_{\sf max} \leq m \vee v_{\sf min} > v_{\sf
  max})$. 

No constants need to be replaced with the term it represents. 

\item[Step 5:] The conjunction of the negation of this formula with
  the assumptions $n {\geq} 1 \wedge v_{\sf min} {\leq} v_{\sf max}$
  is equivalent to $\Gamma = (n {\geq} 1 \wedge v_{\sf min} {\leq}
  v_{\sf max} \wedge v_{\sf min} \leq m)$. 
\end{description}

\noindent Alternatively, we can choose to consider $a_0$ as a
parameter and thus not to eliminate $a_0$, then we obtain 
$$\neg (1 {\leq}  n \rightarrow v_{\sf min} {\leq} a(1) {\leq} v_{\sf
  max})  \vee a(1) \leq m;$$
the conjunction of this formula with $\K$ is equivalent to 
$\K \wedge a(1) \leq m$. 

\medskip
\noindent The invariant $\Phi_1 := \exists k (1 \leq k \leq i \wedge
a(k) = {\sf max})$ can be handled similarly -- all properties are
still satisfied because the formulae we consider will contain the 
universally quantified formula $\neg \Phi_1 = \forall k (1 \leq k \leq
i \rightarrow a(k) \neq {\sf max})$. 
}
\end{example}
\noindent {\bf Comment.}
For fully exploiting the power of extended locality, we can relax 
the assumptions of Theorems~\ref{case2} and \ref{thm:case1a} 
and allow ${\cal K}$ and $\Gamma$ to consist of augmented 
clauses, require that ${\cal T}_0 \subseteq {\cal T}_0 \cup \Gamma$ 
satisfies ${\sf ELoc}$; 
allow that ${\sf Init}({\overline x}), \Phi({\overline x}), 
\neg \Phi({\overline x})$ and 
${\sf Update}({\overline x}, {\overline x}')$ consist of augmented 
clauses in which
arbitrary $\Pi_0$-formulae are allowed to appear (and the extension
terms in ${\sf Update}$ are ground). The decidability 
results still hold if we can guarantee that the formulae we 
obtain with the hierarchical reduction belong to a fragment for which 
satisfiability w.r.t.\ ${\cal T}_0$ is decidable.

\subsection{Case 3: Both variables and functions can be updated} 
\label{sect:case3}

Assume now that both variables in $V$ and functions in $\Sigma$ 
may change their values during the transitions. 
We consider transition constraint systems $T_S$
in which the formulae in ${\sf Update}$ contain variables in 
$X$ and functions in $\Sigma$ and possibly parameters in $\Sigma_P$.
We assume that $\Sigma_P \cap \Sigma = \emptyset$. 
We therefore 
assume that the background theory ${\cal T}_S$ 
is an extension of 
a $\Pi_0$-theory ${\cal T}_0$ with axioms ${\cal K}$
specifying the properties of the functions in 
$\Sigma_S \backslash \Sigma_0$. 
We make the following assumptions:
\begin{description}
\item[Assumption 1:] 
${\cal T}_0 {\cup} {\cal K} {\cup} {\sf Init}$ and 
${\cal T}_0 {\cup} {\cal K} {\cup} \Phi$
are extensions of ${\cal T}_0$ with flat and linear clauses in $\K \cup
{\sf Init}$ resp.\ $\K \cup \Phi$ satisfying ${\sf ELoc}$ (and the additional 
requirements in Thm.~\ref{transfer}), such that all variables occur
below extension functions. 

\vspace{-3mm}
\item[Assumption 2:] For every $f \in \Sigma$, 
${\sf Update}(f, f')$ -- describing the update rules for $f$ 
-- 
is a set of clauses which, for every $\Pi_S$-theory ${\cal T}$,  
defines an extension with a new function $f' \not\in \Sigma$, such that   
${\cal T} {\subseteq} {\cal T} {\cup} {\sf Update}(f, f')$ satisfies ${\sf ELoc}$. 
\footnote{This is always the case if ${\sf Update}(f, f')$ are 
updates by definitions for $f'$ by (disjoint) case distinction  
or updates in which guarded boundedness conditions are specified
depending on case distinctions.} 
\footnote{By the results in \cite{ihlemann-sofronie-ijcar10} then also
  ${\cal T} {\subseteq} {\cal T} {\cup} \bigcup_{f \in \Sigma} {\sf
    Update}(f, f')$ satisfies ${\sf ELoc}$.} 
\end{description}
\begin{theorem}
Under Assumptions 1 and 2 the following hold:
\begin{itemize}
\vspace{-3mm}
\item[(a)] Assume that ground satisfiability of formulae 
in ${\cal T}_0$ is decidable. Let $\Gamma$ be a set of clauses 
expressing constraints on parameters in $\Sigma_P$  
s.t.\ ${\cal T}_0 \subseteq {\cal T}_0 \cup \Gamma$ 
is a local extension and all variables in $\Gamma$ occur below
extension functions and one of the following conditions holds: 
\begin{itemize}
\item[(i)] $\Gamma$ does not contain (non-constant)
functions in $\Sigma_P$ occurring also in $\K$ or ${\sf Init}$ or
$\Phi$, or   
\item[(ii)] ${\cal T}_0 \cup \Gamma \subseteq {\cal T}_0 \cup \Gamma
  \cup \K \cup {\sf Init}$ and ${\cal T}_0 \cup \Gamma \subseteq {\cal T}_0 \cup \Gamma
  \cup \K \cup \Phi$ satisfy condition ${\sf Comp_f}$. 
\end{itemize} 
Then checking whether $\Phi$ is an invariant 
(under conditions $\Gamma$) is decidable.

\vspace{-3mm}
\item[(b)] If the theory ${\cal T}_0$ 
has quantifier elimination, this can be used 
to construct a weakest condition ${\overline \Gamma}$
on the parameters under which 
$\Phi$ is an invariant. 
\end{itemize}
\label{case3}

\vspace{-4mm}
\end{theorem}
{\em Proof:} (a) If (i) holds, then by Theorem~\ref{transfer}, Assumption~1 implies that 
${\cal T}_0 \cup \Gamma  \subseteq {\cal T}_0 \cup {\Gamma} \cup {\cal K} \cup {\sf Init}$ and 
${\cal T}_0 \cup \Gamma \subseteq {\cal T}_0 \cup \Gamma \cup {\cal K}  \cup \Phi$
satisfy condition ${\sf (ELoc)}$, hence (ii) holds. 
We first analyze the problem of showing that initial states satisfy 
$\Phi$ under conditions $\Gamma$.
Since ${\cal T}_0 \cup \Gamma  \subseteq {\cal T}_0 \cup {\Gamma} \cup {\cal K} \cup {\sf Init}$ satisfies condition ${\sf (ELoc)}$, 
the following are equivalent:
\begin{itemize}
\item[(1)] $\exists {\overline x} ({\sf Init} \wedge \neg \Phi)
  \models_{{\cal T}_0 \cup \Gamma \cup {\cal K}} \perp$.
\item[(2)] $\T_0 \cup \Gamma \cup \K \cup {\sf Init} \cup \neg \Phi
  \models \perp$. 
\item[(3)] $\T_0 \cup \Gamma \cup (\K \cup {\sf Init})[G] \cup G
  \models \perp$, \\
where $G$ is the set of clauses obtained from $\neg
  \Phi$ by Skolemization and translation to clause form. 
\item[(4)] $\T_0 \cup \Gamma \cup (\K \cup {\sf Init})_0 \cup G_0 \cup
  {\sf Def} \models \perp$ (with the notation in
  Theorem~\ref{lemma-rel-transl}). 
\item[(5)] $\T_0 \cup \Gamma \cup (\K \cup {\sf Init})_0 \cup G_0 \cup
  {\sf Con}[G]_0 \models \perp$ (again, with the notation in
  Theorem~\ref{lemma-rel-transl}).
\item[(6)] ${\cal T}_0 \cup \Gamma  \cup G' \models \perp$ where 
$G' =  (\K \cup {\sf Init})_0 \cup G_0 \cup
  {\sf Con}[G]_0 $, a ground formula because of the condition that all
  variables in $\K$ and ${\sf Init}$ occur below extension symbols. 
\end{itemize}
${\cal T}_0 \subseteq {\cal T}_0 \cup \Gamma$ is a local extension, so 
(6) can be reduced to a ground 
satisfiability check w.r.t.\ ${\cal T}_0$ (which is decidable): (6) is
equivalent to the following: 
\begin{itemize}
\item[(7)] ${\cal T}_0 \cup \Gamma[G'] \cup G' \models \perp$.
\item[(8)] ${\cal T}_0 \cup \Gamma[G']_0 \cup G'_0 \cup {\sf Def}'
  \models \perp$ (with the notation in
  Theorem~\ref{lemma-rel-transl}). 
\item[(9)] $\T_0 \cup \Gamma[G']_0  \cup G'_0 \cup {\sf Con}[G']_0 \models \perp$ (again, with the notation in
  Theorem~\ref{lemma-rel-transl}).
\end{itemize}
Consider now invariance under updates. We have to show that: 
\begin{itemize}
\item[(10)] $\T_0 \cup \Gamma \cup \K \cup \Phi \cup \bigcup_{f \in
    \Sigma} {\sf Update}(f, f') \cup \neg \Phi'
\models \perp$
\end{itemize}
(where $\Phi'$ is obtained from $\Phi$ by replacing each $f$ with $f'$).
We have the following chain of theory extensions, all satisfying
condition ${\sf (ELoc)}$: 
$$ \T_0 \subseteq \T_0 \cup \Gamma \subseteq \T_0 \cup \Gamma \cup \K
\cup \Phi \subseteq \T_0 \cup \Gamma \cup \K \cup
\Phi \cup \bigcup_{f \in \Sigma} {\sf Update}(f, f').$$
 By using hierarchical reasoning in chains of theory extensions we can
reduce, in 4 steps, the test in (10) to a ground 
satisfiability check w.r.t.\ ${\cal T}_0$ (which is decidable).

\smallskip
\noindent (b) 
Assume that the set $\Gamma$ of constraints referring to functional parameters 
in $\Sigma_p$ is not a priori given. 
We can use the method for symbol elimination presented in
Section~\ref{symbol-elimination} and obtain a constraint $\Gamma_i$ 
on the parameters such that $\Gamma_i \wedge {\sf Init} \wedge \neg
\Phi$ is unsatisfiable. 
Invariance under transitions can be solved similarly
and yields a constraint $\Gamma_u$. 
If all local extensions satisfy condition $({\sf Comp}_{f})$ and all extension
clauses are  flat  and linear then Theorems~\ref{symb-elim-weakest} (for {\sf Init})
and ~\ref{symb-elim-weakest-chains} (for updates) ensure that we can effectively  
construct {\em the weakest condition} 
$\overline{\Gamma} = \Gamma_i \wedge \Gamma_u$ 
under which $\Phi$ is invariant.\QED

\smallskip
\noindent 
{\bf Comment.} We can extend this result to 
fully exploit extended locality by allowing, in 
Assumptions 1 and 2, ${\cal K}$, ${\sf Init}$, $\Phi$, ${\sf Update}$ 
to consist of augmented clauses, requiring  ${\sf ELoc}$ for 
${\cal T}_0 {\subseteq} {\cal T}_0 {\cup} \Gamma$, and decidability of 
${\cal T}_0$-satisfiability for 
the fragment to which the formulae obtained  
after the hierarchical reduction belong. 
\begin{example}
\label{arrays}
{\em 
Consider an algorithm for inserting an element $c$ into a sorted 
array $a$ at a (fixed, but parametric) position $i_0$. 
We want to derive constraints on the value of $c$ which guarantee that 
the array remains sorted after the insertion. 
Let ${\cal T}_S$ be the disjoint 
combination of Presburger arithmetic
(${\mathbb Z}$, sort ${\sf index}$) and 
a theory ${\cal T}_e$  of elements (here, ${\mathbb R}$, sort ${\sf elem}$). 
We model the array $a$ by using a function $a$ of sort 
${\sf index} \rightarrow {\sf elem}$, and a constant 
$ub$ of sort ${\sf index}$ (for the size of the array).    
The safety condition is the condition that the array is sorted, i.e.\ 

\smallskip
\noindent ${\sf Sorted}(a, ub) ~~~~~~~~~~~~~~ \forall i, j : {\sf index} (0 \leq i \leq j \leq ub \rightarrow a(i) \leq a(j)).$

\smallskip
\noindent 
The update rules are described by the following formula: 

\smallskip
\noindent 
$\begin{array}{ll}
{\sf Update}(a, a', ub, ub')  ~~~~ & \forall i : {\sf index} (0 \leq i < i_0 \rightarrow a'(i) = a(i) ) \wedge
~ a'(i_0) = c ~\wedge \\
& \forall i : {\sf index} (i_0 <  i \leq ub' \rightarrow a'(i) = a(i-1) )  \wedge ~ ub' = ub + 1 
\end{array}$

\smallskip
\noindent 
We want to determine conditions on $c$ and $a$ 
s.t.\ sortedness is preserved, i.e.:

\smallskip
$~~~~~ {\sf Sorted}(a, ub) \wedge {\sf Update}(a, a', ub, ub') \wedge 0 {\leq} d {\leq} ub' {-} 1  \wedge  a'(d) {>} a'(d+1) \models \perp,$

\smallskip 
\noindent where $d$ is a Skolem constant intruduced for the existential
quantifier in $\neg {\sf Sorted}(a', ub')$.\footnote{Note that for the
  negation of ${\sf Sorted}$ we use a different formula. This is an
  optimization we used in this example for improving readability.}  

\smallskip
\noindent 
The examples of local extensions in 
Section~\ref{local} show that ${\sf Sorted}$ and ${\sf Update}$ define
local theory extensions (satisfying condition ${\sf ELoc}$). We
instantiate accordingly and perform a two-step hierarchical reduction
motivated by the fact that the following are equivalent: 
\begin{itemize} 
\item ${\sf Sorted}(a, ub) \wedge {\sf Update}(a, a', ub, ub') \wedge
  (0 {\leq} d {\leq} ub' {-} 1) \wedge a'(d) {>} a'(d+1)$ unsatisfiable.
\item ${\sf Sorted}(a, ub) \wedge {\sf Update}(a, a', ub, ub')[G]
  \wedge G$ unsatisfiable,  \\
where $G = (0 {\leq} d {\leq} ub' {-} 1) \wedge a'(d) {>} a'(d+1)$ and \\
$ {\sf Update}(a, a', ub, ub')[G] =   ub' {=} ub {+} 1  \wedge$ \\
$\mbox{\hspace{8mm}}(0 {\leq} d {<} i_0 {\rightarrow} a'(d) {=} a(d)) \wedge$ 
$(i_0 {<}  d {\leq} ub {\rightarrow} a'(d) {=} a(d-1) ) ~~ \wedge ~~ a'(i_0) {=} c ~~ \wedge$ \\
$\mbox{\hspace{8mm}}(0 {\leq} d{+}1 {<} i_0 {\rightarrow} a'(d{+}1) {=} a(d{+}1)) \wedge$ 
$(i_0 {<}  d{+}1 {\leq} ub {\rightarrow} a'(d{+}1) {=} a(d) )$. 
\item ${\sf Sorted}(a, ub) \wedge {\sf Update}(a, a', ub, ub')[G]_0
  \wedge G_0 \wedge {\sf Con}[G]_0$ unsatisfiable  \\
where $G_0 = (0 {\leq} d {\leq} ub' {-}
  1) \wedge c_{a'(d)} {>} c_{a'(d+1)}$ and \\
 ${\sf Update}(a, a', ub,  ub')[G]_0 = ub' {=} ub {+} 1  \wedge$ \\
$\mbox{\hspace{8mm}}(0 {\leq} d {<} i_0 {\rightarrow} c_{a'(d)} {=} a(d)) \wedge$ 
$(i_0 {<}  d {\leq} ub {\rightarrow} c_{a'(d)} {=} a(d-1) ) ~~ \wedge ~~ c_{a'(i_0)} {=} c ~~ \wedge$ \\
$\mbox{\hspace{8mm}}(0 {\leq} d{+}1 {<} i_0 {\rightarrow} c_{a'(d{+}1)} {=} a(d{+}1)) \wedge$ 
$(i_0 {<}  d{+}1 {\leq} ub {\rightarrow} c_{a'(d{+}1)} {=} a(d) )$ and
\\
${\sf Con}[G] = \{ i_0 = d \rightarrow c_{a'(i_0)} = c_{a'(d)},  i_0 =
d + 1\rightarrow c_{a'(i_0)} = c_{a'(d + 1)} \}$. \\
(Since $d$ and $d+1$ cannot be equal, the instances 
$d = d+1 \rightarrow c_{a'(d)} =  c_{a'(d + 1)}$ of the congruence
axiom are always true, hence redundant, and
can therefore be ignored.)

\item ${\sf Sorted}(a, ub)[G'] \wedge G' \models_{{\cal T}_S \cup \Gamma} \perp$, where \\
$G' = (0 {\leq} d {\leq} ub' {-} 1) \wedge c_{a'(d)} {>} c_{a'(d+1)} \wedge   ub' {=} ub {+} 1  \wedge$ \\
$\mbox{\hspace{8mm}} (0 {\leq} d {<} i_0 {\rightarrow} c_{a'(d)} {=} a(d)) \wedge$ 
$(i_0 {<}  d {\leq} ub {\rightarrow} c_{a'(d)} {=} a(d-1) ) ~~ \wedge ~~ c_{a'(i_0)} {=} c ~~ \wedge$ \\
$\mbox{\hspace{8mm}} (0 {\leq} d{+}1 {<} i_0 {\rightarrow} c_{a'(d{+}1)} {=} a(d{+}1)) \wedge$ 
$(i_0 {<}  d{+}1 {\leq} ub {\rightarrow} c_{a'(d{+}1)} {=} a(d) )
\wedge$  \\
$\mbox{\hspace{8mm}} i_0 = d \rightarrow c_{a'(i_0)} = c_{a'(d)},  i_0 =
d + 1\rightarrow c_{a'(i_0)} = c_{a'(d + 1)}. $
\item $a(d {-} 1) \leq a(d) \wedge a(d) \leq a(d {+} 1) \wedge 
a(d{-}1) \leq a(d{+}1) \wedge G'$
  unsatisfiable. 

\smallskip
\noindent 
Note that $a(d {-} 1) \leq a(d{+}1)$ is a consequence of $a(d {-} 1) \leq
a(d) \wedge a(d) \leq a(d {+} 1)$ and therefore can be ignored. \\
No instances of congruence axioms are needed in this case, because 
they are all of the form $d+1 = d \rightarrow a(d+1) = a(d), d-1 = d
\rightarrow a(d-1) = a(d),d-1 = d+1 \rightarrow a(d-1) = a(d+1)$ and
are always true (because the premises are false) hence redundant. 
 
\end{itemize}
\begin{description}
\item[Step 1:] We start with this last formula. 
\item[Step 2:] The constants corresponding to terms containing
  paramaters are $c$, $i_0$, $d, d-1, d+1$, and those which rename
$a(d),a(d-1), a(d+1)$. All the other constants are regarded as
existentially quantified variables. 
\item[Step 3:] We eliminate the existentially quantified variables
  using a method for quantifier elimination in the combination of real
  arithmetic and Presburger arithmetic. However, in many cases this
  can lead to a considerable increase in the size of the formulae. 
  We now present an optimization we proposed in
  \cite{peuter-sofronie-cade2019}. 

\medskip
\noindent Any set of 
clauses  $\bigwedge_{i = 1}^n (\phi_i(x, f) \rightarrow C_i(x, x', f, f'))$, where 
$\phi_i \wedge \phi_j \models_{\cal T} \perp$ for $i \neq j$ and
$\models_{\cal T} \bigvee_{i = 1}^n \phi_i$ is equivalent to 
$\bigvee_{i = 1}^n (\phi_i(x, f) \wedge C_i(x, x', f, f')).$ 
The two instances of the update axioms in $G'$ have this form. 
By the transformation above and distributivity we obtain the 
following equivalent DNF formula\footnote{The transformation to DNF
  is, in fact, part of a procedure that optimizes the algorithm for
  symbol elimination in theory extensions, which is described in \cite{peuter-sofronie-cade2019}.}:

\smallskip
\noindent 
$\begin{array}{ll}
\text{--} ~~ & (\psi_1 \wedge \psi) \vee (\psi_2 \wedge \psi) \vee (\psi_3 \wedge \psi) \vee (\psi_4 \wedge \psi) \models_{{\cal T}_S \cup \Gamma} \perp, \text{ where } \\[1ex]
& \psi = a(d {-} 1) \leq a(d) \wedge a(d) \leq a(d {+} 1) \wedge a'(d) {>} a'(d+1) \wedge  a'(i_0) {=} c \wedge ub' {=} ub {-} 1  \\
 & \psi_1 = (0 {\leq} d < d{+}1 < i_0 \wedge a'(d) {=} a(d) \wedge a'(d {+}1 ) {=} a(d{+}1))  \\ 
& \psi_2 = (0 {\leq} d < d{+}1 = i_0 {\leq} ub \wedge a'(d) {=} a(d) \wedge a'(d {+}1 ) {=} c)  \\ 
& \psi_3 = (0 {\leq} d = i_0 < d{+}1 {\leq} ub \wedge a'(d) {=} c \wedge a'(d {+}1 ) {=} a(d)) \\ 
& \psi_4 = (0 {\leq}  i_0 < d < d{+}1 {\leq} ub \wedge a'(d) {=} a(d{-}1) \wedge a'(d {+}1 ) {=} a(d)).
\end{array}$

\smallskip
\noindent 
$\psi_1 \wedge \psi$ and $\psi_4 \wedge \psi$ are clearly unsatisfiable.
Consider now $\psi_2 \wedge \psi$. 
We purify the formulae and eliminate all constants (i.e.\ 
existentially quantified variables) 
with the exception of $c$, $i_0$, $d, d-1, d+1$, and those which rename
$a(d),a(d-1), a(d+1)$. 

\item[Step 4 ($\psi_2 \wedge \psi$):] We replace the constants renaming $a(d),a(d-1), a(d+1)$
  back with the terms they rename. We regard $d$ as an existentially
  quantified variable. 

\item[Step 5 ($\psi_2 \wedge \psi$):] We negate the formula obtained this way and obtain: 

${\overline \Gamma}_2 = \forall d ( 0 {\leq} d {<} i_0 \wedge a(d) {\leq} a(i_0) \wedge a(d {-} 1) {\leq} a(d) \rightarrow a(d) {\leq} c).$ 
\end{description} 

\noindent 
Under the assumption of sortedness for $a$, we obtain the equivalent condition: 
$\overline{\Gamma}'_2 = \forall x ( x {<} i_0 \rightarrow a(x) {\leq} c)$.
Similarly, from $\psi_3 \wedge \psi$ we obtain condition 
$\overline{\Gamma}'_3 = \forall x (i_0 {\leq} x \rightarrow c {\leq} a(x))$. 
Thus, the weakest condition 
under which $\Phi$ is an invariant (assuming sortedness) is 
${\overline \Gamma}' = \forall x [( x {<} i_0 \rightarrow a(x) {\leq} c) \wedge (i_0 {\leq} x \rightarrow c {\leq} a(x))]$. 

\smallskip
\noindent We also consider the problem of determining conditions on $a$ and $c$ 
under which $a'$ is sorted, without a priori assuming sortedness for $a$. Then  
$\psi_1 \wedge \psi \models \perp$ yields 
$\overline{\Gamma}_1 = \forall x (0 {\leq} x {<} x+1 {<} i_0 \rightarrow a(x) {\leq} a(x+1))$ and 
$\psi_4 \wedge \psi \models \perp$  yields 
$\overline{\Gamma}_4 = \forall x (i_0 {\leq} x {<} x+1 {<} ub \rightarrow a(x) {\leq} a(x+1))$. 
Hence, the overall condition we obtain is in this case 

\smallskip
$~~~~~ \overline{\Gamma} = {\sf Sorted}(a) \wedge \forall x ( (x < i_0 \rightarrow a(x) \leq c)
\wedge (i_0 \leq x \rightarrow c \leq a(x))).$ 

\smallskip
\noindent 
Our tests with Redlog and Qepcad show that Qepcad offers the best
simplification
possibilities, but cannot be used for eliminating a large number of
variables, 
whereas Redlog (system using virtual substitution) performs well if we
need to eliminate many variables, but yields formulae that need to be
simplified further. 
}
\end{example}

\noindent {\bf Comment:} 
All these ideas scale up, in principle, also to bounded model
checking. The verification problems are in general unproblematic: 
we have to check whether 
$${\cal T}_S \wedge {\sf Init}_0 \wedge \bigwedge_{i = 1}^j {\sf
  Update}_i \wedge \neg \Phi_j \models \perp \quad \text{ for all } 0
\leq j \leq k.$$
Under Assumptions 1, 2 and the assumption that 
$\T_0 \subseteq \T_0 \cup \K \cup \Phi \cup {\sf
  Init}_0$ is a local extension
we have the chain of (local) theory extensions: 

\smallskip
\noindent $\T_0 \subseteq \T_0 \cup \K \cup \Phi \cup {\sf Init}_0 \subseteq \T_0 \cup \K
\cup {\sf Init}_0 \cup {\sf Update}_1 \subseteq \dots  \T_0 \cup \K
\cup {\sf Init}_0 \cup {\sf Update}_1\cup \dots \cup {\sf Update}_j$

\smallskip 
\noindent which can be used for reducing the BMC problem above to 
checking ground satisfiability w.r.t.\ $\T_0$. 
However, constraint synthesis is more complicated for bounded model
checking: The 
formulae obtained after quantifier elimination from
the (much longer) formulae stemming from bounded model checking are
difficult to understand by a human. 

\section{Hybrid Automata} 
\label{sect-ha}

In this section we analyze situations in which invariant checking is
decidable for various classes of hybrid automata, and analyze
possibilities of synthesizing constraints on parameters for parametric
hybrid automata. In Section~\ref{lha} we analyze parametric linear
hybrid automata, in Sections~\ref{pha} and ~\ref{iha} we extend the methods developed 
for parametric LHA  to more general HA. 

\begin{example}[Running example]
\label{sect-ex}
{\em 
We consider a temperature controller, modeled as a hybrid
automaton with two modes: a heating mode (in which the environment of the 
object is heated) and a normal mode (heating is switched off).  
The control variable is $x$ (the temperature of the object). 
We assume that the system has two parameters (which can be functional or not). 
\begin{itemize}
\item The temperature of the heated environment (due to the heater): 
a constant $h$ or (if it changes over time) a unary function $h$.  
\item Perturbation of the temperature of the environment due to external
  causes (e.g. external temperature), modeled using 
a constant $f$ or (if it
changes in time) a unary function $f$.  
\end{itemize}
{\bf Invariants and 
flows} in the two modes are described below ($k > 0$ is a constant which 
depends only on the surface of the object which is being heated): 

\smallskip
\noindent 
\begin{tabular}{lll}
{\bf Mode 1} (Heating): &  Invariant: $T_a \leq x(t) \leq T_b$; & Flow:
$\frac{dx}{dt} = - k (x - (h + f))$, \\
{\bf Mode 2} (Normal): &  Invariant: $T_c \leq x(t) \leq T_d$; & Flow:
$\frac{dx}{dt} = - k (x - f)$. 
\end{tabular}

\smallskip
\noindent 
{\bf Control switches.} We have two control switches: 
\begin{description}
\item[$e_{12}${:}] switch from Mode 1 to Mode 2  
(if the temperature of the object becomes too high then heating is switched off):  
${\sf guard}_{e_{12}}{:}~ x \geq T_b$; ${\sf jump}_{e_{12}}{:}~ (x' = x)$. 
\item[$e_{21}${:}] switch from Mode 2 to Mode 1
(if the temperature of the object becomes too low then heating is switched on):  
${\sf guard}_{e_{21}}{:}~ x \leq T_c$; ${\sf jump}_{e_{21}}{:}~ (x' = x)$. 
\end{description}
\noindent Let ${\sf Safe} = T_m \leq x(t) \leq T_M$ be a safety
condition for the heater. Our goals are:
\begin{itemize}
\item[(1)] check that ${\sf Safe}$ is an invariant (or that it holds on all runs of bounded length), 
\item[(2)] generate constraints which guarantee that ${\sf Safe}$ is an
  invariant,  
\end{itemize}
In this section we give methods for solving (1) and (2) for 
increasingly larger classes of parametric hybrid automata.
}
\end{example}

\subsection{Parametric Linear Hybrid Automata}
\label{lha}

In \cite{Henzinger} a class of hybrid automata was introduced in which the
flow conditions, the guards and the invariants have a special form. 

\begin{definition}
Let $X = \{ x_1,
\dots, x_n \}$ be a set of variables. An (atomic) linear predicate on the variables $x_1,
\dots, x_n$ is a linear strict or non-strict
inequality
of the form 
$a_1 x_1 + \dots a_n x_n \rhd a$, 
where $a_1, \dots, a_n,
a \in {\mathbb R}$ and $\rhd \in \{ \leq, < , \geq, > \}$. 
A convex linear predicate is a finite conjunction of linear
inequalities.
\end{definition} 

\begin{definition}[\cite{Henzinger}]
A hybrid automaton $S$  is
a linear hybrid automaton (LHA) if it satisfies the following
two requirements: 

\smallskip
\noindent {\em 1. Linearity:} For every control mode $q \in Q$, 
the flow condition ${\sf flow}_q$, the invariant condition ${\sf Inv}_q$, 
and the initial condition ${\sf Init}_q$ are convex linear 
predicates.
For every control switch $e = (q,q') \in E$, the jump condition
${\sf jump}_e$ and the guard ${\sf guard}_e$ are convex linear 
predicates.
In addition, 
as in \cite{damm-ihlemann-sofronie-hscc11,damm-ihlemann-sofronie-2011}, 
we assume that the 
flow conditions ${\sf flow}_q$ are conjunctions of {\em non-strict} 
linear inequalities. 

\smallskip
\noindent {\em 2. Flow independence:} For every control mode $q \in Q$, 
the flow condition ${\sf flow}_q$ is a predicate over the variables
in ${\dot X}$ only (and does not contain any variables from $X$).
This requirement ensures that the possible flows
are independent from the values of the variables, and only depend
on the control mode.
\end{definition} 

\medskip
\noindent 
We also consider {\em parametric} linear hybrid automata (PLHA), 
defined as  
linear hybrid automata for which a set $\Sigma_P = P_c \cup P_f$ of 
parameters is specified (consisting of parametric constants $P_c$ and 
parametric functions $P_f$)
with the difference that for every control mode $q \in Q$ and every mode 
switch $e$: 
\begin{itemize}
\item[(1)] the linear constraints in the invariant conditions ${\sf Inv}_q$, 
initial conditions ${\sf Init}_q$, and 
guard conditions ${\sf guard}_e$ are of the form: 
$g \leq \sum_{i = 1}^n a_i x_i \leq f$, 

\vspace{-3mm}
\item[(2)] the inequalities in the 
flow conditions ${\sf flow}_q$ are of the form:  
$\sum_{i = 1}^n b_i {\dot x}_i \leq b$, 

\vspace{-3mm}
\item[(3)] the linear constraints in ${\sf jump}_e$ are of the form 
$\sum_{i = 1}^n b_i x_i + c_i x'_i \leq d$,
\end{itemize}
\noindent (possibly relative to an interval $I$) 
where the coefficients $a_i, b_i, c_i$ and the bounds $b, d$ 
are either numerical constants or parametric constants in $P_c$; 
and $g$ and $f$ are (i) constants or parametric constants in $P_c$, or 
(ii) parameteric functions in $P_f$ satisfying 
the convexity (for $g$) resp. concavity condition (for $f$), or 
terms with one free variable $t$ such that the associated functions
have these convexity/concavity properties 
and $\forall t (g(t) \leq f(t))$. 
The flow independence conditions hold as 
in the case of linear hybrid automata.

\smallskip
\noindent {\bf Note:} In the definition of PLHA 
we allow a general form of parametricity, in which the 
bounds in state invariants, guards and jump conditions can be 
expressed using functions with certain properties.
Such parametric descriptions of bounds are useful for instance in 
situations in which we want to verify systems which have non-linear 
behavior and use a parametric approximation for them. 

\begin{example}
{\em 
Consider the hybrid automaton $S$ presented in Example~\ref{sect-ex}.   
If in the heating mode the invariant is $T_a \leq x(t) \leq T_b$ and 
the flow is $\frac{dx}{dt} = - k (x - (h+f))$, where $k > 0$, then 
we can approximate the flow by the linear flow: 

\vspace{-6mm}
\begin{eqnarray}
-k (T_b - (h+f)) \leq \dot{x} \leq -k (T_a - (h+f)). \label{flow1}
\end{eqnarray}

\vspace{-2mm} 
\noindent We can obtain similar bounds for $\dot{x}$ also for mode 2. 
Thus, we can approximate $S$ using a linear hybrid automaton $S'$. 
%
If we can guarantee safety 
in $S'$, then safety is preserved for all possible runs 
which satisfy the flow conditions of $S'$, in particular also for all runs of
$S$, so $S$ is safe. 
%
If a formula $\Phi$ is an inductive invariant of $S'$ it is
also an inductive invariant of $S$, because all possible jumps and
flows of $S$ satisfy, in particular, the conditions of the jumps and
flows of the abstracted system $S'$. 
}
\end{example} 

\medskip
\noindent We provide methods to decide whether a formula $\Phi$ 
is an invariant and to 
derive conditions that guarantee that a PLHA $S$ has a given safety property.
To use Thm.~\ref{inv-trans-qe}, we analyze the possible updates in PLHA
by jumps and flows.

\smallskip
\noindent 
{\bf Jumps.} A jump update can be expressed by the linear inequality 

$~~~~~~~~~{\sf Jump}_e({\overline x}, {\overline x'}) = {\sf guard}_e({\overline x}) \wedge {\sf
  jump}_e({\overline x}, {\overline x'})$. 

\smallskip
\noindent {\bf Flows.}
Assume that 
${\sf flow}_q(t)  =  
\bigwedge_{j = 1}^{n_q} (\sum_{i = 1}^n c^q_{ij} \dot{x}_i(t) \leq_j c_j^q)$. 
We alternatively 
axiomatize flows in mode $q$ in the time interval $[t_0, t_1]$ 
(where $0 {\leq} t_0 {\leq} t_1$) as follows:\\

$~~{\sf Flow}_q(t_0, t_1) = 
\forall t (t_0 {\leq} t {\leq} t_1 {\rightarrow} {\sf Inv}_q({\overline x}(t)))  \wedge   
\forall t, t' (t_0 {\leq} t {<} t' {\leq} t_1 {\rightarrow} {\underline {\sf
    flow}}_q(t, t'))$

\noindent where: 
$\displaystyle{{\underline {\sf flow}}_q(t, t') = 
\bigwedge_{j = 1}^{n_q}  
(\sum_{i = 1}^n c^q_{ij} (x_i(t') - x_i(t)) \leq_j c_j^q (t' - t)).}$

\noindent In \cite{damm-ihlemann-sofronie-hscc11,damm-ihlemann-sofronie-2011} 
we showed that for LHA no precision is lost with this axiomatization
and that we can simplify the axiomatization of flows further by 
suitably instantiating the universal quantifiers in ${\sf
  Flow}_q$. These results are summarized in Theorem~\ref{transl-invar-par}. 

 \begin{theorem}[\cite{damm-ihlemann-sofronie-2011}]
The following are equivalent for any LHA: 
\begin{itemize}
\item[(1)] $\Phi$ is an invariant of the automaton.  
\vspace{-3mm}
\item[(2)] 
For every $q \in Q$ and $e = (q,q') \in E$, 
$\T_0 \cup F_{\sf init}(q) \models \perp$, 
$\T_0 \cup F_{\sf flow}'(q) \models \perp$ and 
$\T_0 \cup F_{\sf jump}(e) \models \perp$, where $\T_0$ is the theory
of real numbers and: 
\end{itemize}

\noindent $\begin{array}{@{}lcl@{}}
F_{\sf Init}(q) & := & 
{\sf Init}_q({\overline x}(t_0)) \wedge \neg \Phi({\overline x}(t_0)) \\
F_{\sf flow}'(q)  & := & {\sf Inv}_q({\overline x}(t_0)) \wedge 
\Phi({\overline x}(t_0)) \wedge {\underline {\sf flow}}_q(t_0, t) \wedge 
 {\sf Inv}_q({\overline x}(t)) \wedge 
\neg \Phi({\overline x}(t)) \wedge t_0 < t \\
F_{\sf jump}(e) &:= &  {\sf Inv}_q({\overline x}(t)) \wedge \Phi({\overline x}(t)) {\wedge} 
{\sf Jump}_e({\overline x}(t), {\overline x}'(0)) {\wedge} 
{\sf Inv}_{q'}({\overline x}'(0))  {\wedge} 
\neg \Phi({\overline x}'(0)).
\end{array}$
\label{transl-invar-par}
\end{theorem}

\noindent 
Let $S$ be a parametric LHA with parameters $\Sigma_P = P_c \cup P_f$.
Assume that the properties of the parameters are expressed as 
$\Gamma_0 \wedge \Gamma_f$, where $\Gamma_0$ is a conjunction
of linear inequalities representing the relationships 
between parameters in $P_c$  
and $\Gamma_f$ is a set of (universally quantified) clauses 
expressing the properties of the 
functional parameters (in $P_f$) -- containing 
the convexity/concavity conditions 
for the bounding functional parameters. 
Let $\Phi$ be a property expressed as convex linear predicate 
over $X$, possibly containing parameters (constants as coefficients; 
either constants or functions as bounds in the linear inequalities).

\noindent 
Since the theory of real numbers allows quantifier elimination, the following 
result is a direct consequence of Thm.\ \ref{inv-trans-qe}. 

\begin{theorem}
Let $S$ be a PLHA and $\Phi$ a property expressed 
as a convex linear predicate over $X$, possibly containing parameters. 
We can effectively derive a set $\Gamma$ of 
(universally quantified) 
constraints on the parameters such that whenever $\Gamma$ holds in an 
interpretation ${\cal A}$,  
$\Phi$ is an invariant w.r.t.\ ${\cal A}$. 
\label{param-funct}
\end{theorem}
This method for constraint synthesis can in particular be used for: 

\noindent {\em 1. Invariant generation:} Let $S$ be a fixed (non-parametric) 
LHA. We consider invariant ``templates''  $\Phi$ 
expressed by linear inequalities with parametric bounds and coefficients
and determine constraints on these parameters which ensure
that $\Phi$ is an invariant. By finding values of the parameters satisfying 
these constraints we can generate concrete invariants.  

\noindent {\em 2. Generation of control conditions:} Assume that 
 $\Phi$ is fixed (non-parametric), but that the mode invariants, 
the flow conditions, the guards, and the jumps are represented 
parametrically (as conjunctions of a bounded number of linear inequalities). 
We can determine constraints on the parameters which ensure
that $\Phi$ is an invariant. By finding values of the parameters satisfying 
these constraints we can determine control conditions which guarantee that 
$\Phi$ is invariant. 

 \begin{example}
{\em 
Consider a variant of the HA in Example~\ref{sect-ex}
in which $T_a, T_b$ are functional 
parameters, ${\sf Inv}_1$ is 
$T_a(t) \leq x(t) \leq T_b(t)$, 
and the flow in mode 1 is described by: $- k (x_b -
g) \leq \dot{x} \leq - k (x_a - g)$ (for simplicity we abbreviated $h +
f$ by $g$). 
Let $\Phi = (T_m {\leq} x(t) {\leq} T_M)$. Assume that the properties of the 
parameters are axiomatized by   
${\cal KF}$: $\{ \forall t (x_a \leq T_a(t)), \forall t (T_b(t)\leq
x_b), ~x_a < x_b, ~T_m < T_M \}$. 
We derive 
a condition $\Gamma$ which guarantees that $\Phi$ is preserved 
under flows in mode 1 using 
Algorithm~\ref{alg-symb-elim}, Section~\ref{symbol-elimination} and 
Thm.~\ref{inv-trans-qe} as follows. 

\noindent By Theorem~\ref{transl-invar-par}, $\Phi$ is preserved under
flows in mode 1 iff the following formula is unsatisfiable in the
extension of real arithmetic with a function symbol $x$:
$$t_0 < t_1 \wedge \Phi(x(t_0)) \wedge {\sf Inv}_1(x(t_0)) \wedge {\sf flow}_1(t_0, t_1)
\wedge {\sf Inv}_1(x(t_1)) \wedge \neg \Phi(x(t_1))$$
(as explained in \cite{damm-ihlemann-sofronie-2011}, no additional
assumptions about continuity and derivability of $x$ need to be made
from now on). 
%
The formula can be written out as: 
 
\smallskip
{\small 
$\begin{array}{ll}
& t_0 < t_1 ~ \wedge~ (T_m \leq x(t_0) \leq T_M)  ~\wedge~ (T_a(t_0)
\leq x(t_0) \leq T_b(t_0)) ~\wedge~ \\
& (x(t_1) \leq x(t_0) - k(t_1-t_0) (x_a - g)  \wedge x(t_0) - k(t_1-t_0) (x_b - g)
\leq x(t_1)) ~\wedge \\
& (T_a(t_1) \leq x(t_1) \leq T_b(t_1)) ~\wedge~ (x(t_1) < T_m ~~ \vee ~~ T_M < x(t_1)). 
\end{array}$  
} 

\smallskip
\noindent We apply Algorithm~\ref{alg-symb-elim} to this formula. The
set of parameters is $\Sigma_P = \{ T_a, T_b, T_m, T_M, k, x_a, x_b, g \}$.

\begin{description}
\item[Step 1:] After purification (with definitions ${\sf Def} = \{
  T_{a0} = 
T_a(t_0), T_{b0} =  T_b(t_0), T_{a1} = 
T_a(t_1), T_{b1} =  T_b(t_1), x_0 =  x(t_0), x_1 = 
x(t_1) \}$), we obtain: 

\smallskip
{\small $\begin{array}{ll}
G_1 := & t_0 < t_1 ~\wedge~ (T_m \leq x_0 \leq T_M)  ~\wedge~ (T_{a0}
\leq x_0 \leq T_{b0}) ~\wedge~ \\
& (x_1 \leq x_0 - k(t_1-t_0) (x_a - g)  \wedge x_0 - k(t_1-t_0) (x_b - g)
\leq x_1) ~\wedge~ \\
& (T_{a1} \leq x_1 \leq T_{b1}) ~\wedge~ (x_1 < T_m ~~ \vee ~~ T_M < x_1). 
\end{array}$
} 

\smallskip
\noindent All instances of the congruence axioms corresponding to
${\sf Def}$ are true (hence redundant) in the presence of $t_0 < t_1$ and can be
omitted. 

\item[Step 2:] Among the constants in $G_1$ we identify the constants
  $T_{a0}, T_{bo},T_{a1}, T_{b1}$ (corresponding to the terms
  $T_a(t_i)$, $T_b(t_i), i = 0, 1$ starting
  with the parameters $T_a, T_b$), the constants $t_0, t_1$ occurring
  as arguments to $T_a$ and $T_b$ and the constants $T_m, T_M, k, x_a,
  x_b, g$ corresponding to the remaining parameters. 
The remaining constants are $x_0$ and $x_1$. We regard these constants
as (existentially quantified) variables. The formula we obtain this
way is: 

\smallskip
{\small $\begin{array}{ll}
\exists x_0, x_1  & t_0 < t_1 ~\wedge~ (T_m \leq x_0 \leq T_M)  ~\wedge~ (T_{a0}
\leq x_0 \leq T_{b0}) ~\wedge~ \\
& (x_1 \leq x_0 - k(t_1-t_0) (x_a - g)  \wedge x_0 - k(t_1-t_0) (x_b - g)
\leq x_1) ~\wedge~ \\
& (T_{a1} \leq x_1 \leq T_{b1}) ~\wedge~ (x_1 < T_m ~~ \vee ~~ T_M < x_1). 
\end{array}$
}

\item[Step 3:] We eliminate the quantifiers $\exists x_0, x_1$ using a
  method for quantifier elimination in the theory of real numbers. 
The result can be further simplified if we assume ${\cal KF}$ 
holds and $T_m \leq T_M$. 

\item[Step 4:]  We then replace 
back $T_{a0}, T_{a1}$ and $T_{b0}, T_{b1}$ and regard $t_0$ and $t_1$
as existentially quantified variables and obtain:

\smallskip
{\small 
$\begin{array}{ll}
\exists t_0, t_1  & (t_0 < t_1 \wedge T_{a}(t_0) \leq T_M \wedge T_m \leq T_{b}(t_0) \wedge T_{a}(t_0) \leq T_{b}(t_0) \wedge T_m \leq T_M
\wedge \\
& T_m \leq T_b(t_1) + k (t_1-t_0) (x_b - g) \wedge T_a(t_0) \leq T_b(t_1) + k
(t_1-t_0) (x_b - g) \wedge \\
& T_a(t_1) + k (t_1-t_0) (x_a - g) \leq T_M \wedge T_a(t_1) + k (t_1-t_0) (x_a - g) \leq T_b(t_0) \wedge \\ 
& [((k (t_1-t_0) (x_b - g) > 0) \wedge T_a(t_1) < T_m \wedge T_a(t_0) < T_m +
    k (t_1  - t_0) (x_b - g)) \vee \\
& ~ ((k (t_1-t_0) (x_a - g) < 0 ) \wedge T_b(t_1) > T_M \wedge T_M + k (t_1  -
    t_0) (x_a - g) < T_b(t_0))]).
\end{array}$
} 

\item[Step 5:] The condition $\Gamma_1$ which ensures that $\Phi$ is an 
invariant under flows in mode 1 is the negation of the formula above: 
 
\smallskip
{\small 
 $\begin{array}{rl}
\forall t_0, t_1 & (t_0 < t_1 \wedge T_{a}(t_0) \leq T_M \wedge T_m
\leq T_{b}(t_0) \wedge T_{a}(t_0) \leq T_{b}(t_0) \wedge T_m \leq T_M
\wedge \\
& T_m \leq T_b(t_1) + k (t_1-t_0) (x_b - g) \wedge T_a(t_0) \leq T_b(t_1) + k
(t_1-t_0) (x_b - g) \wedge \\
& T_a(t_1) + k (t_1-t_0) (x_a - g) \leq T_M \wedge T_a(t_1) + k (t_1-t_0) (x_a
- g) \leq T_b(t_0) \\
\rightarrow & [((k (t_1-t_0) (x_b - g) \leq 0) \vee T_a(t_1) \geq T_m \vee
  T_a(t_0) 
\geq T_m + k (t_1  - t_0) (x_b - g)) \wedge \\
& ~ ((k (t_1-t_0) (x_a - g) \geq 0 ) \vee T_b(t_1) \leq T_M \vee T_M + k (t_1  -
    t_0) (x_a - g) \geq T_b(t_0))]).
\end{array}$
} 
\end{description}

\medskip
\noindent {\em Invariant generation.} Assume that the control of the LHA 
(i.e.\ the functions $T_a, T_b$ and their bounds $x_a, x_b$) is fixed, 
e.g.\ $k = 1$, $T_{a}$ is the constant function $x_a = 15$ and $T_b$ is the 
constant function $x_b = 20$, $h = 25$ and $f = 10$. 
We can use the constraint in $\Gamma_1$ to determine for which values of 
the constants $T_m$ and $T_M$, the formula $\Phi$ is an invariant under flows in mode 1
(we can check that it is, e.g., for $T_m = 15$ and $T_M = 20$).  
The generation of control conditions for guaranteeing that a (non-parametric) 
$\Phi$ is invariant is similar. 
}
\end{example}

\subsection{Parametric Hybrid Automata}
 \label{pha}

We now extend the methods developed above for parametric LHA 
to more general HA.  
For the sake of simplicity, we only consider (parametric) 
hybrid automata $S$ with one continuous variable $x$.\footnote{The case when we have several variables is similar, but 
the presentation is more complicated.} 
Let $\Sigma = \{ x \}$ and $\Sigma' = \{ x' \}$.  
Assume that mode invariants, initial states, guards and jump conditions 
are expressed as sets of clauses in an extension of the theory of real 
numbers with
additional functions in a set $\Sigma_1 = \Sigma \cup \Sigma_P$, 
where $\Sigma_P$ is a set of parameter names (both functions and constants). 
We study the problem of deriving constraints on parameters
which guarantee that a certain formula is an invariant. 
We therefore analyze the possible updates by jumps and flows. 

\smallskip
\noindent 
{\bf Jumps.} Assume that the guards and the jump conditions are given by 
formulae in a certain extension of the theory of real numbers. 
A jump update can be expressed by the formula:  
${\sf Jump}_e(x, x') = {\sf guard}_e(x) \wedge {\sf
  jump}_e(x, x')$. 

\smallskip
\noindent {\bf Flows.} In the case of parametric hybrid automata, 
the flows are described by differential equations. We assume that the
variables $x$ represent differentiable functions during flows. 
As we restrict to  the case of one continuous variable $x$, 
we assume that in mode $q$ the flow is described by:  
$\frac{dx}{dt}(t) = f_q(x(t))$. Thus: 

$~~{\sf Flow}_q(t_0, t_1) = 
\forall t (t_0 {\leq} t {\leq} t_1 {\rightarrow} {\sf Inv}_q(x(t)))  \wedge   
\forall t (t_0 {\leq} t {\leq} t_1 {\rightarrow} \frac{dx}{dt}(t) =
f_q(x(t))).$

\smallskip
\noindent Let $\Phi$ be a set of clauses 
in the signature $\Pi_0 {\cup} \Sigma_P {\cup} \Sigma$ which can contain 
(implicitly universally quantified) variables as arguments of the 
functions in $\Sigma_P \cup \Sigma$.

\begin{theorem}
\begin{itemize}
\item[(1)] For every jump $e$ 
we can construct a universally quantified formula 
$\forall {\overline x} \Gamma_e ({\overline x})$ 
(containing also 
some of the parameters) such that for every structure ${\cal A}$ 
with signature $\Pi_0 \cup \Sigma \cup \Sigma' \cup \Sigma_P$ if 
${\cal A}$ is a model of ${\cal T}_0$ and of $\Gamma_e$ then 
$\Phi$ is an invariant under the jump $e$ (in interpretation ${\cal A}$). 

\vspace{-3mm}
\item[(2)] For every flow in a mode $q$ we can construct a universally quantified formula 
$\forall {\overline x} \Gamma_q ({\overline x})$ (containing also 
some of the parameters) such that for every structure ${\cal A}$ 
with signature $\Pi_0 \cup \Sigma \cup \Sigma' \cup \Sigma_P$ if 
${\cal A}$ is a model of ${\cal T}_0$ and of $\Gamma_q$ then 
$\Phi$ is an invariant under flows in $q$ (in interpretation ${\cal A}$). 
\end{itemize}
\label{qe-non-linear}
\end{theorem}
{\em Proof:} (1) follows from Thm.~\ref{inv-trans-qe} 
for the case of jump updates. 
(2) Assume that ${\cal A}$ is a model in which $\Phi$ is not invariant 
under flows in mode $q$. Then there exist time points 
$t_0, t_1 \in {\mathbb R}$ such that (i) 
${\cal A} \models \forall t (t_0 {\leq} t {\leq} t_1 \rightarrow 
{\sf Inv}_q(x(t)))$ 
and (ii) the interpretation in ${\cal A}$ of the 
function $x$, $x_A : {\mathbb R} \rightarrow {\mathbb R}$ is differentiable 
and has the property that 
$\forall t (t_0 {\leq} t {\leq} t_1 {\rightarrow} \frac{dx}{dt}(t) =
f_q(x(t))).$
Then, by the mean value theorem:  

\medskip
$\begin{array}{@{}ll}
{\cal A} \models & \Phi(t_0) \wedge \forall t (t_0 {\leq} t {\leq} t_1 \rightarrow 
{\sf Inv}_q(x(t))) \wedge \\
& \forall t, t' (t_0 {\leq} t {<} t' {\leq t_1}  {\rightarrow} \exists c ( t
{\leq} c {\leq} t' \wedge \frac{x(t') -  x(t)}{t' - t} = f_q(x(c)))) \wedge \neg
\Phi(x(t_1)).
\end{array}$

\medskip
\noindent 
Therefore, ${\cal A}$ is a model of any set of instances of the formula
above, in particular of those instances in which the universally quantified
variables are instantiated with the constants $\{ t_0, t_1 \}$ 
(we can take more or fewer instances, depending on how strong we want 
the condition $\Gamma_q$ to be). 
For every choice of instances for the pair of variables $t, t'$ in the 
flow description above we need to replace the
existentially quantified variable $c$ with a new constant. 
We can now use Steps 1-5 of the symbol elimination algorithm in Section~\ref{symbol-elimination}
(Alg.~\ref{alg-symb-elim}) and Thm.~\ref{inv-trans-qe} and obtain the formula $\Gamma_q$. \QED

\begin{example}
{\em 
Consider a variant of the HA in 
Example~\ref{sect-ex} in which $f$ and $h$ are unary functions, 
and the invariants and flow in the two modes are described by: 

\smallskip
\begin{tabular}{ll}
{\bf Mode 1} (Heating): &  Invariant: 
${\sf Inv}_1(x(t)) := x(t) \leq T_M$ \\
& Flow: $\frac{dx}{dt}(t) = - k (x(t) - (h(t) + f(t)))$\\
{\bf Mode 2} (Normal): &  Invariant: ${\sf Inv}_1(x(t)) := T_m \leq x(t)$\\
& Flow: $\frac{dx}{dt}(t) = - k (x(t) - f(t))$
\end{tabular}

\noindent Let $\Phi(t) = T_m \leq x(t) \leq T_M$. 
We use Algorithm~\ref{alg-symb-elim} to derive constraints which guarantee that this $\Phi$ is invariant 
under flows in interval $[t_0, t_1]$, where $t_0 < t_1$, in mode 2, i.e.\ such that the
conjunction of the following formulae is unsatisfiable:

\begin{itemize}
\item[(a)] $\forall t (t_0 \leq t \leq t_1 \rightarrow T_m \leq x(t))$

\item[(b)] 
$\forall t', t'' (t_0 \leq t' < t'' \leq t_1 \rightarrow  \exists y (t' \leq y
  \leq t'' \wedge \frac{x(t'')
  {-} x(t')}{t''  {-} t'} = -k (x(y) - f(y))))$

\item[(c)] 
$(T_m \leq x(t_0) \wedge x(t_0) \leq T_M) ~~ \wedge ~~ (x(t_1) > T_M \vee x(t_1) < T_m)$

\item[(d)] $t_0 < t_1$.
\end{itemize}

\noindent 
Let $T = \{ x(t_1), x(t_0) \}$. We instantiate $t$ with $t_0$, and
$t_1$ in (a);  $t'$ with $t_0$ and $t''$ with $t_1$ in (b), and
obtain the following instances:

\medskip
\noindent 
$\begin{array}{ll} 
 & (T_m \leq x(t_0)) \wedge 
(T_m \leq x(t_1)) \wedge \exists y (t_0 \leq y \leq t_1 \wedge \frac{x(t_1)
  {-} x(t_0)}{t_1  {-} t_0} = -k (x(y) - f(y))) \wedge \\[1ex]
& (T_m \leq x(t_0) \wedge x(t_0) \leq T_M) \wedge
(x(t_1) > T_M \vee x(t_1) < T_m) \wedge t_0 < t_1.
\end{array}$

\medskip
\noindent 
We introduce a new constant $c$ for the existentially quantified
variable $y$. We obtain: 

\medskip
\noindent 
$\begin{array}{ll} 
 & t_0 < t_1 \wedge t_0 \leq c \leq t_1 \wedge (T_m {\leq} x(t_0)) ~\wedge~ 
(T_m \leq x(t_1)) ~\wedge~ \frac{x(t_1)
  {-} x(t_0)}{t_1  {-} t_0} = -k (x(c) - f(c)) \wedge \\[1ex]
& (T_m {\leq} x(t_0) \wedge x(t_0) {\leq} T_M) ~~ \wedge ~~
(x(t_1) {>} T_M \vee x(t_1) {<} T_m)
\end{array}$

\medskip
\noindent Consider this last formula.

\smallskip
\noindent 
We use a variant of Algorithm~\ref{alg-symb-elim} -- in which some of
the constants related to parameters are eliminated -- for obtaining a constraint on the
parameters in the set $\Sigma_P = \{ f, x, k, c, T_m, T_M \}$ under which this formula is
unsatisfiable.

\begin{description} 
\item[Step 1:] 
We purify the formula introducing the abbreviations: $c_0 = x(t_0)$, 
$c_1 = x(t_1)$, $d = x(c)$ and $d_f = f(c)$. 
We obtain the following set of constraints: 

\smallskip
$\begin{array}{ll} 
 & t_0 < t_1 \wedge t_0 \leq c \leq t_1 \wedge \\[1ex]
&  [T_m {\leq} c_0 \wedge T_m {\leq} c_1 \wedge  \frac{c_1 
  {-} c_0}{t_1  {-} t_0} = -k (d - d_f) \wedge (T_m {\leq} c_0 \wedge c_0 {\leq} T_M) ~~ \wedge ~~
(c_1 {>} T_M \vee c_1 {<} T_m)]
\end{array}$

\item[Step 2:] We distinguish the variables $c_0, c_1, d, d_f$
  introduced for terms starting with the parameters $x$ and $f$,
  the variables $t_0, t_1, c$ used as arguments to the parameters $x$ and
  $f$, and the variables $k, T_m, T_M$ corresponding to the remaining parameters. 

We can for instance choose to eliminate $t_0, t_1,c_0$ and $c_1$. We regard these constants as existentially quantified
variables and obtain: 

\smallskip
$\begin{array}{ll} 
\exists t_0, t_1, c_0, c_1 & t_0 < t_1 \wedge t_0 \leq c \leq t_1
\wedge [T_m {\leq} c_0 \wedge T_m {\leq} c_1 \wedge  \frac{c_1 
  {-} c_0}{t_1  {-} t_0} = -k (d - d_f) \wedge \\[1ex]
& ~ (T_m {\leq} c_0 \wedge c_0 {\leq} T_M) ~~ \wedge ~~
(c_1 {>} T_M \vee c_1 {<} T_m)]
\end{array}$

\item[Step 3:] 
We eliminate the variables 
$t_0, t_1, c_0,$ and $c_1$ (we used Redlog \cite{redlog}) under 
the assumption that $t_0 < t_1, t_0 \leq c \leq t_1$ and $T_m \leq T_M$ and
obtain the equivalent formula:  
$$ ( - k (d - d_f) > 0 ~~ \wedge ~~ T_m < T_M)$$

\item[Step 4:] We now replace again the constants $d$ and $d_f$ with the terms they 
represent and replace $c$ by an existentially quantified variable and
obtain: 
$$\exists c (-k (x(c) - f(c)) > 0 ~~ \wedge ~~
T_m < T_M).$$
\item[Step 5:] The negation of this formula is $ \forall c (-k (x(c) - f(c)) \leq 0 ~~ \vee ~~
T_M \leq T_m).$
\end{description}

\medskip
We can conclude that if $T_M > T_m$ then 
$\Phi$ is invariant under all flows in mode $2$ 
if condition $\forall c ~~k{*}(f(c){-}x(c)) {\leq} 0 $ holds (i.e.\ if the system 
does not heat in this state because of the external temperature).
Of course this is only a sufficient condition, and it is not the
weakest condition under which $\Phi$ is an invariant because we 
used a partial (incomplete) instantiation.

\noindent This example can thus be
seen as an illustration of the fact that also incomplete
instantiations or instantiations for non-local theories can be used
for obtaining constraints on parameters which entail inductive
invariance of given formulae if we are not interested in generating a
weakest condition on parameters. 
}
\label{ex-heater1} 
\end{example}

\section{Interconnected Families of Hybrid Automata}
\label{iha}

We can also consider systems of interconnected 
parametric hybrid automata $\{ S_1, \dots, S_n \}$ with a parametric number of components
under the assumptions:
\begin{itemize}
\item[(1)] The invariants, guards, jump and flow conditions of $S_1, \dots, S_n$ 
can all be expressed similarly (and can be written globally, 
using indices);  
\item[(2)] The relationships between the hybrid automata are uniform
(and can again be expressed globally using indices); 
\item[(3)] The topology of the system can be represented using data structures 
(e.g.\ arrays, lists, trees). 
\end{itemize}
A general formalization of such situations is out of the scope of this paper. 
We here present the ideas on an example. 

\begin{example} 
{\em 
Consider a family of $n$ water tanks with a uniform description, 
each modeled by the hybrid automaton $S_i$. 
Assume that every $S_i$ has one continuous variable $L_i$ 
(representing the water level in $S_i$), and that the input and output 
in mode $q$ are described by parameters ${\sf in}_i$ and ${\sf out}_i$. 
Every $S_i$ has one mode 
in which the water level evolves according to rule 
$\dot{L_i} = {\sf in}_i - {\sf out}_i $. 
We write $L(i, t), {\sf in}(i)$ and ${\sf out}(i)$ instead of 
$L_i(t), {\sf in}_i$ and ${\sf out}_i$, respectively.

\noindent Assume that the water tanks are interconnected in such a way that 
the input of system $S_{i+1}$ is the output of system $S_i$. 
A global constraint describing the communication of the systems is
therefore:
 
\medskip
$~~~~~\forall i [2 \leq i \leq n  \rightarrow ({\sf in}(i) = {\sf out}(i-1))]
~~~~ \wedge ~~~ {\sf in}(1) = {\sf in}.$

\medskip
\noindent An example of a ``global'' update describing the evolution of the
systems $S_i$ during a flow in interval $[t_0, t_1]$:  

\medskip
$~~~~~\forall i  (L(i, t_1) = L(i, t_0) + ({\sf in}(i) - {\sf out}(i)) (t_1 -
t_0)).$

\medskip
\noindent 
Let $\Phi(t) = \forall i (L(i, t) \leq L_{\sf overflow})$. Assume that 
$\forall i ({\sf in}(i) \geq 0 \wedge {\sf out}(i) \geq 0)$. 
We generate a formula which guarantees that $\Phi$ is an invariant 
using Steps 1-5 in the symbol elimination method in
Section~\ref{symbol-elimination} 
and Thm.~\ref{inv-trans-qe}. 
We start with the following formula (for simplicity of presentation 
we already replaced ${\sf
  in}(i)$ with ${\sf out}(i-1)$): 

\medskip
$\begin{array}{ll}
 & t_0 < t_1 \wedge (\forall i (L(i, t_0) \leq
L_{\sf overflow}) \wedge \exists j (L(j, t_1) > L_{\sf overflow})) \wedge \\[1ex]
& \forall i ((i = 1 \wedge L(1, t_1) = L(1,t_0) + ({\sf in} - {\sf out}(1))(t_1
- t_0)) \vee \\[1ex]
& ~~~~ (i > 1 \wedge L(i, t_1) = L(i,t_0) + ({\sf out}(i-1) - {\sf
  out}(i))(t_1 - t_0))). 
\end{array}$

\medskip
\noindent We use Algorithm~\ref{alg-symb-elim} with set of parameters $\Sigma_P
= \{ {\sf in}, {\sf out}, L_{\sf overflow} \}$. 

\begin{description}
\item[Step 1:] We skolemize (replacing $j$ with the constant $i_0$) 
and instantiate all universally quantified variables $i$ in the formula 
with $i_0$. 
After 
replacing $L(i_0, t_j)$ with  $c_j$,   
${\sf out}(i_0-1)$ with $d_1$, and ${\sf out}(i_0)$ with $d_2$  
we obtain: 

\medskip
$\begin{array}{ll}
 &  [t_0 < t_1 \wedge (c_0 \leq L_{\sf overflow}) \wedge c_1 > L_{\sf
    overflow}  \wedge \\[1ex]
& ((i_0 = 1 \wedge c_1 = c_0 + ({\sf in} - d_2)(t_1- t_0)) \vee (i_0 > 1 \wedge c_1 = c_0 + (d_1 - d_2)(t_1 - t_0)))]. \\
\end{array}$

\medskip

\item[Step 2:] We distinguish the following type of constants: $d_1,
  d_2$ introduced for terms starting with the parameter ${\sf out}$;
  $i_0$ occurring below a parameter; and  the remaining
  parameters $\{ {\sf in}, L_{\sf
    overflow} \}$ on the one hand, and $\{ t_0, t_1, c_0, c_1 \}$ the
  rest of constants, which are regarded as existentially quantified
  variables. 
We obtain: 
\end{description}

\smallskip
$\begin{array}{ll}
\exists t_0, t_1 \exists c_0, c_1  &  [t_0 < t_1 \wedge (c_0 \leq L_{\sf overflow}) \wedge c_1 > L_{\sf
    overflow}  \wedge \\[1ex]
& ((i_0 = 1 \wedge c_1 = c_0 + ({\sf in} - d_2)(t_1- t_0)) \vee (i_0 > 1 \wedge c_1 = c_0 + (d_1 - d_2)(t_1 - t_0)))]. \\[1ex]
\end{array}$

\begin{description} 
\item[Step 3:] We eliminate  $c_1$ and $c_0$ using quantifier
elimination and obtain: 

\noindent $\begin{array}{ll}
  & \exists t_0, t_1 [t_0 {<} t_1 \wedge ((i_0 = 1 \wedge 
 - ({\sf in} - d_2)(t_1- t_0) {<} 0) \vee 
(i_0 > 1 \wedge -(d_1 - d_2)(t_1 - t_0) {<} 0))]. 
\end{array}$

\noindent 
This is equivalent (after eliminating also $t_0, t_1$) with: 
 
$(i_0 = 1 \wedge ({\sf in} - d_2) > 0) \vee (i_0 > 1 \wedge 
(d_1 - d_2) > 0).$

\item[Step 4:] 
We replace $d_1, d_2$ back and regard $i_0$ as  existentially
quantified variable: 

$\exists i_0 ((i_0 = 1 \wedge 
({\sf in} - {\sf out}(i_0)) > 0) \vee (i_0 > 1 \wedge 
({\sf out}(i_0-1) - {\sf out}(i_0)) > 0)).$

\item[Step 5:] 
The negation is $\forall i ((i {=} 1 \rightarrow ({\sf in} {-} {\sf out}(i_0))
{\leq} 0) \wedge  (i {>} 1 \rightarrow 
({\sf out}(i{-}1) {-} {\sf out}(i)) {\leq} 0))$. \\
This condition guarantees that $\Phi$ is an invariant for the family of systems.
\end{description} 

\noindent 
Similar results can be obtained if updates are caused by changes in 
topology (insertion or deletion of water tanks in the system). 
}
\end{example}

\noindent We formally defined and studied a class of interconnected linear hybrid automata 
in \cite{damm-horbach-sofronie-frocos15} where we proved that locality results allow us to prove
``small model properties'' for such systems, and to reduce the
verification of general interconnected systems in the class to the
verification of a finite number of finite such systems.

\section{Conclusions}
\label{conclusions}
In this paper we studied certain classes of verification problems for 
parametric reactive and simple hybrid systems. 
We identified some deductive problems 
which need to be solved, and  
properties of the underlying theories which ensure that these 
verification problems are decidable. 
We gave examples of theories with the desired properties, and 
illustrated the methods on several examples. 

Parametricity in hybrid systems was addressed before 
in e.g.\ \cite{Henzinger,Platzer09,frehse,farn-wang}. 
Some approaches to invariant generation use a 
parametric form for the invariants and use constraint solving for 
generating invariants with a certain shape \cite{rybal,tiwari}. 
In all these approaches, the parameters are constants occurring in the 
description of the systems or in the invariants. 
In e.g.\ \cite{Ghilardi,Cimatti} also functions are used in the description 
of reactive or hybrid systems. In this paper we go one step further: we allow both 
functions and data to be parametric, and present ways of constructing 
(weakest) constraints on such parameters which guarantee safety -- 
which turns out to be very useful. 
We tackled  
some examples (e.g.\ a temperature controller in which the 
continuous variable is the temperature, 
and the evolution of the external temperature is a functional 
parameter) by 
generating abstractions 
and identified situations in which constraints on these parametric 
functions which {\em imply} safety can be derived. (We showed e.g.\ 
that the ``cooling'' state of the temperature controller 
is safe provided the outside temperature is lower than the 
interior temperature.) Since we use an abstraction,
we cannot always guarantee that these constraints are ``weakest''. 
We then analyzed the applicability of these ideas 
to increasingly more complex hybrid automata 
(parametric linear hybrid automata, parametric hybrid automata, 
interconnected families of hybrid automata). More details on such
problems can be found in \cite{damm-horbach-sofronie-frocos15}.
In recent work we analyzed the applicability of these ideas for
invariant generation in \cite{peuter-sofronie-cade2019}.

\medskip
\noindent {\bf Acknowledgments.} We thank Werner Damm for helpful
discussions on the verification of hybrid automata, and M{\u a}d{\u
  a}lina Era{\c s}cu for performing quantifier elimination for the
examples from \cite{sofronie-ijcar2010} using Mathematica, Redlog and
Qepcad. We thank the reviewers for their helpful comments.

\bibliographystyle{plain}


\end{document}